\documentclass[aps,prl,reprint,nofootinbib,superscriptaddress]{revtex4-2}
\usepackage[T1]{fontenc}
\usepackage{amsmath,amssymb,mathtools,bm}
\usepackage{mathrsfs}
\usepackage{microtype,xcolor,graphicx}
\usepackage[colorlinks=true,citecolor=blue!45!black,linkcolor=blue!45!black,urlcolor=blue!45!black]{hyperref}
\hypersetup{pdftitle={Quantum Cost of Emptiness in a One-Dimensional Bose Gas},pdfauthor={Bin Huang and Ziyan Zhu}}
\newcommand{\dd}{\,\mathrm d}
\newcommand{\ii}{\mathrm i}

\newcommand{\Tr}{\operatorname{Tr}}

\newcommand{\Mc}{M_c}

\newcommand{\Ret}{\mathsf T_L}

\newcommand{\XB}{X_{\mathrm B}}
\newcommand{\srcwin}{\mathfrak a}

\makeatletter
\@booleantrue\acknowledgments@sw
\makeatother
\begin{document}
\title{Quantum Cost of Emptiness in a One-Dimensional Bose Gas}
\author{Bin Huang}
\thanks{These authors contributed equally to this work}
\email[Email: ]{huangbin26@mails.ucas.ac.cn}
\affiliation{Institute of Theoretical Physics, Chinese Academy of Sciences, Beijing 100190, China}
\affiliation{University of Chinese Academy of Sciences, Beijing 100049, China}
\author{Ziyan Zhu}
\thanks{These authors contributed equally to this work}
\email[Email: ]{zhuziyan26@mails.ucas.ac.cn}
\affiliation{Institute of Theoretical Physics, Chinese Academy of Sciences, Beijing 100190, China}
\affiliation{University of Chinese Academy of Sciences, Beijing 100049, China}
\begin{abstract}
The probability \(P_0(R)\) of an empty interval of radius \(R\) probes the far tail of density fluctuations in the thermodynamic ground state of the repulsive Lieb--Liniger gas. Here \(\rho\) is the bulk density and \(c\) the repulsion. As \(\ell = \rho R \to \infty\) at fixed \(\gamma = c/\rho > 0\), we find \(-\log P_0 = f(\gamma)\ell^2 + B(\gamma)\sqrt{\ell} - \frac{1}{4}\log\ell + O_\gamma(1)\), with no linear term and \(B = 2\zeta(3/2)\sqrt{f/(\pi\gamma)}\). Exact Gaussian completion of the auxiliary-field determinant reduces the leading collective response to an electrostatic slit problem, yielding \(f(\gamma)\) in elliptic integrals. Its dipole polarization tail drives auxiliary scattering returns whose local density saturates while the active rapidity range expands as \(\sqrt{R}\), producing the square-root boundary cost. The full return spectrum and the occupied-rapidity endpoints together fix the \(-1/4\) logarithmic coefficient. Through the contact identity, widening the void probes a dense healing scale at the boundary whose growing characteristic occupation supports an emergent classical-field description even in a strongly interacting bulk.

\end{abstract}
\maketitle
\emph{Introduction.---}A macroscopic empty interval in the ground state of the repulsive Lieb--Liniger gas probes a collective density fluctuation beyond harmonic-fluid theory
\cite{Haldane1981,CazalillaEtAl2011,AbanovKorepin2002}.
Its probability measures the cost of excluding an extensive number of particles from an otherwise homogeneous fluid.
Measurements of density fluctuations and atom-resolved correlations in cold gases
\cite{EsteveEtAl2006,ArmijoEtAl2010,Jacqmin2011,YaoEtAl2025}, together with counting-statistics results for short intervals
\cite{BastianelloPiroliCalabrese2018} and weakly interacting thermal gases
\cite{ArzamasovsGangardt2019}, motivate a quantitative description of rare number fluctuations.
For a large empty interval, this requires understanding how interactions determine both the leading collective cost and the corrections associated with its boundaries.

In the impenetrable Tonks--Girardeau limit, Bose--Fermi mapping gives a sine-kernel determinant that determines the quadratic evacuation cost and its logarithmic correction
\cite{Girardeau1960,DeiftItsKrasovskyZhou2006}.
Hydrodynamic instantons \cite{Abanov2005} determine the leading quadratic cost in weakly interacting Bose and polytropic liquids
\cite{YehKamenev2020,YehGangardtKamenev2022,AbanovGangardt2025}.
Weak-coupling numerics and a perturbative argument also indicated the absence of a linear term
\cite{YehKamenev2020}, while quantum-pressure estimates suggested logarithmic corrections
\cite{YehGangardtKamenev2022}.
Singular endpoints of the dispersionless instanton complicate these estimates.
At general finite repulsion, a microscopic determination of the leading rate and the order of its first correction therefore remains essential.

The auxiliary-field Fredholm representation derived by Korepin, Its, and Waldron provides a microscopic starting point; they also conjectured its large-interval asymptotics
\cite{KIW1995,KBI1993}.
Extracting the probability from this representation requires controlling the field average: terms subleading at fixed fields can become leading after vacuum averaging, as found for quantum nonlinear Schr\"odinger two-point functions
\cite{Slavnov1998}.
More recently, Khanikati and Abanov used a joint dual-field saddle to conjecture a leading-rate integral equation, and separately proposed the absence of a linear term
\cite{KhanikatiAbanov2026}.
They tested the rate against weak- and strong-coupling expansions and hydrodynamic numerics, while their microscopic derivation assumes uniform determinant asymptotics.
Solving the leading-rate equation explicitly at general coupling and controlling the remaining field average to determine the subquadratic terms therefore remain open.

{\clubpenalty=10000
Here we obtain an explicit elliptic-integral expression for the leading rate \(f(\gamma)\) at fixed finite repulsion.
We also determine the large-interval expansion through logarithmic order, with no linear term, a square-root correction whose amplitude is fixed by \(f(\gamma)\) and the bulk coupling, and a coupling-independent logarithmic coefficient of \(-1/4\), leaving an \(O(1)\) remainder.
Correlated Gaussian completion identifies an auxiliary charge response, which we solve by an electrostatic image construction.
The same charge moment determines the leading rate and the polarization tail governing the collective return spectrum, linking the quadratic and square-root costs.
An exact boundary-density identity connects this response to the cost of enlarging an existing void and identifies a dense healing scale with increasing occupation and decreasing effective coupling.
The resulting expansion differs from that obtained by taking the impenetrable limit first.
Finite-window determinant calculations agree closely with the quadratic and square-root coefficients and, upon subtracting the exact leading terms, yield conditional logarithmic estimates near \(-1/4\), with finite-window shifts.
\par}

\emph{The emptiness law.---}We use the repulsive Lieb--Liniger Hamiltonian
\cite{LiebLiniger1963}
\begin{equation}
 H=\int\dd x\left[\frac12\partial_x\psi^\dagger\partial_x\psi
             +\frac c2(\psi^\dagger)^2\psi^2\right],\qquad c>0 .
 \label{eq:H}
\end{equation}
Here \(\hbar=m=1\), and the Bose fields obey
\([\psi(x),\psi^\dagger(y)]=\delta(x-y)\).
The bulk density is \(\rho\), and \(\gamma=c/\rho\).
For \([-R,R]\), define
\(\Pi_R=\mathbf1_{\{\widehat N_{[-R,R]}=0\}}\),
\(P_0(R)=\langle\Pi_R\rangle_0=\langle:\!e^{-\widehat N_{[-R,R]}}\!:\rangle_0\) and
\(S(\ell;\gamma)=-\log P_0(\ell/\rho)\), where \(\ell=\rho R\), \(\widehat N_I=\int_I\psi^\dagger\psi\,\dd x\) counts particles
and the expectation is in the ground state. Normal ordering enforces the zero-particle
projector. Taking the thermodynamic ground-state limit at fixed
\(L=2R\), then \(\ell=\rho R\to\infty\) at fixed \(c,\rho>0\),
we obtain the following asymptotic expansion. The coefficients
\(f,A,B\) depend on \(\gamma\), and \(\zeta\) is the Riemann zeta function:
\begin{align}
 S(\ell;\gamma)&=f\ell^2+A\ell+B\sqrt\ell-\tfrac14\log\ell+O_\gamma(1),\nonumber\\
 A&=0,\qquad B=\frac{2\zeta(3/2)}{\sqrt\pi}\sqrt{\frac f\gamma}.
 \label{eq:main}
\end{align}
The rate \(f(\gamma)\) is given explicitly in Eq.~\eqref{eq:explicitrate}. The vanishing linear term
rules out a separate cost per excluded particle, while \(f\) and
\(\gamma\) fix the growing boundary cost. The logarithmic coefficient
is coupling independent at finite repulsion.

\emph{From the exact field average to a charge problem.---}The ground-state
rapidities \(\lambda\) occupy \([-q,q]\); their density \(\sigma_q\) integrates to
\(\rho\) and fixes \(q\) through the Lieb equation.
For an auxiliary rapidity source \(j\), write
\(\mathcal D_c[j]=\det(I+V_c[j])\). At each fixed \(L\), we realize the
dual-field average of Ref.~\cite{KIW1995} as an expectation over a
centered real Gaussian field, obtaining
\begin{equation}
 P_0(R)=\frac{\mathbb E_C\mathcal D_c[\ii L\lambda+\phi]}{\mathcal D_c[0]}.
 \label{eq:microscopic}
\end{equation}
The field \(\phi\) represents scattering factors. Its anchored covariance
\(C(\lambda,\mu)=C_0(\lambda-\mu)-C_0(\lambda)-C_0(\mu)\) is generated by
\(C_0(p)=-\log(1+p^2/c^2)\). The operator \(V_c\) on \([-q,q]\) retains both shifts
\(\pm\ii c\), including their common diagonal. The positive Gaudin
determinant \(\mathcal D_c[0]\) fixes the normalization. At zero
length the quotient equals one, fixing \(P_0(0)=1\)
(SM \cite{SM}, Sec.~S1).

To isolate the collective response before analyzing the boundary,
we introduce a reference kernel obtained by taking the infinite-coupling
limit at fixed \(q\) and source \(j\), retaining their values from the
finite-coupling problem. Its deformed sine-kernel determinant is
\(\mathcal D_{\rm ref}[j]=\det(I+V_{\rm ref}[j])
= e^{Q[j]}\mathcal W_{\rm ref}[j]\).
Integrating the scalar term in its exact source variation from zero
source gives \(Q[j]=\tfrac12\langle j,K_0^{-1}j\rangle\), where
\(K_0(p)=-2\log|p|\) acts on neutral charges. The form is continued
bilinearly to complex sources: \(Q[\ii p]=-Q[p]\), hence
\(Q[\ii L\lambda]=-L^2Q[\lambda]<0\).
Writing \(\mathcal D_c=e^Q\mathcal W_c\) in Eq.~\eqref{eq:microscopic}
exposes the quadratic field weight \(e^{Q[\phi]}\), which changes the
covariance to \(\Sigma\). Completing the linear cross term also shifts
the residual source to \(\ii Lp_*+\eta\), with \(\eta\) centered and of
covariance \(\Sigma\). Factoring \(L\) from the phase \(\ii L\lambda\)
leaves the unit driving \(\lambda\). For its induced auxiliary response
charge \(h_c\), the dressed source obeys
\begin{equation}
 p_*=\lambda+Ch_c=K_0h_c.
 \label{eq:dressed}
\end{equation}
In neutral pairings, the anchoring terms and additive constants drop out,
giving \((K_0-C)h_c=\mathcal L_ch_c=\lambda\), with
\(\mathcal L_c(p)=\log(1+c^2/p^2)\).
The two-body Bethe phase-shift covariance thus screens the
infinite-coupling logarithmic Coulomb interaction. The resulting kernel
is the potential of a charge on a slit at height \(c/2\) above a grounded
image plane.
The even kernel and odd driving give an odd, neutral response. On real neutral
finite-energy charges supported in \([-q,q]\), the positive interaction
defines the strictly concave functional

\begin{align}
 \mathcal F[h]&=\int_{-q}^q\lambda h(\lambda)\dd\lambda\nonumber\\
 &\quad-\frac12\int_{-q}^q\!\int_{-q}^q
 h(\lambda)\mathcal L_c(\lambda-\mu)h(\mu)\dd\lambda\dd\mu.
 \label{eq:response}
\end{align}
Its unique maximum solves the charge equation:
\begin{align}
 \int_{-q}^q\mathcal L_c(\lambda-\mu)h_c(\mu)\dd\mu&=\lambda,\nonumber\\
 F_c=\mathcal F[h_c]&=\frac12\int_{-q}^q\lambda h_c(\lambda)\dd\lambda.
 \label{eq:charge}
\end{align}
The charge equation holds for \(|\lambda|<q\). The normalization \(f=4F_c/\rho^2\) makes \(F_cL^2=f\ell^2\),
in agreement with Eqs.~(3)--(5) of Ref.~\cite{KhanikatiAbanov2026}.
The full Gaussian completion (End Matter A) now reads
\begin{equation}
 P_0(R)=\frac{Z_C}{\mathcal D_c[0]}e^{-F_cL^2}\,
 \mathbb E_\Sigma\mathcal W_c[\ii Lp_*+\eta],
 \label{eq:fullcompletion}
\end{equation}
where \(Z_C\) is the source-independent Gaussian normalization.
The charge moment supplies the quadratic exponent; the retained average
determines the corrections to this term.

\emph{The image solution.---}Equation~\eqref{eq:charge} asks for a
charge distribution with prescribed potential \(\lambda\) on a finite
slit. Introduce the complex rapidity coordinate \(z=x+\ii y\).
A source at \(\mu+\ii c/2\) and its opposite image at
\(\mu-\ii c/2\) produce
\(2\log(r_{\rm image}/r_{\rm source})=\mathcal L_c(\lambda-\mu)\)
at \(z=\lambda+\ii c/2\). Integrating these pairs against \(h_c\)
turns the integral equation into a Dirichlet problem: the harmonic
potential \(\varphi\) vanishes on the real axis and equals \(x\) on the slit.

Oddness also makes the imaginary axis an equipotential. In the first
quadrant, subtract the screening response from the imposed field:
\(\Psi=2xy/c-\varphi\). The quadrant edges and the horizontal fin
\(0<x<q,\ y=c/2\) all have \(\Psi=0\), while
\(\Psi\sim2xy/c\) at infinity. The physical data \(c,q\) fix its
height and length.

\begin{figure}[t]
 \centering
 \typeout{FIG1-ACTUAL-COLUMNWIDTH=\the\columnwidth; TEXTWIDTH=\the\textwidth; TEXTHEIGHT=\the\textheight}
 \includegraphics[width=\columnwidth]{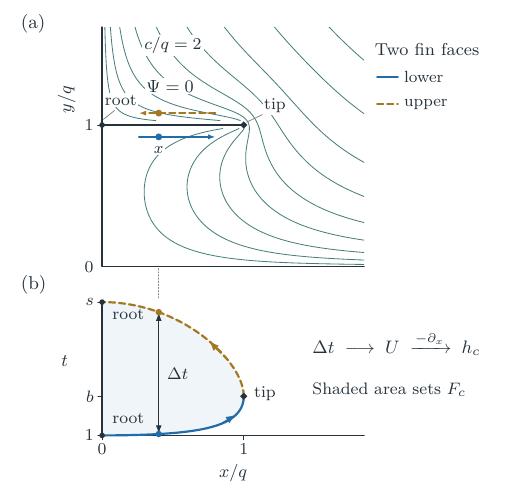}
 \caption{Image geometry and auxiliary charge at $c/q=2$.
(a) The grounded fin has two faces; markers denote the same $x$,
offset for visibility. (b) Their inverse-map separation $\Delta t=t_+-t_-$
determines the primitive $U$ of the auxiliary charge $h_c=-U'$.}
 \label{fig:images}
\end{figure}

A Schwarz--Christoffel map \(z(t)\) from the lower half \(t\)-plane solves
the fin geometry in Fig.~\ref{fig:images}(a), sending \(t=1,b,s\)
(\(1<b<s\)) to the lower-face root, tip, and upper-face root. The normal-field
jump relates the inverse-branch separation \(t_+-t_-\)
[Fig.~\ref{fig:images}(b)] to the charge primitive
\(U(\lambda)=-\int_{-q}^{\lambda}h_c(\mu)\dd\mu\).
Its area gives \(F_c=\int_0^q U(x)\dd x\), with full-slit integral
\(2F_c\). Closure makes both roots coincide.
Write \(m_{\rm e}=1-s^{-1}\), \(b=sE(m_{\rm e})/K(m_{\rm e})\), and
\(\alpha=\arcsin\sqrt{(b-1)/(s-1)}\). The aspect ratio determines \(s\) through
\begin{equation}
 \frac cq=\frac{\pi\sqrt s}
 {K(m_{\rm e})[bF(\alpha|1-s)-E(\alpha|1-s)]}.
 \label{eq:imageaspect}
\end{equation}
Here \(F,E\) use Legendre's elliptic parameter convention, with
\(K(m)=F(\pi/2|m)\) and \(E(m)=E(\pi/2|m)\).
Integrating the branch separation gives the explicit rate
\begin{equation}
 f(\gamma)=\frac{\gamma^2K(m_{\rm e})^4}{3\pi^4s^2}
              [2b(1+s)-s-3b^2].
 \label{eq:explicitrate}
\end{equation}
With \(q\) fixed by the Lieb equation, this gives \(f(\gamma)\) for
all \(\gamma>0\), with limits
\begin{equation}
 f\sim\frac{16\sqrt\gamma}{3\pi}\ (\gamma\to0),\qquad
 f\to\frac{\pi^2}{2}\ (\gamma\to\infty).
 \label{eq:limits}
\end{equation}
The full crossover curves are shown in SM Fig.~S1.
These limits reproduce the weak-coupling hydrodynamic result
\cite{YehGangardtKamenev2022,AbanovGangardt2025} and the
Tonks--Girardeau sine-kernel result \cite{Girardeau1960,DeiftItsKrasovskyZhou2006}.
The conformal map and charge moment are derived in End Matter B.

\emph{Polarization and the return scale.---}The same charge also fixes
the boundary scale through its distant polarization.
At stationarity, \(\langle h_c,\mathcal L_ch_c\rangle=\langle\lambda,h_c\rangle\):
half the driving work is stored in the electrostatic response.
Neutrality makes the first moment, \(2F_c\), control the exterior tail.
For \(|\lambda|>q\), the exterior logarithmic potential gives
\(J_{\rm tail}=-\tfrac12\partial_\lambda(K_0h_c)\).
Since \(h_c=-U'\) and the positive primitive vanishes at \(\pm q\),
integration by parts gives

\begin{align}
 \Mc&=\int_{-q}^q U(\mu)\dd\mu=\frac{\rho^2 f}{2},\nonumber\\
 J_{\rm tail}(\lambda)&=\int_{-q}^q\frac{U(\mu)\dd\mu}{(\lambda-\mu)^2}
 =\frac{M_c}{\lambda^2}+O(\lambda^{-4}).
 \label{eq:tail}
\end{align}
Local saturation for \(cLJ_{\rm tail}\gg1\)
[Eq.~\eqref{eq:localdensity}] begins at \(cLJ_{\rm tail}\sim1\), setting
\begin{equation}
 \lambda_* = \sqrt{c\rho^2 fR},\qquad
 N_*:=\frac{\lambda_*}{c}=\sqrt{\frac{f\ell}{\gamma}}.
 \label{eq:returnscale}
\end{equation}
The dimensionless range \(N_*\) grows as \(\sqrt R\) at every fixed finite coupling.
For Fourier wave number \(k\) conjugate to rapidity, the scattering kernel
\(P_c=(2\pi)^{-1}\partial_c\mathcal L_c\) has the Fourier pair
\begin{equation}
 P_c(p)=\frac{c}{\pi(p^2+c^2)},\qquad
 \widehat P_c(k)=e^{-c|k|}.
 \label{eq:poissonbridge}
\end{equation}

\emph{Saturation and collective returns.---}To analyze the remaining
factor in Eq.~\eqref{eq:fullcompletion}, first divide
\(\mathcal D_c[\ii Lp_*]\) by its reference determinant, isolating
\(\det(I-\mathsf T_L)\).
The positive contraction \(\Ret\) describes propagation across the
imaginary-rapidity separation \(c\) by Eq.~\eqref{eq:poissonbridge},
followed by projection onto the polarized
response subspace, retaining both channels and the endpoint background.
Writing \(\mathcal W_{\rm ref}[j]=e^{-Q[j]}\mathcal D_{\rm ref}[j]\),
restoring the Gaussian average gives

\begin{equation}
 P_0(R)=\frac{Z_C}{\mathcal D_c[0]}e^{-F_cL^2}e^{r_{\rm end}(L)}
 \det(I-\mathsf T_L)Z_{\rm G}(L).
 \label{eq:separation}
\end{equation}
Here \(r_{\rm end}=\log\mathcal W_{\rm ref}[\ii Lp_*]\), and
\(Z_{\rm G}=\mathbb E_\Sigma\mathcal W_c[\ii Lp_*+\eta]/
\mathcal W_c[\ii Lp_*]\) retains the residual field average.
The return cost is \(S_{\rm ret}=-\log\det(I-\mathsf T_L)\)
(SM, Secs.~S5--S6
\cite{ItsKozlowski2013,GIK2019}).

In the leading exterior phase-space approximation, \(LJ_{\rm tail}\)
sets the local wave-number range, while \(cLJ_{\rm tail}\) controls
saturation. With \(k\) conjugate to rapidity and \(m\) counting repeated
returns, the local logarithmic return density is
\begin{align}
 \mathcal F_R(\lambda)
 &=\frac1\pi\int_0^{LJ_{\rm tail}(\lambda)}
                    \log(1-e^{-ck})\dd k\nonumber\\
 &=-\frac1{\pi c}\sum_{m\ge1}
        \frac{1-e^{-mcLJ_{\rm tail}(\lambda)}}{m^2}.
 \label{eq:localdensity}
\end{align}
The nonpositive density \(\mathcal F_R\) enters \(S_{\rm ret}\)
with the opposite sign. At fixed exterior rapidity it saturates to
\(-\pi/(6c)\), so increasing \(R\) expands the active region rather
than its saturated local magnitude. The inverse-square dipole tail
sets the crossover at \(\lambda_*\propto\sqrt R\), producing a square-root
cost instead of a cost proportional to the hole length. A matching scale
\(\max(c,q)\ll\Lambda\ll\lambda_*\) separates the inner and exterior
approximations; their common overlap must be subtracted.
In \(\log\det(I-\mathsf T_L)\), the inner contribution and exterior
overlap subtraction cancel at each return order, summing to
\(-\pi\Lambda/(3c)\) and \(+\pi\Lambda/(3c)\), respectively.
The remaining range is \(\lambda\sim\lambda_*\propto\sqrt R\).

The matched leading term is the full-line integral of the equal-mass
point-charge model, \(J_{\rm tail}\sim M_c/\lambda^2\).
Each return order contributes
\begin{equation}
 \int_{\mathbb R}\!\bigl(1-e^{-mcLM_c/\lambda^2}\bigr)\dd\lambda
       =2\sqrt{\pi mcLM_c}.
 \label{eq:cyclemoment}
\end{equation}
Combining this \(m^{1/2}\) factor with \(m^{-2}\) in
Eq.~\eqref{eq:localdensity} gives
\(2\zeta(3/2)\sqrt{LM_c/c}/\sqrt\pi=2\zeta(3/2)N_*/\sqrt\pi\).
The connected logarithm weights each closed cycle by \(1/m\),
and wave-number integration supplies the
second inverse power. Thus \(\zeta(3/2)\) sums all collective returns.
The logarithmic correction depends on the long-cycle spectrum,
including its boundary conditions.

We establish both the square-root and logarithmic terms rigorously from the complete return operator. Removing
the reference-resolvent matrix dressing and replacing the charge by an
equal-mass point charge yields two scalar channels. Inversion and Fourier
transformation reduce them to Bessel heat propagation, regular at the
origin and projected onto \((0,2N_*)\). These steps and a cycle-summable
comparison with continuous absorption change the cost by \(O(1)\),
preserving the logarithm. With absorption at \(2N_*\), the comparison
spectrum is set by the positive zeros \(j_{0,n}\) of \(J_0\),
whose asymptotics are \(j_{0,n}=\pi(n-\tfrac14)+O(n^{-1})\).
Retaining this shift in the spectral sum gives (\mbox{End Matter C})
\begin{equation}
 S_{\rm ret}=\frac{2N_*}{\sqrt\pi}\zeta(3/2)-\log N_*+O_{c,q}(1).
 \label{eq:zeta}
\end{equation}
The original occupied-rapidity endpoints supply
\(-r_{\rm end}=+\tfrac14\log\ell+O(1)\); combined with
\(-\log N_*=-\tfrac12\log\ell+O(1)\), they give
\(-\tfrac14\log\ell\). For the charge profile \(h_c\), the positive
mixed-source representation bounds the remaining average by
\(0\le-\log Z_{\rm G}\le C_{c,q}/L\). Combining these terms in
Eq.~\eqref{eq:separation} gives the square-root and logarithmic
corrections, with \(A=0\) and an \(O(1)\) remainder. Saturation and the full return spectrum explain
this replacement of a linear cost; the Calogero--Sutherland gas,
by comparison, has a linear term at generic coupling \cite{Abanov2005}.

\emph{Physical response and emergent weak coupling.---}The return
calculation identifies \(N_*\) as the scale of the boundary cost.
To establish its physical meaning, we examine the conditional cost of
enlarging an existing void. Increasing the total empty length by
\(\Delta L\) changes the leading quadratic cost by
\(f\rho^2R\,\Delta L+O((\Delta L)^2)\). Through the contact identity
below, this increment identifies a boundary density scale and its
associated healing length. The same \(f(\gamma)\) thus fixes both the
rapidity return range and the real-space contact response:
\begin{align}
 n_*&=f\rho^2R,\qquad d_*=(cn_*)^{-1/2}=\lambda_*^{-1},\nonumber\\
 n_*d_*&=N_*,\qquad \gamma_*:=\frac{c}{n_*}
             =\frac{\gamma}{f\ell}=N_*^{-2}.
 \label{eq:front}
\end{align}
Extending the empty interval by total length \(d_*u\) gives
\begin{equation}
 -\log\frac{P_0(R+d_*u/2)}{P_0(R)}=N_*u+o(N_*),
 \label{eq:hardshape}
\end{equation}
for bounded fixed \(u\ge0\), by finite differences of
Eq.~\eqref{eq:main}. Translation invariance identifies the ratio with
the conditional probability of additional exterior gaps, giving
\(n_*,d_*\) their physical meaning as void-extension scales.
To relate \(n_*\) to the conditional boundary density, we use the
general void/conditional-intensity identity \cite{TorquatoEtAl2008},
which here reads \(\rho\,\partial_\ell S(\rho r;\gamma)=2n_r(r^+)\), where \(n_r\) is the density
conditioned on the radius-\(r\) void.

With \(d_*\) fixed at the base radius \(R\), set \(R_u=R+d_*u/2\).
For every continuous compactly supported
test function~\(\chi\),
\begin{equation}
 \int\!\chi(u)\frac{n_{R_u}(R_u^+)}{n_*}\dd u
 \longrightarrow\int\!\chi(u)\dd u.
 \label{eq:edgedensity}
\end{equation}
Convexity of \(S\), proved from Weyl-alcove log-concavity
(SM, Sec.~S7), strengthens this relation to the pointwise contact limit
\begin{equation}
 \frac{n_R(R^+)}{n_*}\longrightarrow1.
 \label{eq:pointwise-contact}
\end{equation}
Indeed, the convex secant bounds applied to \(S(\ell)=f\ell^2+o(\ell^2)\)
give \(\partial_\ell S/(2f\ell)\to1\) (SM, Sec.~S7).

With the density and length scales identified by this contact response,
we express the physical Bose field in the corresponding boundary units.
For dimensionless positions \(X,Y\),
\(\psi_*(X)=\psi(R+d_*X)/\sqrt{n_*}\) obeys
\begin{equation}
 [\psi_*(X),\psi_*^\dagger(Y)]=N_*^{-1}\delta(X-Y).
 \label{eq:commutator}
\end{equation}
Since the density scale \(n_*\propto R\) rises faster than the healing
length \(d_*\propto R^{-1/2}\) contracts, the characteristic occupation
\(N_*=n_*d_*\propto\sqrt R\) grows with the void. The ratio
\(c/\lambda_*=N_*^{-1}\) controls both the rescaled commutator and the
separation of scattering and active-rapidity scales, while the local
coupling \(\gamma_*=N_*^{-2}\to0\).

\begin{figure}[t]
 \centering
 \includegraphics[width=\columnwidth]{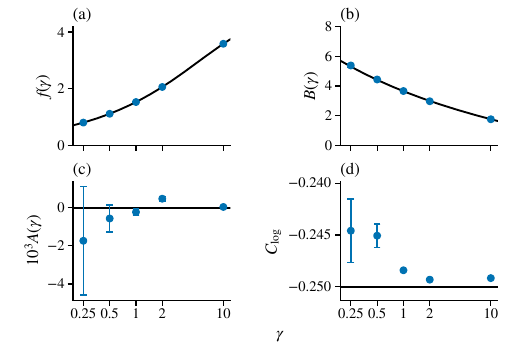}
 \caption{Analytic predictions (black) and finite-window fit estimates
(blue) at $\gamma=0.25,0.5,1,2,10$ on logarithmic horizontal axes.
(a)--(c) Nine-window free fits at $\ell=64,80,96,112,128,160,192,224,256$.
(d) Four-window conditional fits with
analytic $f,A=0,B$ fixed; fit details are given in the text.
Bars are empirical MC standard errors, excluding systematic and
correction-model uncertainties.}
 \label{fig:finitewindow}
\end{figure}

\emph{Finite-window comparison.---}We evaluate \(P_0(R)\) by Monte Carlo averaging
Eq.~\eqref{eq:microscopic} and fit the resulting emptiness cost over nine windows,
\(\ell=64,\allowbreak80,\allowbreak96,\allowbreak112,\allowbreak128,\allowbreak160,\allowbreak192,\allowbreak224,\allowbreak256\).
The fit includes \(f\ell^2+A\ell+B\sqrt\ell+C_{\log}\log\ell+C_0+C_1/\sqrt\ell\),
with all six coefficients free. Across \(\gamma=0.25,0.5,1,2,10\),
the fitted \(f\) differs from the image prediction by less than \(0.80\) parts per million,
while \(B\) differs by less than \(1.44\%\) [Fig.~\ref{fig:finitewindow}(a)--(c)].
These fits compare the bulk and boundary coefficients without imposing their predicted relation.

{\widowpenalty=10000
To extract the logarithmic coefficient, we fix the analytic \(f,A=0,B\)
and fit \(C_{\log},C_0,C_1\) over the four largest windows,
\(\ell=160,192,224,256\). The estimates lie near \(-1/4\) [Fig.~\ref{fig:finitewindow}(d)].
Error bars show propagated MC statistical uncertainty; corrections omitted from the fitting form
can produce additional finite-window shifts. The numerical procedure, convergence checks,
and window dependence are detailed in the SM.
\par}

\newpage

\emph{Discussion.---}The elliptic solution makes the leading evacuation rate \(f(\gamma)\) explicit at arbitrary fixed finite repulsion.
The ratio \(c/q\) fixes the auxiliary electrostatic geometry and its response energy, while the Lieb equation supplies the equilibrium normalization.
Together they determine the dominant void cost across the interaction crossover, connecting the weak-coupling hydrodynamic and Tonks--Girardeau limits.

The same charge moment fixes the inverse-square polarization tail and hence \(B(\gamma)\), without an independent boundary parameter.
Despite having only two spatial endpoints, the void drives collective returns over an expanding rapidity range \(\lambda_*\propto\sqrt R\).
The local return density saturates, while this growing range produces a square-root cost between the quadratic and logarithmic orders.
This mechanism shows why a correction near singular instanton endpoints depends on both its explicit prefactor and the range over which the response accumulates.
The square-root term revises earlier weak-coupling estimates of logarithmic corrections \cite{YehKamenev2020,YehGangardtKamenev2022}, while preserving the leading hydrodynamic cost.
Although subleading, it suppresses the emptiness probability by \(\exp[-B\sqrt\ell]\).

The boundary-density identity gives the leading rate a further physical role: it determines the response to enlarging an existing void through the scales \(n_*=f\rho^2R\) and \(d_*=(cn_*)^{-1/2}\).
Their growing characteristic occupation \(N_*=n_*d_*\) controls both the boundary cost and the conditional cost of adding a total empty length \(d_*\), suggesting a dense healing region at the void boundary.
The decreasing local coupling \(\gamma_*=N_*^{-2}\) and rescaled field commutator proportional to \(N_*^{-1}\) favor an emergent classical-field description, even in a strongly interacting bulk.
This contrasts with a dilute edge in a smooth trap, where decreasing density increases the local coupling and leads to free-fermion edge behavior \cite{Stephan2019}.

{\clubpenalty=10000
These asymptotics require \(\lambda_*\gg\max(c,q)\).
At weak coupling, this becomes \(\ell\sqrt\gamma\gg1\), placing the void beyond the bulk healing length, with \(B\sim8\zeta(3/2)\gamma^{-1/4}/(\sqrt3\,\pi)\) and \(B\sqrt\ell/(f\ell^2)\propto(\ell\sqrt\gamma)^{-3/2}\ll1\).
Decreasing the interaction enlarges that length and shifts a fixed interval toward the Poisson regime of the Poisson-to-Gaussian crossover \cite{YehKamenev2020}.
At strong coupling, \(\ell\gg\gamma\) and \(B\sim\sqrt{2\pi}\,\zeta(3/2)\gamma^{-1/2}\): increasing repulsion reduces the amplitude and delays the square-root regime to larger voids.
Taking the impenetrable limit first instead gives \(S_{\mathrm{TG}}=(\pi^2/2)\ell^2+\tfrac14\log(\pi\ell)+O(1)\), without a square-root term and with the opposite logarithmic coefficient \cite{DeiftItsKrasovskyZhou2006}.
Thus, a leading evacuation rate close to its Tonks value does not imply a fermionized boundary response: at any fixed finite repulsion, a sufficiently large void develops a locally weak-coupling contact scale.
\par}

\par\smallskip
\begin{acknowledgments}
Under the authors' direction, OpenAI's GPT-6 Astra assisted with
manuscript preparation, analytical derivations and proof checks, and
cross-checks of the analytic and numerical coefficient evaluations.
The authors independently verified all analytical identities, operator
bounds, and high-precision numerical results, and take full
responsibility for the content of the paper.
\end{acknowledgments}
\par\smallskip
\section*{Data availability}The accompanying data and code package
contains the finite-window estimates and empirical Monte Carlo covariance
matrices underlying Fig.~2, tabulated analytic coefficients, and the
fitting and plotting scripts. Sec.~S8 of the Supplemental Material
describes the sampling and fitting procedures.

\clearpage
\appendix
\section*{End Matter}
\par\medskip
\setcounter{equation}{0}
\renewcommand{\theequation}{A\arabic{equation}}
\renewcommand{\theHequation}{appendix.A.\arabic{equation}}
\phantomsection\label{app:response}
\emph{Appendix A: Reference response and full Gaussian completion.---}
The Lieb equation and normalization fix \(q\):
\begin{equation}
 \sigma_q(\lambda)-\int_{-q}^q
 \frac{c\,\sigma_q(\mu)\dd\mu}{\pi[c^2+(\lambda-\mu)^2]}
 =\frac1{2\pi}.
 \label{eq:ELieb}
\end{equation}
Here \(\int_{-q}^q\sigma_q\dd\lambda=\rho\).
On \(L^2([-q,q])\), with \(d=\lambda-\mu\) and
\(\Delta j=j(\lambda)-j(\mu)\), the finite-coupling kernel is
\[
 V_c[j](\lambda,\mu)=-\frac{c}{2\pi d}
 \left[\frac{e^{\Delta j/2}}{d+\ii c}
             +\frac{e^{-\Delta j/2}}{d-\ii c}\right].
\]
Its confluent diagonal is
\(V_c[j](\lambda,\lambda)=\ii j'(\lambda)/(2\pi)-1/(\pi c)\),
and \(V_c[0]=-P_c\), with \(P_c\) from Eq.~\eqref{eq:poissonbridge}.
The infinite-coupling limit at fixed \(q,j\) gives
\(V_{\rm ref}[j](x,y)=\ii\sinh([j(x)-j(y)]/2)/[\pi(x-y)]\).
Modulo constants, write
\(j(q\cos\theta)=\sum_{n\ge1}a_n\cos(n\theta)\) and
\(g_j(q\cos\theta)=\sum_{n\ge1}a_n\sin(n\theta)\).
The exact source identity \cite{GIK2019} gives the scalar variation
\(-\int\delta j\,g_j'/(2\pi)=\tfrac14\sum n a_n\delta a_n\), hence
\begin{equation}
 \begin{gathered}
 Q[j]=\frac18\sum_{n\ge1}na_n^2
     =\frac12\langle j,K_0^{-1}j\rangle,\\
 \mathcal W_c[j]=e^{-Q[j]}\mathcal D_c[j].
 \end{gathered}
 \label{eq:Equadratic}
\end{equation}
Here \(K_0(d)=-2\log|d|\) acts on neutral charges and
\(Q\) is bilinear, so \(Q[\ii p]=-Q[p]\).
With \(M_f\) denoting multiplication by \(f\), the exact reference remainder is
\[
 \begin{gathered}
 \mathcal W_{\rm ref}[j]=\det(I+M_{e^{\ii g_j}-1}\mathsf K_*),\\
 \mathsf K_*(x,y)=\frac{1}{2\pi^2}
 \frac{\log\frac{q+x}{q-x}-\log\frac{q+y}{q-y}}{x-y},
 \qquad 0\le\mathsf K_*\le\tfrac12 I.
 \end{gathered}
\]
The diagonal is continuous; the weighted determinant is trace class
(SM, Sec.~S2.2).

The anchored covariance
\(C(\lambda,\mu)=C_0(\lambda-\mu)-C_0(\lambda)-C_0(\mu)\),
\(C_0(d)=-\log(1+d^2/c^2)\), obeys \(K_0-C=\mathcal L_c\)
in neutral pairings, up to constants.
Tilting by \(\dd P_\Sigma=Z_C^{-1}e^{Q[\phi]}\dd P_C\),
\(Z_C=\mathbb E_Ce^{Q[\phi]}\), and completing the cross term gives
\begin{align}
 \Sigma&=(C^{-1}-K_0^{-1})^{-1},\nonumber\\
 p_*&=\lambda+\Sigma K_0^{-1}\lambda=\lambda+Ch_c,\nonumber\\
 h_c&=K_0^{-1}p_*=\mathcal L_c^{-1}\lambda.
 \label{eq:completion}
\end{align}
For the full centered field \(\eta\) of covariance \(\Sigma\),
\begin{equation}
 P_0(R)=\frac{Z_C}{\mathcal D_c[0]}e^{-F_cL^2}
       \mathbb E_\Sigma\mathcal W_c[\ii Lp_*+\eta].
 \label{eq:Efullcompletion}
\end{equation}
The Gaussian characteristic identity applied to the positive source
representation below retains the cross covariance and all field modes.
The inverses in \eqref{eq:completion} are defined first on finite projections.
For neutral charges \(h=-U'\) of finite energy, with \(U\) extended by zero,
\(\langle h,\mathcal L_ch\rangle\ge
\tfrac12(1-e^{-c/q})\langle h,K_0h\rangle\).
This strict margin and analytic covariance decay make \(Z_C\) finite,
permit the projection limit, and fix the unique response (SM, Sec.~S3).

\par\medskip
\setcounter{equation}{0}
\renewcommand{\theequation}{B\arabic{equation}}
\renewcommand{\theHequation}{appendix.B.\arabic{equation}}
\phantomsection\label{app:image}
\emph{Appendix B: Image solution and charge moment.---}
The potential
\begin{equation}
 \varphi(z)=2\Re\int_{-q}^q h_c(\mu)
 \log\frac{z-\mu+\ii c/2}{z-\mu-\ii c/2}\dd\mu
 \label{eq:Spotential}
\end{equation}
vanishes on \(y=0\) and equals \(x\) on the slit.
Since \(\Delta\varphi=-4\pi h_c(x)\delta(y-c/2)\),
\(F_c=(8\pi)^{-1}\int_{y>0}|\nabla\varphi|^2\dd x\dd y\).
Oddness makes \(x=0\) an equipotential; \(\Psi=2xy/c-\varphi\)
vanishes on both quadrant axes and fin faces.
Finite energy admits the tip singularity but excludes an endpoint point charge.
The lower-half-plane map is
\begin{equation}
 z'(t)=\ii a\frac{t-b}{\sqrt{t(t-1)(t-s)}},\qquad 1<b<s.
 \label{eq:Smap}
\end{equation}
Continue the roots from \(t>s\) through the lower half-plane,
with \(z\sim2\ii a\sqrt t\) and zero translation at infinity.
The points \(0,1,b,s\) map to the corner, lower root, tip and upper root.
Closure gives
\begin{equation}
 \int_1^s\frac{t-b}{\sqrt{t(t-1)(s-t)}}\dd t=0,
 \qquad b=\frac{sE(m_{\rm e})}{K(m_{\rm e})}.
 \label{eq:Sclosure}
\end{equation}
Set \(m_{\rm e}=1-s^{-1}\),
\(\alpha=\arcsin\sqrt{(b-1)/(s-1)}\), and
\begin{align}
 H_{\rm g}&=\frac{\pi\sqrt s}{K(m_{\rm e})},\nonumber\\
 W_{\rm g}&=2[bF(\alpha|1-s)-E(\alpha|1-s)],\nonumber\\
 P_{\rm g}&=\frac{2b(1+s)-s}{3}-b^2.
 \label{eq:imageparameters}
\end{align}
The elliptic parameter convention is that of Eq.~\eqref{eq:imageaspect}.
Height \(aH_{\rm g}=c/2\) and length \(aW_{\rm g}=q\) give
\(c/q=2H_{\rm g}/W_{\rm g}\); SM Sec.~S4 fixes the unique real branch.
For \(t=1+(s-1)\sin^2\theta\),
\(x_{\rm fin}(t)=2a[bF(\theta|1-s)-E(\theta|1-s)]\)
runs from zero to \(q\) and back, with inverse branches \(t_-<t_+\).
From \(\Psi(z(t))=-4a^2\Im t/c\) and the normal-field jump,
\begin{equation}
 \begin{gathered}
 U(x)=\frac{a^2}{\pi c}(t_+-t_-),\qquad h_c=-U',\\
 F_c=\int_0^q U(x)\dd x.
 \end{gathered}
 \label{eq:Eimagecharge}
\end{equation}
Integrating this branch area using closure yields
\begin{equation}
 f=\frac{4q^2}{\rho^2}\frac{P_{\rm g}}{H_{\rm g}^2W_{\rm g}^2}.
 \label{eq:ellipticrate}
\end{equation}
Since \(q=cW_{\rm g}/(2H_{\rm g})\), this is
\(f=\gamma^2P_{\rm g}/H_{\rm g}^4\), giving Eq.~\eqref{eq:explicitrate}.

\par\medskip
\setcounter{equation}{0}
\renewcommand{\theequation}{C\arabic{equation}}
\renewcommand{\theHequation}{appendix.C.\arabic{equation}}
\phantomsection\label{app:returns}
\emph{Appendix C: Return spectrum and the remaining field.---}
Let \(\mathrm{Ext}:L^2([-q,q])\to L^2(\mathbb R)\) extend by zero,
and let \(P_\pm\) project onto the \(\mathbb R_\pm\) Fourier subspaces. Define
\[
 \mathsf B_L f=\bigl(P_+[e^{-\ii Lp_*/2}\mathrm{Ext}f],
                         P_-[e^{\ii Lp_*/2}\mathrm{Ext}f]\bigr).
\]
With \(\mathsf E_c=e^{-c|D_\lambda|}\oplus e^{-c|D_\lambda|}\),
\(\mathsf B_L^\dagger\mathsf B_L=\mathsf A_{\rm ref}
=I+V_{\rm ref}[\ii Lp_*]\) and
\(I+V_c[\ii Lp_*]=\mathsf B_L^\dagger(I-\mathsf E_c)\mathsf B_L\).
Thus \(\mathcal V=\mathsf B_L\mathsf A_{\rm ref}^{-1/2}\) is an isometry and
\(\mathsf T_L=\mathcal V^\dagger\mathsf E_c\mathcal V\) is a positive
trace-class strict contraction on the interval space, with
\(\mathcal D_c[\ii Lp_*]/\mathcal D_{\rm ref}[\ii Lp_*]=\det(I-\mathsf T_L)\).

Set \(\tau=LM_c/c=N_*^2\), \(\XB=2\sqrt\tau\).
In the shifted representation \cite{ItsKozlowski2013}, removing the matrix
dressing, including mixed channels, changes the log determinant by
\(O(\tau^{-1/2})\); an equal-mass point-charge replacement costs \(O(1)\)
(SM, Sec.~S5).
On the positive-frequency space \(\mathcal H_+\), the point-charge return is
\(e^{-D_x/2}\Pi_we^{-D_x/2}\), where \(D_x=-\ii\partial_x\ge0\),
\(w(x)=e^{-\ii\tau/x}\), and \(\Pi_w=I-M_wM_w^*\) projects onto
\((\operatorname{Ran}M_w)^\perp\).
The unitary inversion \(\mathcal I f(x)=x^{-1}f(-1/x)\) and Fourier transform
map \(\operatorname{Ran}\Pi_w\) to \(L^2(0,\tau;\dd s)\), where \(s\) is frequency after inversion:
\(D_x\mapsto x^2D_x-\ii x\mapsto-\partial_s(s\partial_s)\).
Then \(x=2\sqrt s\), with unitary weight
\((\mathcal U f)(x)=\sqrt{x/2}f(x^2/4)\), gives
\begin{equation}
 \begin{gathered}
 T_{\XB}=P_{\XB}e^{-\mathsf{H}_{\mathrm{B}}}P_{\XB},\qquad \mathsf{H}_{\mathrm{B}}=-\partial_x^2-\frac1{4x^2},\\
 S_{\rm ret}=-2\log\det(I-T_{\XB})+O(1).
 \end{gathered}
 \label{eq:EBessel}
\end{equation}
The origin has the inherited regular branch \(\sqrt x\), without
\(\sqrt x\log x\); \(P_{\XB}\) projects onto \((0,\XB)\).
This Bessel operator \cite{BasorEhrhardt2003} has singular-symbol asymptotics
consistent with Ref.~\cite{BelitskyKorchemsky2021}'s conjecture.
For continuous absorption at \(\XB\), the comparison operator
\(\mathsf H_{\mathrm B,\XB}^{\mathrm D}\) obeys, uniformly for \(\XB,m\ge1\),
\begin{equation}
 0\le\Tr T_{\XB}^m-\Tr e^{-m\mathsf{H}_{\mathrm{B},\XB}^{\mathrm{D}}}
 \le C\frac{\sqrt{\log(2+m)}}{\sqrt m}.
 \label{eq:Ecycles}
\end{equation}
Summation with cycle weight \(1/m\) costs \(O(1)\).
The absorbed eigenvalues are \(j_{0,n}^2/\XB^2\), with
\(j_{0,n}=\pi(n-\tfrac14)+O(n^{-1})\).
Replacing their squares by \(\pi^2(n-\tfrac14)^2\) costs
\(O(\sum n^{-2})\). Retaining the shift in the Gamma product gives
\begin{equation}
 S_{\rm ret}=\frac{\XB}{\sqrt\pi}\zeta(3/2)-\log \XB+O(1).
 \label{eq:Ereturn}
\end{equation}

At each finite auxiliary-source window \(\srcwin\), the remainder has a
positive representation \(\Phi_{\srcwin}[j]=\mathbb E_{\srcwin}e^{\ii J_{\srcwin}(j)}\).
Suppressing \(\srcwin\), full mixed-channel configurations obey
\(Y=J(p_*)\ge0\), \(H=\tfrac12\operatorname{Var}_\Sigma J(\eta)\le KY^2\).
Regular positivity of \(g_{p_*}=2\pi U\) makes \(K\) finite and window-independent.
Let \(Z=\mathbb E e^{-LY}\), \(I=-\log Z\), and
\(\dd\nu_L=Z^{-1}e^{-LY}\dd\nu\). Gaussian averaging gives
\(G_{\srcwin}=\mathbb E_L e^{-H}\).
Using \(\Pr_L(Y>y)\le\min(1,e^{I-Ly})\) and Jensen, then taking the
window limit at fixed \(L\) before \(L\to\infty\), with \(I(L)=O(\sqrt L)\), yields
\begin{equation}
 0\le-\log Z_{\rm G}\le K\mathbb E_LY^2
 \le K\frac{I^2+2I+2}{L^2}=O(L^{-1}).
 \label{eq:Eresidual}
\end{equation}
Bounded Fourier--Laplace transforms justify passing the full Gaussian
integral through this limit (SM, Sec.~S6).

Finally, \(g_{p_*}(x)=2\pi U(x)=\sqrt{q^2-x^2}\,u(x)\),
\(0<u_{\min}\le u(x)\le u_{\max}<\infty\).
Monotonicity of the positive-kernel determinant in its damping profile
sandwiches \(\mathcal W_{\rm ref}[\ii Lp_*]\) between
\(\mathcal W_{\rm ref}[\ii Lu_{\max}x]\) and
\(\mathcal W_{\rm ref}[\ii Lu_{\min}x]\), sine-kernel gaps with their
quadratic exponents removed. The Widom--Dyson asymptotic
\cite{DeiftItsKrasovskyZhou2006} gives
\(r_{\rm end}=-\tfrac14\log(qL)+O(1)\).
Equation~\eqref{eq:Efullcompletion} then gives
\(S=F_cL^2+S_{\rm ret}-r_{\rm end}-\log Z_{\rm G}+O(1)\).
Since \(\XB=2\sqrt{f\ell/\gamma}\), the logarithms combine as
\(-\tfrac12\log\ell+\tfrac14\log\ell\).
Thus \(A=0\), \(B\) is fixed by \(f,\gamma\), and \(C_{\log}=-1/4\),
with an \(O(1)\) remainder.


\begin{thebibliography}{0}%
\makeatletter
\providecommand \@ifxundefined [1]{%
 \@ifx{#1\undefined}
}%
\providecommand \@ifnum [1]{%
 \ifnum #1\expandafter \@firstoftwo
 \else \expandafter \@secondoftwo
 \fi
}%
\providecommand \@ifx [1]{%
 \ifx #1\expandafter \@firstoftwo
 \else \expandafter \@secondoftwo
 \fi
}%
\providecommand \natexlab [1]{#1}%
\providecommand \enquote  [1]{``#1''}%
\providecommand \bibnamefont  [1]{#1}%
\providecommand \bibfnamefont [1]{#1}%
\providecommand \citenamefont [1]{#1}%
\providecommand \href@noop [0]{\@secondoftwo}%
\providecommand \href [0]{\begingroup \@sanitize@url \@href}%
\providecommand \@href[1]{\@@startlink{#1}\@@href}%
\providecommand \@@href[1]{\endgroup#1\@@endlink}%
\providecommand \@sanitize@url [0]{\catcode `\\12\catcode `\$12\catcode
  `\&12\catcode `\#12\catcode `\^12\catcode `\_12\catcode `\%12\relax}%
\providecommand \@@startlink[1]{}%
\providecommand \@@endlink[0]{}%
\providecommand \url  [0]{\begingroup\@sanitize@url \@url }%
\providecommand \@url [1]{\endgroup\@href {#1}{\urlprefix }}%
\providecommand \urlprefix  [0]{URL }%
\providecommand \Eprint [0]{\href }%
\providecommand \doibase [0]{https://doi.org/}%
\providecommand \selectlanguage [0]{\@gobble}%
\providecommand \bibinfo  [0]{\@secondoftwo}%
\providecommand \bibfield  [0]{\@secondoftwo}%
\providecommand \translation [1]{[#1]}%
\providecommand \BibitemOpen [0]{}%
\providecommand \bibitemStop [0]{}%
\providecommand \bibitemNoStop [0]{.\EOS\space}%
\providecommand \EOS [0]{\spacefactor3000\relax}%
\providecommand \BibitemShut  [1]{\csname bibitem#1\endcsname}%
\let\auto@bib@innerbib\@empty
\end{thebibliography}%


\begin{thebibliography}{99}

\bibitem{Haldane1981}
F.~D.~M. Haldane,
Effective Harmonic-Fluid Approach to Low-Energy Properties of
One-Dimensional Quantum Fluids,
\href{https://doi.org/10.1103/PhysRevLett.47.1840}
{Phys. Rev. Lett. \textbf{47}, 1840--1843 (1981)}.

\bibitem{CazalillaEtAl2011}
M.~A. Cazalilla, R. Citro, T. Giamarchi, E. Orignac, and M. Rigol,
One dimensional bosons: From condensed matter systems to ultracold gases,
\href{https://doi.org/10.1103/RevModPhys.83.1405}
{Rev. Mod. Phys. \textbf{83}, 1405--1466 (2011)}.

\bibitem{AbanovKorepin2002}
A.~G. Abanov and V.~E. Korepin,
On the probability of ferromagnetic strings in antiferromagnetic spin chains,
\href{https://doi.org/10.1016/S0550-3213(02)00899-4}
{Nucl. Phys. B \textbf{647}, 565--580 (2002)}.

\bibitem{EsteveEtAl2006}
J. Esteve, J.-B. Trebbia, T. Schumm, A. Aspect, C.~I. Westbrook,
and I. Bouchoule,
Observations of Density Fluctuations in an Elongated Bose Gas:
Ideal Gas and Quasicondensate Regimes,
\href{https://doi.org/10.1103/PhysRevLett.96.130403}
{Phys. Rev. Lett. \textbf{96}, 130403 (2006)}.

\bibitem{ArmijoEtAl2010}
J. Armijo, T. Jacqmin, K.~V. Kheruntsyan, and I. Bouchoule,
Probing Three-Body Correlations in a Quantum Gas Using the
Measurement of the Third Moment of Density Fluctuations,
\href{https://doi.org/10.1103/PhysRevLett.105.230402}
{Phys. Rev. Lett. \textbf{105}, 230402 (2010)}.

\bibitem{Jacqmin2011}
T. Jacqmin, J. Armijo, T. Berrada, K.~V. Kheruntsyan,
and I. Bouchoule,
Sub-Poissonian fluctuations in a 1D Bose gas: from the quantum
quasi-condensate to the strongly interacting regime,
\href{https://doi.org/10.1103/PhysRevLett.106.230405}
{Phys. Rev. Lett. \textbf{106}, 230405 (2011)},
\href{https://arxiv.org/abs/1103.3028v4}{arXiv:1103.3028v4}.

\bibitem{YaoEtAl2025}
R. Yao, S. Chi, M. Wang, R.~J. Fletcher, and M. Zwierlein,
Measuring Pair Correlations in Bose and Fermi Gases via Atom-Resolved Microscopy,
\href{https://doi.org/10.1103/PhysRevLett.134.183402}
{Phys. Rev. Lett. \textbf{134}, 183402 (2025)}.

\bibitem{BastianelloPiroliCalabrese2018}
A. Bastianello, L. Piroli, and P. Calabrese,
Exact local correlations and full counting statistics for arbitrary
states of the one-dimensional interacting Bose gas,
\href{https://doi.org/10.1103/PhysRevLett.120.190601}
{Phys. Rev. Lett. \textbf{120}, 190601 (2018)},
\href{https://arxiv.org/abs/1802.02115v3}{arXiv:1802.02115v3}.

\bibitem{ArzamasovsGangardt2019}
M. Arzamasovs and D.~M. Gangardt,
Full Counting Statistics and Large Deviations in a Thermal 1D Bose Gas,
\href{https://doi.org/10.1103/PhysRevLett.122.120401}
{Phys. Rev. Lett. \textbf{122}, 120401 (2019)},
\href{https://arxiv.org/abs/1807.09381}{arXiv:1807.09381}.

\bibitem{Girardeau1960}
M. Girardeau,
Relationship between Systems of Impenetrable Bosons and Fermions
in One Dimension,
\href{https://doi.org/10.1063/1.1703687}
{J. Math. Phys. \textbf{1}, 516--523 (1960)}.

\bibitem{DeiftItsKrasovskyZhou2006}
P. Deift, A. Its, I. Krasovsky, and X. Zhou,
The Widom--Dyson constant for the gap probability in random matrix theory,
\href{https://doi.org/10.1016/j.cam.2005.12.040}
{J. Comput. Appl. Math. \textbf{202}, 26--47 (2007)},
\href{https://arxiv.org/abs/math/0601535v1}{arXiv:math/0601535v1}.

\bibitem{Abanov2005}
A.~G. Abanov,
Hydrodynamics of correlated systems. Emptiness Formation Probability
and Random Matrices,
in \emph{Applications of Random Matrices in Physics},
edited by E. Br\'ezin et al. (Springer, Dordrecht, 2006), pp. 139--161,
\href{https://doi.org/10.1007/1-4020-4531-X_5}{doi:10.1007/1-4020-4531-X\_5},
\href{https://arxiv.org/abs/cond-mat/0504307v1}{arXiv:cond-mat/0504307v1}.

\bibitem{YehKamenev2020}
H.-C. Yeh and A. Kamenev,
Emptiness formation probability in one-dimensional Bose liquids,
\href{https://doi.org/10.1103/PhysRevA.101.023623}
{Phys. Rev. A \textbf{101}, 023623 (2020)},
\href{https://arxiv.org/abs/1910.08054}{arXiv:1910.08054}.

\bibitem{YehGangardtKamenev2022}
H.-C. Yeh, D.~M. Gangardt, and A. Kamenev,
Emptiness formation in polytropic quantum liquids,
\href{https://doi.org/10.1088/1751-8121/ac47b1}
{J. Phys. A: Math. Theor. \textbf{55}, 064002 (2022)},
\href{https://arxiv.org/abs/2109.09916v2}
{arXiv:2109.09916v2}.

\bibitem{AbanovGangardt2025}
A.~G. Abanov and D.~M. Gangardt,
Emptiness Instanton in Quantum Polytropic Gas,
\href{https://doi.org/10.21468/SciPostPhys.18.4.122}
{SciPost Phys. \textbf{18}, 122 (2025)},
\href{https://arxiv.org/abs/2412.11686}{arXiv:2412.11686}.

\bibitem{KIW1995}
V. Korepin, A. Its, and A. Waldron,
Probability of phase separation for the Bose gas with delta interactions,
\href{https://arxiv.org/abs/cond-mat/9510068}
{arXiv:cond-mat/9510068 (1995)}.

\bibitem{KBI1993}
V.~E. Korepin, N.~M. Bogoliubov, and A.~G. Izergin,
\emph{Quantum Inverse Scattering Method and Correlation Functions}
(Cambridge University Press, Cambridge, 1993),
\href{https://doi.org/10.1017/CBO9780511628832}
{doi:10.1017/CBO9780511628832}.

\bibitem{Slavnov1998}
N.~A. Slavnov,
Asymptotics of the Fredholm Determinant Associated with the Correlation Functions of the Quantum Nonlinear Schr\"odinger Equation,
\href{https://doi.org/10.1023/A:1011396806846}
{J. Math. Sci. \textbf{104}, 1135--1143 (2001)},
\href{https://arxiv.org/abs/solv-int/9810016}{arXiv:solv-int/9810016}.

\bibitem{KhanikatiAbanov2026}
B.~A. Khanikati and A.~G. Abanov,
Emptiness formation in the Lieb--Liniger gas:
hydrodynamic instantons and a conjectured rate function,
\href{https://arxiv.org/abs/2608.26640v1}
{arXiv:2608.26640v1 (2026)}.

\bibitem{LiebLiniger1963}
E.~H. Lieb and W. Liniger,
Exact Analysis of an Interacting Bose Gas. I. The General Solution
and the Ground State,
\href{https://doi.org/10.1103/PhysRev.130.1605}
{Phys. Rev. \textbf{130}, 1605--1616 (1963)}.

\bibitem{SM}
See the accompanying Supplemental Material for the published thermodynamic source identity, ordinary-Gaussian realization and integrable-envelope proof, charge-domain and endpoint checks, relative Hardy comparison, summable Bessel heat comparison, endpoint logarithm, complete mixed-source Gaussian bound, Weyl-alcove log-concavity and action convexity, radius-averaged contact response, and finite-window determinant numerics,
which includes main-text Refs.~\cite{IzerginCokerKorepin1992,MininPronkoTarasov2023,FrizOberhauser2010,Korepin1982,KorepinSlavnov1997,GohmannKozlowskiMinin2026,DerridaGerschenfeld2009,BothnerDeiftItsKrasovsky2018,Olshanii1998,BasorEhrhardt2003,BelitskyKorchemsky2021,Emsiz2010,Prekopa1973}.

\bibitem{ItsKozlowski2013}
A.~R. Its and K.~K. Kozlowski,
On determinants of integrable operators with shifts,
\href{https://doi.org/10.1093/imrn/rnt191}
{Int. Math. Res. Not. \textbf{2014}, 6826--6838 (2014)},
\href{https://arxiv.org/abs/1301.2221v1}{arXiv:1301.2221v1}.

\bibitem{GIK2019}
R. Gharakhloo, A.~R. Its, and K.~K. Kozlowski,
Riemann--Hilbert approach to a generalized sine kernel,
\href{https://doi.org/10.1007/s11005-019-01218-3}
{Lett. Math. Phys. \textbf{110}, 297--325 (2020)},
\href{https://arxiv.org/abs/1905.04907v1}{arXiv:1905.04907v1}.

\bibitem{TorquatoEtAl2008}
S. Torquato, A. Scardicchio, and C.~E. Zachary,
Point processes in arbitrary dimension from fermionic gases, random matrix theory, and number theory,
\href{https://doi.org/10.1088/1742-5468/2008/11/P11019}
{J. Stat. Mech. (2008) P11019}.

\bibitem{Stephan2019}
J.-M. St\'ephan,
Free fermions at the edge of interacting systems,
\href{https://doi.org/10.21468/SciPostPhys.6.5.057}
{SciPost Phys. \textbf{6}, 057 (2019)}.

\bibitem{IzerginCokerKorepin1992}
A.~G. Izergin, D.~A. Coker, and V.~E. Korepin,
Determinant formula for the six-vertex model,
\href{https://doi.org/10.1088/0305-4470/25/16/010}
{J. Phys. A: Math. Gen. \textbf{25}, 4315--4334 (1992)}.

\bibitem{MininPronkoTarasov2023}
M.~D. Minin, A.~G. Pronko, and V.~O. Tarasov,
Construction of determinants for the six-vertex model with domain wall
boundary conditions,
\href{https://doi.org/10.1088/1751-8121/acdd37}
{J. Phys. A: Math. Theor. \textbf{56}, 295204 (2023)},
\href{https://arxiv.org/abs/2304.01824v1}{arXiv:2304.01824v1}.

\bibitem{FrizOberhauser2010}
P. Friz and H. Oberhauser,
A generalized Fernique theorem and applications,
\href{https://arxiv.org/abs/1004.1923v2}{arXiv:1004.1923v2 (2010)}.

\bibitem{Korepin1982}
V.~E. Korepin,
Calculation of norms of Bethe wave functions,
\href{https://doi.org/10.1007/BF01212176}
{Commun. Math. Phys. \textbf{86}, 391--418 (1982)}.

\bibitem{KorepinSlavnov1997}
V.~E. Korepin and N.~A. Slavnov,
Normal ordering in the theory of correlation functions of exactly solvable models,
\href{https://doi.org/10.1088/0305-4470/30/24/022}
{J. Phys. A: Math. Gen. \textbf{30}, 8623--8633 (1997)},
\href{https://arxiv.org/abs/hep-th/9709204v1}{arXiv:hep-th/9709204v1}.

\bibitem{GohmannKozlowskiMinin2026}
F. G\"ohmann, K.~K. Kozlowski, and M.~D. Minin,
Asymptotic analysis of a Fredholm determinant occurring in the description
of the dynamical correlation functions of the Lieb--Liniger Bose gas,
\href{https://arxiv.org/abs/2608.16727v2}{arXiv:2608.16727v2 (2026)}.

\bibitem{DerridaGerschenfeld2009}
B. Derrida and A. Gerschenfeld,
Current Fluctuations of the One Dimensional Symmetric Simple
Exclusion Process with Step Initial Condition,
\href{https://doi.org/10.1007/s10955-009-9772-7}
{J. Stat. Phys. \textbf{136}, 1--15 (2009)},
\href{https://arxiv.org/abs/0902.2364v1}{arXiv:0902.2364v1}.

\bibitem{BothnerDeiftItsKrasovsky2018}
T. Bothner, P. Deift, A. Its, and I. Krasovsky,
The sine process under the influence of a varying potential,
\href{https://doi.org/10.1063/1.5050394}
{J. Math. Phys. \textbf{59}, 091414 (2018)},
\href{https://arxiv.org/abs/1807.11387v1}{arXiv:1807.11387v1}.

\bibitem{Olshanii1998}
M. Olshanii,
Atomic Scattering in the Presence of an External Confinement
and a Gas of Impenetrable Bosons,
\href{https://doi.org/10.1103/PhysRevLett.81.938}
{Phys. Rev. Lett. \textbf{81}, 938--941 (1998)}.

\bibitem{BasorEhrhardt2003}
E.~L. Basor and T. Ehrhardt,
Asymptotics of determinants of Bessel operators,
\href{https://doi.org/10.1007/s00220-002-0769-1}
{Commun. Math. Phys. \textbf{234}, 491--516 (2003)},
\href{https://arxiv.org/abs/math/0204304}{arXiv:math/0204304}.

\bibitem{BelitskyKorchemsky2021}
A.~V. Belitsky and G.~P. Korchemsky,
Crossing bridges with strong Szeg\H{o} limit theorem,
\href{https://doi.org/10.1007/JHEP04(2021)257}
{J. High Energy Phys. 04 (2021) 257},
\href{https://arxiv.org/abs/2006.01831v2}{arXiv:2006.01831v2}.

\bibitem{Emsiz2010}
E. Emsiz,
Completeness of the Bethe ansatz on Weyl alcoves,
\href{https://doi.org/10.1007/s11005-009-0359-7}
{Lett. Math. Phys. \textbf{91}, 61--70 (2010)},
\href{https://arxiv.org/abs/1009.4487}{arXiv:1009.4487}.

\bibitem{Prekopa1973}
A. Pr\'ekopa,
On logarithmic concave measures and functions,
\href{https://rutcor.rutgers.edu/Prekopa/pdf/SCIENT2.pdf}
{Acta Sci. Math. (Szeged) \textbf{34}, 335--343 (1973)}.

\end{thebibliography}
\end{document}


\maketitle
\begingroup
\renewcommand{\thefootnote}{\fnsymbol{footnote}}
\footnotetext[1]{These authors contributed equally to this work.}
\footnotetext[2]{Email: \href{mailto:huangbin26@mails.ucas.ac.cn}{huangbin26@mails.ucas.ac.cn}}
\footnotetext[3]{Email: \href{mailto:zhuziyan26@mails.ucas.ac.cn}{zhuziyan26@mails.ucas.ac.cn}}
\endgroup
\noindent
This Supplemental Material presents the auxiliary-field formulation,
electrostatic image solution, and asymptotic analysis underlying the Letter.
It also derives the physical contact response and details the numerical methods.
\begingroup
\small
\setcounter{tocdepth}{1}
\tableofcontents
\endgroup
\clearpage
\noindent\textbf{Core quantities and conventions.}
\begin{center}
\begin{tabular}{@{}p{.35\linewidth}p{.59\linewidth}@{}}
\toprule
Symbols & Meaning\\
\midrule
$R,L=2R,\ell=\rho R$ & Physical half-width, full width, dimensionless half-width\\
$\psi(x),\phi(\lambda)$ & Physical Bose field and auxiliary Gaussian field\\
$\rho,\sigma_q,h_c=-U'$ & Bulk density, Bethe root density, neutral response charge\\
$F_c,\ M_c=2F_c$ & Charge energy and primitive mass; $M_c=\int\lambda h_c$, not $\int h_c$\\
$f=4F_c/\rho^2,\ A=0,\ B$ & Quadratic rate, absent linear term, square-root coefficient\\
$p_*,\Sigma$ & Dressed source and completed Gaussian covariance\\
$n_*,d_*,N_*$ & Physical response density, length and particle-number scales\\
$\tau=LM_c/c=N_*^2$ & Scale of the auxiliary return problem\\
$\mathcal D_{\rm ref},V_{\rm ref}$ & Reference at the same occupied endpoint $q$ and source\\
$\mathsf T_L,\ Z_{\rm G}(L)$ & Full return operator and residual Gaussian factor\\
$\mathcal V,\ \mathsf E_c$ & Normalized two-channel isometry and propagation multiplier\\
$\XB=2N_*,\ T_{\XB}$ & Auxiliary Bessel interval length and projected propagator\\
$X,Y$ & Dimensionless physical positions in the rescaled Bose field\\
$\srcwin=(J,\Lambda),\ \eta_{\rm det}$ & Auxiliary source-window index and detection efficiency\\
\bottomrule
\end{tabular}
\end{center}
The image map uses auxiliary electrostatic coordinates; the Bessel
coordinates arise after inversion and Fourier transformation. Neither
is a physical real-space coordinate, and the Bessel modes are not
physical Lieb--Liniger excitations. The physical
response is derived in Sec.~\ref{phys:response}, without identifying
$h_c$ or $U$ with a real-space density. Matrices and local operators are
defined in the sections that use them.

This supplement uses the Letter's determinant and return notation,
with $L=2R$.
In Sec.~\ref{sec:numerics}, $n$ counts quadrature nodes and
$M_{\rm feat}$ counts Gaussian feature pairs.

\section{Observable, exact input, and limits}
\label{main:framework}
We use the repulsive Lieb--Liniger Hamiltonian
\begin{equation}
 H=\int\dd x\left[\frac12\partial_x\psi^\dagger\partial_x\psi
             +\frac c2(\psi^\dagger)^2\psi^2\right],\qquad c>0 .
 \label{main:hamiltonian}
\end{equation}
In units \(\hbar=m=1\), \(\psi\) is the Bose annihilation field,
\([\psi(x),\psi^\dagger(y)]=\delta(x-y)\).
The local density is \(\hat n(x)=\psi^\dagger(x)\psi(x)\), and
\(\widehat N_I=\int_I\hat n(x)\,\dd x\).
The thermodynamic ground state is taken at fixed density $\rho$ before the empty interval is enlarged. Its half-width is $R$, its full length is $L=2R$, and $\ell=\rho R$, $\gamma=c/\rho$. The observable and its cost are
\begin{equation}
 P_0(R)=\langle\Pi_R\rangle_0
       =\langle{:}\exp[-\widehat N_{[-R,R]}]{:}\rangle_0,
 \qquad S(\ell;\gamma)=-\log P_0(\ell/\rho).
 \label{main:observable}
\end{equation}
Here \(\Pi_R=\mathbf1_{\{\widehat N_{[-R,R]}=0\}}\).
The occupied rapidities are $\lambda\in[-q,q]$. Their density $\sigma_q(\lambda)$ obeys
\begin{equation}
 \sigma_q(\lambda)-\int_{-q}^q\frac{\dd\mu}{2\pi}
 \frac{2c\,\sigma_q(\mu)}{c^2+(\lambda-\mu)^2}=\frac1{2\pi},
 \qquad \int_{-q}^q\sigma_q(\lambda)\,\dd\lambda=\rho.
 \label{main:lieb}
\end{equation}
This fixes the equilibrium normalization \cite{suppLiebLiniger1963,suppKBI1993}. The auxiliary response profile introduced below is a different quantity from $\sigma_q$.

The result at each fixed finite $\gamma>0$ is
\begin{equation}
 \displaystyle
 S(\ell;\gamma)=f(\gamma)\ell^2+
 \frac{2\zeta(3/2)}{\sqrt\pi}\sqrt{\frac{f(\gamma)}{\gamma}}\sqrt\ell
 -\frac14\log\ell+O_\gamma(1).
 \label{main:result}
\end{equation}
The following sections establish the factorization
\begin{equation}
 P_0(R)=\frac{Z_C}{\mathcal D_c[0]}\,
 e^{-F_cL^2}\,e^{r_{\rm end}(L)}\,
 \Det(I-\mathsf T_L)\,Z_{\rm G}(L).
 \label{main:factorization}
\end{equation}
Here $Z_C/\mathcal D_c[0]$ is independent of $L$, and $\mathsf T_L$ sums the two auxiliary return channels. Their contributions to $S$ are:
\begin{center}
\begin{tabular}{@{}p{.40\linewidth}p{.52\linewidth}@{}}
\toprule
Contribution to $S$ & Size and origin\\
\midrule
$F_cL^2$ & Exact quadratic completion of the auxiliary field\\
$-\log\Det(I-\mathsf T_L)$ & $\sqrt L$ and $-\tfrac12\log L$: all return cycles\\
$-r_{\rm end}(L)$ & $+\tfrac14\log L$: the original rapidity endpoints\\
$-\log Z_{\rm G}(L)$ & $O(L^{-1})$: the remaining field average\\
\bottomrule
\end{tabular}
\end{center}
The stronger bound in the last row is proved in Sec.~\ref{gauss:section}. It does not determine the bounded terms in the other factors. All logarithms are natural. Constants below may depend on fixed $c$ and $q$; no uniform weak-coupling limit is implied.

\subsection{Exact auxiliary-field integral}
For an analytic auxiliary source $j(\lambda)$, define
\begin{align}
 \mathcal D_c[j]&=\Det(I+V_c[j]),\nonumber\\
 V_c[j](\lambda,\mu)&=-\frac{c}{2\pi d}
 \left[\frac{e^{\Delta j/2}}{d+\ii c}
       +\frac{e^{-\Delta j/2}}{d-\ii c}\right],
 \quad d=\lambda-\mu,\quad\Delta j=j(\lambda)-j(\mu),\nonumber\\
 V_c[j](\lambda,\lambda)&=\frac{\ii j'(\lambda)}{2\pi}-\frac1{\pi c}.
 \label{main:kernel}
\end{align}
The diagonal is the limit of the sum of both channels. The function $j$ is an auxiliary source on rapidity space; it is not a source for the physical density $\psi^\dagger\psi$.

The normalized thermodynamic formula of Korepin, Its, and Waldron \cite{suppKIW1995}, Sec.~2, Eqs.~(5)--(12), especially Eq.~(6), with their length $x=L$, together with the ordinary-Gaussian representation established below, gives
\begin{equation}
 P_0(R)=\frac{\E_C\mathcal D_c[\ii L\lambda+\phi]}{\mathcal D_c[0]}.
 \label{main:gaussian-input}
\end{equation}
The real auxiliary field $\phi(\lambda)$ has covariance
\begin{equation}
 C(\lambda,\mu)=C_0(\lambda-\mu)-C_0(\lambda)-C_0(\mu),
 \qquad C_0(d)=-\log(1+d^2/c^2).
 \label{main:covariance}
\end{equation}
Its value at zero is fixed to zero; the removed constant mode does not enter any source difference. A useful construction is
\begin{equation}
 C(\lambda,\mu)=2\Re\sum_{n\ge1}
 \frac{w_c(\lambda)^n\overline{w_c(\mu)^n}}n,
 \qquad w_c(\lambda)=\frac{\lambda}{c-\ii\lambda}.
 \label{main:features}
\end{equation}
Thus $\phi$ is analytic on compact subsets of $|\Im\lambda|<c/2$. The field exactly represents the scattering factors in the microscopic determinant; it is not a Gaussian approximation to the Bose-gas density.

For any real coefficients $a_i$ with $\sum_i a_i=0$, normal ordering of the commuting dual field in Ref.~\cite{suppKIW1995} gives the same moment as
\begin{equation}
 \E_C\exp\!\left[\sum_i a_i\phi(\lambda_i)\right]
 =\exp\!\left[\frac12\sum_{i,j}a_i a_j C_0(\lambda_i-\lambda_j)\right].
 \label{main:neutral-moments}
\end{equation}
Every term of each Fredholm minor is neutral, including each choice of scattering channels. Differentiation gives the moments needed on the diagonal. The absolute-series bound in Sec.~\ref{env:section} permits the expectation, the rapidity integrals, and the infinite Fredholm sum to be exchanged, so the coefficient identity gives the actual integral \eqref{main:gaussian-input}. This step is taken at fixed $L$. The thermodynamic limit is taken at fixed interval, before the large-interval limit. Normal-ordering context is given in Ref.~\cite{suppKorepinSlavnov1997}, Sec.~3.

The normalization is positive. With
\begin{equation}
 P_c(\lambda,\mu)=\frac{c}{\pi[c^2+(\lambda-\mu)^2]},\qquad
 0\le P_c,\quad \|P_c\|\le r_c=\frac2\pi\arctan(q/c)<1,
 \label{main:normalization}
\end{equation}
one has $\mathcal D_c[0]=\Det(I-P_c)>0$. In particular,
$\log\mathcal D_c[0]\ge-2q/[\pi c(1-r_c)]$. This fixed determinant is retained in every normalized expression.

\subsection{Absolute convergence of the microscopic Gaussian average}
\label{env:section}

We use the rational Izergin--Korepin determinant identity
\cite{suppIzerginCokerKorepin1992}, in the form of
Ref.~\cite{suppMininPronkoTarasov2023}, Eq.~(2.3), with the normalized vertex
matrices $R_\pm$ specified in Eq.~\eqref{env:scattering} below.
For each subdeterminant, the identity is applied to its selected row and
column rapidities; coincident arguments are taken by continuity.
This subsection justifies averaging the published Fredholm series term by
term at fixed $c,q,L$.  The constants may depend on $L$; the growing-interval
limit is treated separately.  A bound on the complete determinant alone
would not justify this interchange.

Let $\mu$ be any positive measure on $[-a,a]$ with mass at most $W$.
For $s(x)=\phi(x)+\ii Lx$, set $b_0=1$ and
\[
 b_n(\mu,\phi)=\frac1{n!}\int
 \left|\det[V_c[s](x_i,x_j)]_{i,j=1}^n\right|\prod_{i=1}^n\dd\mu(x_i).
\]
The complete kernel includes its confluent diagonal. For atomic
$\mu=\sum_iw_i\delta_{x_i}$, $b_n$ is the absolute principal-minor sum of
$A_{\mu,ij}=\sqrt{w_iw_j}\,V_c[s](x_i,x_j)$; repeated rows vanish.
Choose $Q>a$ and a filled ellipse $E_T$, whose boundary is
$z=Q\cosh(T+\ii\theta)$, inside $|\Im z|<c/2$.  Set
\begin{equation}
 U_{\rm an}=\sup_{E_T}|\phi|,\qquad
 \phi(x)=a_0+\sum_{k\geq1}a_kT_k(x/Q),\qquad
 Q_{\rm buf}[\phi]=\frac18\sum_{k\geq1}ka_k^2.
 \label{env:source}
\end{equation}
Analyticity gives $|a_k|\leq2U_{\rm an}e^{-kT}$.  We prove the uniform
envelope
\begin{equation}
 \displaystyle
 \sum_{n\geq0}b_n(\mu,\phi)
 \leq C_1\exp\!\left\{Q_{\rm buf}[\phi]
 +C_2(1+U_{\rm an})\log(2+U_{\rm an})\right\}
 =:\mathcal E(\phi),\qquad \E_C\mathcal E<\infty.
 \label{env:envelope}
\end{equation}
There is no lower bound on node spacing.  The buffer form $Q_{\rm buf}$ is used
only in this estimate; it does not replace the quadratic source form $Q$ on $[-q,q]$.

\subsubsection{A sharp bound for the low-order minors}
For a neutral real signed measure $\sigma$ in the plane with finite logarithmic energy, write
$\mathcal C(\sigma)=-2\iint\log|z-w|\,\dd\sigma(z)\dd\sigma(w)$.
Its potential satisfies $-\Delta v_\sigma=4\pi\sigma$, so integration by
parts and completion of the square give
\begin{equation}
 \mathcal C(\sigma)=\frac1{4\pi}\int_{\mathbb C}|\nabla v_\sigma|^2\,\dd^2z\geq0,
 \qquad
 \langle\sigma,F\rangle-\frac12\mathcal C(\sigma)
 \leq\frac1{8\pi}\int_{\mathbb C}|\nabla F|^2\,\dd^2z.
 \label{env:energy}
\end{equation}
Take $F_t=\Re\phi$ inside $E_t$ and its finite-energy harmonic extension
outside, $0<t\leq T/2$.  In the coordinates of \eqref{env:source}, each
mode is $a_k\cosh(ku)\cos(k\theta)$ inside and
$a_k\cosh(kt)e^{-k(u-t)}\cos(k\theta)$ outside.  Orthogonality gives
\begin{equation}
 \frac1{8\pi}\int|\nabla F_t|^2\,\dd^2z
 =Q_{\rm buf}[\phi]+\frac1{16}\sum_{k\geq1}ka_k^2(e^{2kt}-1)
 \leq Q_{\rm buf}[\phi]+C tU_{\rm an}^2.
 \label{env:sharp-energy}
\end{equation}
This fixes the coefficient of the critical quadratic term.

Let $m_{z,b}$ be unit mass uniformly distributed on a circle of radius
$b$ about $z$.  For row-to-column distances greater than $2b$, the Cauchy
determinant formula and the logarithmic mean-value inequality give
\begin{equation}
 \log\left|\det[(x_i-z_j)^{-1}]_{i,j=1}^k\right|
 \leq-\frac14\mathcal C(A)-k\log b,
 \qquad A=\sum_i m_{x_i,b}-\sum_jm_{z_j,b}.
 \label{env:cauchy-energy}
\end{equation}
Opposite-sign interactions are unchanged by this smoothing; same-sign
smoothing lowers their negative logarithmic energy.  Circles of the same
sign may overlap.  Repeated points follow by continuity.

The scattering factors require a bound that preserves their cancellations.
Set
\begin{align}
 t_\pm(d)=\frac{\pm\ii c}{d(d\pm\ii c)},\qquad
 &R_\pm(d)=\frac{dI\pm\ii c P}{d\pm\ii c},\notag\\
 V_c[s](x,z)&=\frac{\ii}{2\pi}
 \left[e^{[s(x)-s(z)]/2}t_+(x-z)
 -e^{-[s(x)-s(z)]/2}t_-(x-z)\right],
 \label{env:scattering}
\end{align}
where $P$ exchanges two auxiliary two-state spaces.  The rational
Izergin--Korepin identity \cite{suppMininPronkoTarasov2023} writes the ratio of a $k\times k$
$t_\pm$ determinant to its Cauchy determinant as a matrix element of
$k^2$ normalized $R_\pm$ matrices between unit boundary states.
The two eigenvalues on the symmetric and antisymmetric subspaces imply,
for $|\Im d|\leq h<c/2$,
\begin{equation}
 \|R_\pm(d)\|\leq\frac{c+h}{c-h}\leq e^{4h/c},\qquad
 |\det[t_\pm(x_i-z_j)]|
 \leq e^{4hk^2/c}|\det[(x_i-z_j)^{-1}]|.
 \label{env:Rbound}
\end{equation}
This is a norm bound on the scattering product, not a sum of absolute
vertex weights.

Let $\Gamma_h$ be the rectangle with vertices $\pm(a+h)\pm\ii h$.
For fixed sufficiently small $h_0$, the rectangle and radius-$h/4$ circles
about all row and column points lie in $E_{\kappa h}$ when $h\leq h_0$;
the fixed buffer $Q>a$ supplies this geometric margin.  Before separating
the channels, Cauchy's formula gives the exact identity
\begin{equation}
 \det[V_c[s](x_i,x_j)]_{i,j=1}^n
 =\oint_{\Gamma_h}\!\cdots\!\oint_{\Gamma_h}
 \det[V_c[s](x_i,z_j)]_{i,j=1}^n
 \prod_{j=1}^n\frac{\dd z_j}{2\pi\ii(z_j-x_j)}.
 \label{env:contour}
\end{equation}
The diagonal pole of the complete kernel is removable, with value
$\ii s'(x)/(2\pi)-1/(\pi c)$; the shifted poles lie outside the rectangle.
Thus \eqref{env:contour} retains the full diagonal and its derivative terms.

Expand the displaced determinant by the first-channel row and column
sets $I,J$, $|I|=|J|$.  Its at most $\binom{2n}{n}\leq4^n$ terms are
products of two scattering determinants and the source factor
$e^{\langle(A-B)/2,s\rangle}$, where $A$ contains the smoothed rows in
$I$ minus columns in $J$, and $B$ contains their complements.
Both charges are neutral.  Put $q_1=(A-B)/2$ and $q_0=(A+B)/2$.
Since $\mathcal C(A)+\mathcal C(B)
=2\mathcal C(q_0)+2\mathcal C(q_1)$, discard
$\mathcal C(q_0)\geq0$ and apply
\eqref{env:energy}--\eqref{env:Rbound} to $q_1$.
Circle averages of the analytic source equal its point values; the extra
source $\ii Lz$ contributes at most $|L|nh/2$ to the real exponent.
The Cauchy denominators supply another $h^{-n}$.  Consequently
\begin{equation}
 b_n(\mu,\phi)
 \leq\frac{(C W/h^2)^n}{n!}
 \exp\{Q_{\rm buf}[\phi]+ChU_{\rm an}^2+Chn^2\},
 \qquad 0<h\leq h_0.
 \label{env:low-order}
\end{equation}
Here the $n$-fold measure has mass at most $W^n$.

\subsubsection{The high-order tail and the Gaussian integral}
\emph{High orders.} The complete kernel has the analytic form
\begin{equation}
 V_c[s](x,y)=-\frac{c}{\pi[(x-y)^2+c^2]}
 \left[\cosh\frac{s(x)-s(y)}2
 -\ii c\,\frac{\sinh([s(x)-s(y)]/2)}{x-y}\right].
 \label{env:analytic-kernel}
\end{equation}
On a fixed smaller ellipse it is bounded by $M=Ce^{CU_{\rm an}}$.
Its uniformly convergent expansion
$V_c[s](x,y)=\sum_{k\geq0}T_k(x/Q)v_k(y)$ has
$\sup_{y\in[-a,a]}|v_k(y)|\leq Me^{-\beta k}$ for fixed $\beta>0$.
Apply Cauchy--Binet to finite partial sums and pass their uniform limit.
Each term selects $n$ distinct modes. Since $|T_k(x/Q)|\leq1$,
Hadamard bounds the two determinants by $n^{n/2}$ and
$n^{n/2}M^ne^{-\beta\sum k_j}$. Integration therefore gives
\begin{equation}
 \begin{split}
 b_n&\leq\frac{n^n(WM)^n}{n!}
       \sum_{0\leq k_1<\cdots<k_n}e^{-\beta\sum_j k_j}\\
 &\leq\left(\frac{eWM}{1-e^{-\beta}}\right)^n
       e^{-\beta n(n-1)/2}.
 \end{split}
 \label{env:high-order}
\end{equation}
For the last inequality, write $k_j=j-1+m_j$, drop the ordering constraints
on $m_j\geq0$, and use $n^n/n!\leq e^n$.

\emph{Summing the orders.} Choose $n_0=\lceil K(1+U_{\rm an})\rceil$
with fixed $K$ sufficiently large. Equation~\eqref{env:high-order} gives a
uniformly bounded tail above $n_0$. Below $n_0$, insert
$h=h_0/(1+U_{\rm an})$ in \eqref{env:low-order}:
$h(U_{\rm an}^2+n^2)=O(1+U_{\rm an})$ and the prefactor costs at most
$O((1+U_{\rm an})\log(2+U_{\rm an}))$ in the exponent.
Summing gives \eqref{env:envelope} for every such $\mu$.

\emph{Gaussian integrability.} The analytic Gaussian field of
\eqref{main:features}, restricted to $E_T$, is a Gaussian random element
of the continuous-function space. Fernique's theorem
\cite{suppFrizOberhauser2010} gives
\begin{equation}
 \E_Ce^{\eta U_{\rm an}^2}<\infty\qquad\text{for some }\eta>0.
 \label{env:fernique}
\end{equation}
Let $X_k=\sqrt{k}\,a_k/2$ and $T_{\mathrm G}=\operatorname{Cov}(X)$.
Analytic decay gives $\Tr T_{\mathrm G}<\infty$ and
$Q_{\rm buf}=\|X\|^2/2$. The coefficient functionals have zero mass and
finite-energy zero-exterior antiderivatives, so the gap
\eqref{main:gap-bound} extends to them by density on $[-Q,Q]$:
$\|T_{\mathrm G}\|\leq1-\delta_{c,Q}<1$. Hence, for some $r>1$,
\begin{equation}
 \E_Ce^{rQ_{\rm buf}}=\Det(I-rT_{\mathrm G})^{-1/2}<\infty.
 \label{env:quadratic-moment}
\end{equation}
Since $(1+u)\log(2+u)=o(u^2)$, H\"older's inequality with
\eqref{env:fernique} and \eqref{env:quadratic-moment} gives
$\E_C\mathcal E<\infty$.

Taking $\dd\mu=\dd x$ on $[-q,q]$ gives directly
\begin{equation}
 \sum_{n\geq0}\frac1{n!}\int_{[-q,q]^n}
 \left|\det[V_c[\phi+\ii Lx](x_i,x_j)]_{i,j=1}^n\right|
 \,\dd^n x\leq\mathcal E(\phi).
 \label{env:fredholm}
\end{equation}
This permits termwise averaging and the neutral-moment identification,
including confluent derivatives, in Sec.~\ref{main:framework}.
For convergent positive quadratures, fixed-order convergence and
\eqref{env:high-order} give pointwise determinant convergence; the same
integrable envelope passes their Gaussian averages to the limit.

\section{Reference response and quadratic subtraction}
\label{pos:section}
We extract a source-dependent factor that can be averaged exactly over
the scattering field, using a reference kernel obtained by taking the
infinite-coupling limit at fixed $q$ and source $j$, retaining their
values from the finite-coupling problem. In particular, $q$ is determined
by the finite-$c$ Lieb equation. The reference retains the occupied
rapidity interval of the finite-coupling problem.
\subsection{Source modes and the origin of the quadratic factor}

In this section $x,y$ are rapidity variables, distinct from the physical position in Eq.~\eqref{main:hamiltonian}.
The reference determinant uses the unshifted kernel
\begin{equation}
 \mathcal D_{\rm ref}[j]=\Det(I+V_{\rm ref}[j]),\qquad
 V_{\rm ref}[j](x,y)=\frac{\ii\sinh\{[j(x)-j(y)]/2\}}{\pi(x-y)}.
 \label{pos:unshifted-kernel}
\end{equation}
The kernel depends only on source differences. Remove the constant mode
and use cosine modes on the interval. The corresponding sine modes form
the harmonic partner $g_j$; the quadratic form $Q$ written alongside them
follows from the reference source variation derived below:
\begin{equation}
 j(x)=\sum_{n\geq1}a_nT_n(x/q),\qquad
 Q[j]=\frac18\sum_{n\geq1}na_n^2,\qquad
 g_j(q\cos\theta)=\sum_{n\geq1}a_n\sin(n\theta).
 \label{pos:coordinates}
\end{equation}
Here $T_n(\cos\theta)=\cos(n\theta)$. The continuation of $Q$ to complex
sources is bilinear, so $Q[\ii p]=-Q[p]$ for real $p$.

The scalar part of the reference boundary problem is fixed by its source
jump and decay at infinity. In the exterior coordinate
$\omega(z)=(z-\sqrt{z^2-q^2})/q$, with the square root fixed by
$\sqrt{z^2-q^2}\sim z$, it is
\begin{equation}
 k_j(z)=-\frac12\sum_{n\ge1}a_n\omega(z)^n,
 \qquad k_{j,-}+k_{j,+}=-j,
 \qquad k_{j,-}-k_{j,+}=-\ii g_j.
 \label{derive:scalar-response}
\end{equation}
Here $+$ and $-$ are the upper and lower boundary values, and $\Gamma$
below encircles the cut counterclockwise.
The exact source differential identity in Sec.~\ref{pos:reference-proof}
separates this scalar contribution from the remaining matrix response.
Its scalar term is
\begin{equation}
 \begin{split}
 \delta\log\mathcal D_{\rm ref}\big|_{\rm scalar}
 &=\frac{1}{2\pi\ii}\oint_\Gamma\delta j\,k_j'\,\dd z
   =-\frac{1}{2\pi}\int_{-q}^q\delta j(x)g_j'(x)\,\dd x\\
 &=\frac{1}{2\pi}\int_0^\pi
   \left(\sum_{m\ge1}\delta a_m\cos m\theta\right)
   \left(\sum_{n\ge1}n a_n\cos n\theta\right)\dd\theta
   =\frac14\sum_{n\ge1}n a_n\delta a_n.
 \end{split}
 \label{derive:Q-selection}
\end{equation}
Cosine orthogonality gives the last equality. Integrating from zero source
fixes $Q[j]=\frac18\sum n a_n^2$, with no additive constant. This scalar-gauge contribution is fixed by the exact source identity
for finite source amplitude. The residual matrix factor retains its own
source dependence.

There is also a useful charge interpretation of the same form. On neutral
charges, the logarithmic kernel $K_0(x-y)=-2\log|x-y|$ satisfies, for $n\ge1$,
\begin{equation}
 \int_{-q}^q K_0(x-y)
 \frac{T_n(y/q)}{\sqrt{q^2-y^2}}\,\dd y
 =\frac{2\pi}{n}T_n(x/q).
 \label{derive:log-modes}
\end{equation}
This follows by expanding
$K_0(q\cos\theta-q\cos\theta')
=\text{constant}+4\sum_{m\ge1}\cos(m\theta)\cos(m\theta')/m$
and applying cosine orthogonality; the constant disappears for $n\ge1$.
Consequently the reference response charge is
\begin{equation}
 h_j^{\rm ref}(x)=K_0^{-1}j(x)
 =\frac{\sum_{n\ge1}n a_nT_n(x/q)}{2\pi\sqrt{q^2-x^2}}
 =-\frac{g_j'(x)}{2\pi},
 \qquad Q[j]=\frac12\int_{-q}^q j(x)h_j^{\rm ref}(x)\,\dd x.
 \label{derive:reference-charge}
\end{equation}
The same orthogonality verifies the normalization $1/8$. In particular,
the bare source $j=\ii Lx$ gives $Q[j]=-L^2q^2/8$. At finite coupling the
field average changes this charge response, as the next section shows.

\subsection{Exact reference remainder and uniform exterior bounds}
\label{pos:reference-proof}
After extracting the scalar contribution, the remaining source identity is
\begin{align}
 \mathcal W_{\rm ref}[j]
 :=e^{-Q[j]}\mathcal D_{\rm ref}[j]
 &=\Det\!\left(I+M_{e^{\ii g_j}-1}\mathsf K_*\right),\notag\\
 \mathsf K_*(x,y)&=\frac{1}{2\pi^2}
 \frac{\log\frac{q+x}{q-x}-\log\frac{q+y}{q-y}}{x-y}.
 \label{pos:kernel}
\end{align}
Here $M_f$ denotes multiplication by $f$. The diagonal is understood by
continuity. The change $x=q\tanh s$, including
its unitary Jacobian, gives
\begin{equation}
 \mathsf K_*^{(s)}(s,s')=
 \frac{s-s'}{\pi^2\sinh(s-s')},\qquad
 \widehat{\mathsf K_*^{(s)}}(k)
 =\frac{1}{2\cosh^2(\pi k/2)},\qquad
 0\leq\mathsf K_*\leq\tfrac12 I.
 \label{pos:contraction}
\end{equation}
For analytic $j$, $g_j(q\tanh s)$ decays at both ends.  The weighted
determinant therefore exists although the unweighted trace on $\R$ is infinite.
It is the characteristic functional
\begin{equation}
 \mathcal W_{\rm ref}[j]=\E_*e^{\ii J_*(j)},\qquad
 J_*(j)=\sum_{s\in X_*}g_j(q\tanh s)
 \label{pos:unshifted-law}
\end{equation}
of the determinantal process with kernel \eqref{pos:contraction}.

The resolvent proof below fixes the normalization at zero source. Equations
\eqref{derive:Q-selection} and \eqref{pos:kernel} therefore specify both
pieces of the subtraction: a charge quadratic form and a full, nonconstant
characteristic functional.

\paragraph{Reference identity and exterior estimates.}
The normalization of \eqref{pos:kernel} follows from the
auxiliary $2\times2$ matrix boundary-value problem for the unshifted kernel.
Let $\chi_j$ be its solution normalized at infinity, with interval jump
$\bigl(\begin{smallmatrix}2&-e^j\\ e^{-j}&0\end{smallmatrix}\bigr)$.
Use the exterior coordinate $\omega$ and scalar potential $k_j$ of
Eq.~\eqref{derive:scalar-response}, and set
$\mathbf R_j=\chi_j e^{-k_j\sigma_3}$ with
$\sigma_3=\operatorname{diag}(1,-1)$. The boundary jumps in that equation
transform the matrix jump to
$\bigl(\begin{smallmatrix}2e^{\ii g_j}&-1\\1&0\end{smallmatrix}\bigr)$.
At zero source it is solved by
\begin{equation}
 \mathbf R_0(z)=I+\frac{1}{2\pi\ii}
 \log\frac{z+q}{z-q}
 \begin{pmatrix}1&-1\\1&-1\end{pmatrix}.
 \label{pos:Fzero}
\end{equation}
We retain the boundary-side and counterclockwise-contour conventions
of Eq.~\eqref{derive:scalar-response}.
The exact source differential identity of Ref.~\cite{suppGIK2019}, Eq.~(2.1),
with its Riemann--Hilbert resolvent formulas (1.3)--(1.6), reads
\begin{equation}
 \delta\log\mathcal D_{\rm ref}[j]
 =\frac1{4\pi\ii}\oint_\Gamma\delta j\,
       \Tr(\chi_j'\sigma_3\chi_j^{-1})\,\dd z.
 \label{pos:variation-input}
\end{equation}
It is used initially near zero source, where the determinant is invertible.
Set $T=\Tr(\mathbf R_j'\sigma_3\mathbf R_j^{-1})$. The gauge part of
the trace is $2k_j'$; its contour integral is $\delta Q[j]$ by
Eq.~\eqref{derive:Q-selection}.
The square-root endpoint arcs of this integral vanish. For the remaining
term first use $\oint\delta k_jT=0$, since the integrand is analytic outside
the slit and decays at infinity. Thus replace $\delta j$ by
$\delta j+2\delta k_j=O(|z\mp q|^{1/2})$. In the prescribed logarithmic
endpoint class $T$ has at most a simple pole times logarithms, so these
modified endpoint arcs vanish. Shrinking the contour gives
$\oint\delta jT=-\ii\int\delta g_j(T_-+T_+)\,\dd x$.
Put $\mathfrak m=e^{\ii g_j}$, $J_{\mathfrak m}=\left(
\begin{smallmatrix}2\mathfrak m&-1\\1&0\end{smallmatrix}\right)$ and
$\mathsf Y=\mathbf R_{j,+}^{-1}\mathbf R_{j,+}'$. Direct multiplication gives
$J_{\mathfrak m}\sigma_3J_{\mathfrak m}^{-1}=-\sigma_3+4\mathfrak m E_{12}$,
$\Tr(J_{\mathfrak m}'\sigma_3J_{\mathfrak m}^{-1})=0$, and therefore
\begin{equation}
 \delta\log\mathcal D_{\rm ref}-\delta Q
 =-\frac1\pi\int_{-q}^q\delta g_j\,\mathfrak m\,\mathsf Y_{21}\,\dd x.
 \label{pos:source-variation}
\end{equation}
For comparison set $a=\mathbf R_{0,+}e_1$, $b^{\mathsf T}=e_2^{\mathsf T}\mathbf R_{0,+}^{-1}$,
$\beta=\mathfrak m-1$, and $\mathsf Z=\mathbf R_j\mathbf R_0^{-1}$. Then
$\mathsf Z_-=\mathsf Z_+(I+2\beta ab^{\mathsf T})$ and
$b(x)^{\mathsf T}a(y)/[\pi\ii(x-y)]=\mathsf K_*(x,y)$.
The integrable-resolvent formula gives, for
$V_*=M_\beta\mathsf K_*$,
\begin{equation}
 [V_*(I+V_*)^{-1}](x,x)
 =-\frac{\beta}{\pi\ii}
 [b^{\mathsf T}\mathsf Z^{-1}\mathsf Z'a+b^{\mathsf T}a']
 =-\frac{\beta}{\pi\ii}\mathsf Y_{21}.
 \label{pos:dual-resolvent}
\end{equation}
Its trace derivative, with $\delta\beta=\ii\mathfrak m\delta g_j$, is exactly
Eq.~\eqref{pos:source-variation}; removable zeros of $\beta$ are taken by
continuity. Both determinant sides of Eq.~\eqref{pos:kernel} equal one at
zero source. Integration proves their equality near zero, and analytic
continuation in a source-scaling parameter proves the determinant identity
beyond that neighborhood. No logarithm is continued through its zeros.
The source variation and resolvent identity thus fix the gauge term
and zero-source normalization of the weighted trace-class determinant.
For real analytic $j$, the unshifted kernel is $V_{\rm ref}[j]=\ii B_j$, where
$B_j(x,y)=\sinh([j(x)-j(y)]/2)/[\pi(x-y)]$ is real symmetric and trace
class.  Thus $\|(I+V_{\rm ref}[j])^{-1}\|\leq1$ for every real source.
The Laplace-branch inverse used next follows instead from the positive
kernel \eqref{pos:contraction}.

The same positive kernel supplies the uniform exterior estimate needed
in the return calculation. For the dressed source $p_*$ and
$v=g_{p_*}>0$ of Eqs.~\eqref{derive:dressed-source-definition} and
\eqref{pos:profile}, write $\mathbf R_L=\mathbf R_{\ii Lp_*}$,
$m=e^{-Lv}$, $\beta_L=m-1$, $a=\mathbf R_{0,+}e_1$ and
$b^{\mathsf T}=e_2^{\mathsf T}\mathbf R_{0,+}^{-1}$.  With the interval oriented left to right,
$\mathbf R_{L,-}=\mathbf R_{L,+}\bigl(\begin{smallmatrix}2m&-1\\1&0\end{smallmatrix}\bigr)$.
Its Cauchy representation is
\begin{align}
 f_L&=(I+\mathsf K_*M_{\beta_L})^{-1}a,
 &\|f_L\|_2&\leq2\|a\|_2,\notag\\
 \mathbf R_L(z)&=\left[I+\frac{1}{\pi\ii}
 \int_{-q}^{q}\frac{\beta_L(x)f_L(x)b(x)^{\mathsf T}}{z-x}\,\dd x\right]\mathbf R_0(z).
 \label{pos:exterior-representation}
\end{align}
The inverse bound follows directly from
$\|\mathsf K_*M_{\beta_L}\|\leq1/2$, without assuming that this product
is self-adjoint.  The local Cauchy term vanishes since $b^{\mathsf T}a=0$.
Using $\|a\|_2^2=\|b\|_2^2=4q/3$ gives, for
$\operatorname{dist}(z,[-q,q])\geq\delta>0$,
\begin{equation}
 \|\mathbf R_L(z)\|,\ \|\mathbf R_L(z)^{-1}\|
 \leq\left(1+\frac{8q}{3\pi\delta}\right)
       \left(1+\frac{2q}{\pi\delta}\right),
 \qquad \mathbf R_L^{\pm1}=I+O(z^{-1}),
 \label{pos:exterior}
\end{equation}
uniformly in $L$.  The prescribed logarithmic endpoint class makes the
small endpoint arcs vanish; the determinant has no jump and tends to one,
so $\det \mathbf R_L=1$.  In dimension two this also bounds the inverse.  Notice
that the zero-length matrix is $\mathbf R_0$, not $I$.

\subsection{The finite-coupling remainder}
Define $\mathcal W_c[j]=e^{-Q[j]}\mathcal D_c[j]$. Wherever the reference
determinant is nonzero, its exact relative factor is
\begin{equation}
 \mathcal D_c[j]=e^{Q[j]}\mathcal W_{\rm ref}[j]\,
 \frac{\mathcal D_c[j]}{\mathcal D_{\rm ref}[j]}.
 \label{round1:relative-source}
\end{equation}
The quotient follows directly by factoring
$I+V_c=(I+V_{\rm ref})[I+(I+V_{\rm ref})^{-1}(V_c-V_{\rm ref})]$
where the reference inverse exists. Analytic continuation applies to the product, including points where
the quotient alone may have poles. This factorization is exact at finite $c$. Inserting this identity into
\eqref{main:gaussian-input} leaves the correlated product
$e^{Q[\ii L\lambda+\phi]}\mathcal W_c[\ii L\lambda+\phi]$ inside the
same Gaussian average. Its two factors are not independent, and the
definition of $\mathcal W_c$ alone supplies no asymptotic bound on it.
The endpoint evaluation of the reference is deferred until the dressed
source is known; see Sec.~\ref{ret:endpoint}.

\section{Full Gaussian completion and the charge equation}
\label{main:charge}
\subsection{Source decomposition}
Section~\ref{pos:section} obtained $Q$ from the exact reference response.
Its operator form and the finite-$c$ remainder are
\begin{equation}
 Q[j]=\frac12\langle j,K_0^{-1}j\rangle,
 \qquad \mathcal W_c[j]=e^{-Q[j]}\mathcal D_c[j],
 \qquad Z_C=\E_C e^{Q[\phi]}.
 \label{main:quadratic}
\end{equation}

The pairings are between sources modulo constants and neutral charges.
Additive constants in $K_0$ and the anchoring terms of $C$ have no effect
on these pairings. Write $u_0(\lambda)=\lambda$. The source dependence of
the integrand splits as
\begin{equation}
 Q[\phi+\ii Lu_0]=Q[\phi]
 +\ii L\langle K_0^{-1}u_0,\phi\rangle
 -\frac{L^2}{2}\langle u_0,K_0^{-1}u_0\rangle.
 \label{derive:quadratic-expansion}
\end{equation}
The first term changes the Gaussian measure. The second is a
characteristic-function insertion in that new measure, and is correlated
with the source dependence of $\mathcal W_c$. Keeping this mixed term
is what changes the bare reference response into the finite-coupling
charge equation.
\subsection{Stability of the field integral}

\label{env:gap}
Let $K_0(d)=-2\log|d|$ and write a neutral charge supported in $[-a,a]$ as
$h=-W'$, with $W$ extended by zero. In this sector the difference
$K_0-C$ is the kernel $\mathcal L_c(d)=\log(1+c^2/d^2)$.
With Fourier convention $\widehat W(k)=\int e^{-\ii kx}W(x)\dd x$,
\begin{equation}
 \frac{\langle h,\mathcal L_c h\rangle}{\langle h,K_0h\rangle}
 =\frac{\int |k|(1-e^{-c|k|})|\widehat W(k)|^2\dd k}
 {\int |k||\widehat W(k)|^2\dd k}
 \ge \delta_{c,a}:=\frac{1-e^{-c/a}}2>0.
 \label{main:gap-bound}
\end{equation}
Indeed, the interactions of the zero extension with the exterior give
$\int |k||\widehat W|^2\dd k\ge(4/a)\|W\|_2^2$, whereas
$\int_{|k|<1/a}|k||\widehat W|^2\dd k\le(2/a)\|W\|_2^2$,
using $|\widehat W(k)|^2\le\|W\|_1^2\le2a\|W\|_2^2$.
At least half of the energy lies at $|k|\ge1/a$, proving the bound.
Completion in this energy norm includes endpoint-singular charges.
In operator terms, $K_0-C$ remains a strictly positive fraction of $K_0$
on a fixed interval. It therefore defines a unique finite-energy response
and leaves a strict margin for the Gaussian moment. We use it at $a=q$ here and at the larger buffer interval in
Sec.~\ref{env:section}.

\subsection{The tilted covariance and the dressed charge}
The gap and analytic covariance ensure $1\le Z_C<\infty$, including all
modes. More precisely, the covariance of the normalized Chebyshev
coefficients is trace class by analytic decay, while
\eqref{main:gap-bound} gives the corresponding norm strictly below one.
The trace-class condition and the gap are both needed for the Gaussian
Fredholm determinant. Thus
\begin{equation}
 \dd P_\Sigma(\phi)=Z_C^{-1}e^{Q[\phi]}\dd P_C(\phi),
 \qquad \Sigma=(C^{-1}-K_0^{-1})^{-1}.
 \label{derive:tilted-Gaussian}
\end{equation}
The inverse notation can first be read on finite Gaussian projections;
the moment margin permits their limit. It does not assume a bounded
inverse of the continuum covariance on all of $L^2$.

Define the dressed source and its charge by
\begin{equation}
 p_*:=u_0+\Sigma K_0^{-1}u_0,
 \qquad h_c:=K_0^{-1}p_*.
 \label{derive:dressed-source-definition}
\end{equation}
Applying $C^{-1}-K_0^{-1}$ to $\Sigma K_0^{-1}u_0$ gives
$\Sigma K_0^{-1}u_0=C K_0^{-1}(u_0+\Sigma K_0^{-1}u_0)=Ch_c$.
Therefore $K_0h_c=u_0+Ch_c$, or
\begin{equation}
 \mathcal L_c(d)=\log(1+c^2/d^2),\qquad
 \int_{-q}^q\mathcal L_c(\lambda-\mu)h_c(\mu)\,\dd\mu=\lambda,
 \qquad F_c=\frac12\int_{-q}^q\lambda h_c(\lambda)\,\dd\lambda.
 \label{main:charge-equation}
\end{equation}
The response is odd by reflection symmetry and uniqueness, and is neutral.
The useful covariance identities can now be collected without assuming
that $C$ commutes with $K_0$:
\begin{align}
 \Sigma&=(C^{-1}-K_0^{-1})^{-1}=C+C\mathcal L_c^{-1}C,
 &p_*&=\lambda+Ch_c,\nonumber\\
 K_0^{-1}+K_0^{-1}\Sigma K_0^{-1}&=\mathcal L_c^{-1}.
 \label{main:completed-covariance}
\end{align}
Since $p_*=K_0h_c$, Eq.~\eqref{derive:reference-charge} also gives
$g_{p_*}'=-2\pi h_c$. Its endpoint value is zero, so
$g_{p_*}=2\pi U$ with $U$ constructed in Sec.~\ref{sec:image}.
The same response fixes the entire quadratic exponent:
\begin{equation}
 \frac12\left[\langle u_0,K_0^{-1}u_0\rangle
 +\langle K_0^{-1}u_0,\Sigma K_0^{-1}u_0\rangle\right]
 =\frac12\langle u_0,h_c\rangle=F_c>0.
 \label{derive:completed-action}
\end{equation}

\subsection{Completion with the residual functional retained}
For a real linear source evaluation $J$, let $\eta$ have the centered
real Gaussian law of covariance $\Sigma$. Equations
\eqref{derive:quadratic-expansion} and \eqref{derive:tilted-Gaussian} give
\begin{align}
 &\E_C\!\left[e^{Q[\phi+\ii Lu_0]}
                  e^{\ii J(\phi+\ii Lu_0)}\right]\notag\\
 &\quad=Z_Ce^{-L^2\langle u_0,K_0^{-1}u_0\rangle/2-LJ(u_0)}
 \E_\Sigma e^{\ii\{L\langle K_0^{-1}u_0,\eta\rangle+J(\eta)\}}\notag\\
 &\quad=Z_Ce^{-F_cL^2}
           e^{-LJ(p_*)-\frac12\Var_\Sigma J(\eta)}.
 \label{derive:correlated-characteristic}
\end{align}
The Gaussian cross covariance is
$\operatorname{Cov}_\Sigma(\langle K_0^{-1}u_0,\eta\rangle,J(\eta))
=J(\Sigma K_0^{-1}u_0)$; it changes $J(u_0)$ into $J(p_*)$.
The finite-window positive-source construction in
Sec.~\ref{pos:source-proof} represents the whole residual as a positive
mixture of these evaluations. Section~\ref{gauss:section} takes the
source-window limit at fixed $L$. Applied to the complete integrand,
the identity is
\begin{equation}
 P_0(R)=\frac{Z_C}{\mathcal D_c[0]}e^{-F_cL^2}
 \E_\Sigma\mathcal W_c[\ii Lp_*+\eta].
 \label{round1:full-completion}
\end{equation}
The real Gaussian characteristic integral keeps the residual correlated
with the field and shifts its source to $\ii Lp_*+\eta$ through the cross
covariance. This completion requires neither a complex field-contour
shift nor a saddle approximation. The deterministic return estimate in
Sec.~\ref{ret:section} and the remaining field estimate in
Sec.~\ref{gauss:section} establish that the extracted quadratic exponent
is the leading cost.

\section{Image solution and polarization scales}
\label{sec:image}
\subsection{The charge primitive and its moment}
The odd function $h_c$ is the auxiliary response charge. Its primitive is
the profile needed for both the image construction and the return calculation:
\begin{equation}
 \begin{gathered}
 U(\lambda)=-\int_{-q}^{\lambda}h_c(\mu)\,\dd\mu,
 \qquad U=0\ \text{outside }[-q,q],\\
 |D_\lambda|\bigl(1-e^{-c|D_\lambda|}\bigr)U=\frac1{2\pi}\quad (|\lambda|<q),
 \quad D_\lambda=-\ii\partial_\lambda.
 \end{gathered}
 \label{main:U-equation}
\end{equation}
To obtain the second line, extend $U$ by zero and differentiate
$\mathcal L_c h_c=x$ in the open interval. Since $h_c=-U'$ and
$\widehat{\mathcal L_c}(k)=2\pi(1-e^{-c|k|})/|k|$, the differentiated
left side is $2\pi|D_\lambda|(1-e^{-c|D_\lambda|})U$. Neutrality gives $U(q)=U(-q)=0$,
so the extension introduces no endpoint delta charge in $h_c$.
The nonlocal operator in \eqref{main:U-equation} has jump kernel
\begin{equation}
 \nu_c(d)=\frac{c^2(c^2+3d^2)}{\pi d^2(c^2+d^2)^2}>0.
 \label{main:jump}
\end{equation}
Its maximum principle gives $U>0$ in the interval. The regular branch selected by the image construction below has
\begin{equation}
 2\pi U(\lambda)=\sqrt{q^2-\lambda^2}\,u(\lambda),
 \qquad u\text{ analytic and strictly positive on }[-q,q].
 \label{main:endpoint-profile}
\end{equation}
These two properties will later control the residual Gaussian average.
The explicit image solution is used here to evaluate the coefficient
and select this regular branch. In particular, $h_c$ may have an
inverse-square-root endpoint singularity. The finite-energy domain
consists of neutral charges $h=-W'$ whose zero-extended primitive obeys
$\int |k|\,|\widehat W(k)|^2\dd k<\infty$; bounded endpoint charge is not
assumed.

Integration by parts gives
$\int_{-q}^q\lambda h_c(\lambda)\,\dd\lambda
=-[\lambda U]_{-q}^{q}+\int_{-q}^qU(\lambda)\,\dd\lambda$.
It fixes all scale conversions:
\begin{equation}
 M_c=\int_{-q}^q U(\lambda)\,\dd\lambda=2F_c,
 \qquad f=\frac{4F_c}{\rho^2},\qquad
 \tau=\frac{LM_c}{c}=\frac{f\ell}{\gamma},\qquad F_cL^2=f\ell^2.
 \label{main:scales}
\end{equation}
Thus $M_c$ is the mass of the positive primitive $U$ and the first moment
of $h_c$. It is not the total charge: $\int h_c=0$.
For numerical work, Eqs.~\eqref{main:lieb} and \eqref{main:charge-equation}
already define $f(\gamma)$. The following construction evaluates the response as a grounded-boundary
problem.

\subsection{Image boundary problem and the two-face map}

For \(z=x+\ii y\), the
image potential is
\begin{equation}
 \varphi(z)=2\Re\int_{-q}^q h_c(\mu)
 \log\frac{z-\mu+\ii c/2}{z-\mu-\ii c/2}\dd\mu.
 \label{image:eq:Spotential}
\end{equation}
The denominator represents the real charge at $y=c/2$; the numerator is
its opposite image below the grounded line. On the real-charge line,
$2\Re\log[(x-\mu+\ii c)/(x-\mu)]
=\log[1+c^2/(x-\mu)^2]$. Hence the trace is exactly
$\mathcal L_ch_c=x$, while the two distances agree on $y=0$ and the
potential vanishes there. In the upper half-plane,
\(\Delta\varphi=-4\pi h_c(x)\delta(y-c/2)\), with \(h_c\) extended by zero
outside the slit. Green's identity therefore gives the finite response
energy
\begin{equation}
 F_c=\frac1{8\pi}\int_{y>0}|\nabla\varphi|^2\dd x\dd y
    =\frac12\int_{-q}^q xh_c(x)\dd x.
 \label{image:eq:Eenergy}
\end{equation}
The equality follows by integrating
$\int|\nabla\varphi|^2=-\int\varphi\Delta\varphi$; the grounded boundary
term vanishes and the slit trace is $x$. Oddness also makes $x=0$ an
equipotential. Subtract the background harmonic potential with that
same trace: $\Psi=2xy/c-\varphi$ solves Laplace's equation in the
first quadrant slit along \(0<x<q,\ y=c/2\), with zero values
on both faces and both axes and quadrupolar growth at infinity.
The finite-energy response excludes additional endpoint sources.

For the right-hand fin, map the \emph{lower} half of the \(t\)-plane by
\begin{equation}
 z'(t)=\ii a\frac{t-b}{\sqrt{t(t-1)(t-s)}},\qquad 1<b<s.
 \label{image:eq:Smap}
\end{equation}
Here \(a>0\) is a scale, \(z(t)\sim2\ii a\sqrt t\); each square root
is continued from
\(t>s\) through the lower half-plane, and the translation at infinity
is zero. The quadrant corner is \(t=0\). The successive prevertices
\(1,b,s\) are root, tip and
root. The intervals $(1,b)$ and $(b,s)$ traverse the lower and upper faces.
The zero of $z'$ at $b$ turns the map at the fin tip; the square-root
singularities at its roots give the two right-angle attachments to the
axis. Requiring both roots to coincide gives
\begin{equation}
 \int_1^s\frac{t-b}{\sqrt{t(t-1)(s-t)}}\dd t=0,
 \qquad b=\frac{sE(m_{\rm e})}{K(m_{\rm e})}.
 \label{image:eq:Sclosure}
\end{equation}
Here \(m_{\rm e}=1-1/s\). We use Legendre parameter conventions,
\(F(\theta|m_{\rm e})=\int_0^\theta(1-m_{\rm e}\sin^2 u)^{-1/2}\dd u\) and
\(E(\theta|m_{\rm e})=\int_0^\theta(1-m_{\rm e}\sin^2 u)^{1/2}\dd u\);
\(K(m_{\rm e})=F(\pi/2|m_{\rm e})\), \(E(m_{\rm e})=E(\pi/2|m_{\rm e})\).
Writing \(\alpha=\arcsin\sqrt{(b-1)/(s-1)}\), define
\begin{align}
 H_{\rm g}&=\frac{\pi\sqrt s}{K(m_{\rm e})},\qquad
 W_{\rm g}=2[bF(\alpha|1-s)-E(\alpha|1-s)],\nonumber\\
 P_{\rm g}&=\frac{2b(1+s)-s}{3}-b^2.
 \label{image:eq:HWP}
\end{align}
The map gives \(c/2=aH_{\rm g}\), \(q=aW_{\rm g}\).
Thus \(c/q=2H_{\rm g}/W_{\rm g}\) fixes \(s>1\); its strict
monotonicity and endpoint regularity are checked below. All elliptic
integrals above are real, including the parameter \(1-s<0\).

In the lower half-plane the harmonic function that vanishes on the real
axis and has the required growth is
$\Psi(z(t))=-4a^2\Im t/c$. Its normalization follows from
$z\sim2\ii a\sqrt t$, hence $2xy/c=\Im(z^2)/c\sim-4a^2\Im t/c$.
With
\(t=1+(s-1)\sin^2\theta\), the coordinate on both faces is
\begin{equation}
 x_{\rm fin}(t)=2a[bF(\theta|1-s)-E(\theta|1-s)].
 \label{image:eq:Sfin}
\end{equation}
It increases from zero to $q$ on the lower face and decreases to zero
on the upper face. Write its inverse branches as
$t_-(x)\in(1,b)$ and $t_+(x)\in(b,s)$.

\paragraph{The two normal fields determine the charge.}
Use $\partial_y$ in the same upward direction on both faces, rather than
two oppositely oriented outward normals. On either face the
Cauchy--Riemann equations give $\partial_y\Im t=\dd t_\pm/\dd x$.
The smooth background $2xy/c$ has no derivative jump, so
\begin{equation}
 \begin{split}
 (\partial_y\varphi)_+-(\partial_y\varphi)_-&=-4\pi h_c(x),\\
 (\partial_y\Psi)_\pm&=-\frac{4a^2}{c}\,t_\pm'(x),\\
 h_c(x)&=-\frac{a^2}{\pi c}\frac{\dd}{\dd x}
                    [t_+(x)-t_-(x)].
 \end{split}
 \label{derive:normal-field-jump}
\end{equation}
Both inverse branches meet at $b$ when $x=q$. Integrating the last line
with $U(q)=0$ therefore yields
\begin{equation}
 U(x)=\frac{a^2}{\pi c}[t_+(x)-t_-(x)]>0,
 \qquad 0<x<q.
 \label{derive:image-primitive}
\end{equation}
The positive primitive is the separation of the two face coordinates.
Both normal fields contribute to the derivative jump that determines
the charge and its moment.

\paragraph{The charge moment is an area in the map plane.}
To evaluate it, define
\[
 I_j=\int_1^s\frac{t^j\,\dd t}{\sqrt{t(t-1)(s-t)}}.
\]
Closure gives $I_1=bI_0$. Since $U$ is even,
$F_c=\frac12\int_{-q}^qU=\int_0^qU$. The area between the inverse
branches can be integrated in either order:
\begin{equation}
 \begin{split}
 F_c&=\frac{a^2}{\pi c}\int_0^q[t_+(x)-t_-(x)]\,\dd x
     =\frac{a^2}{\pi c}\int_1^s x_{\rm fin}(t)\,\dd t\\
 &=-\frac{a^2}{\pi c}\int_1^s t\,x_{\rm fin}'(t)\,\dd t
   =\frac{a^3}{\pi c}(I_2-bI_1).
 \end{split}
 \label{derive:moment-area}
\end{equation}
Here $x_{\rm fin}'=a(b-t)/\sqrt{t(t-1)(s-t)}$ and both endpoint
values of $x_{\rm fin}$ vanish. Also $I_0=2\pi/H_{\rm g}$.
Integrating the derivative of $\sqrt{t(t-1)(s-t)}$ from $1$ to $s$
gives $3I_2=2(1+s)I_1-sI_0$. Substituting these relations gives
\begin{equation}
 F_c=\frac{a^3}{\pi c}(I_2-b^2I_0),\qquad
 f=\frac{4q^2}{\rho^2}\frac{P_{\rm g}}{H_{\rm g}^2W_{\rm g}^2}.
 \label{image:eq:ellipticrate}
\end{equation}
The combination $I_2-b^2I_0=\int_1^s(t-b)^2\dd t/
\sqrt{t(t-1)(s-t)}$ is positive, giving a positive charge energy and rate.

\subsection{Branch selection and coefficient evaluations}
\label{sec:charge}
The preceding construction fixes $a,s,b,H_{\rm g},W_{\rm g},P_{\rm g}$ and inverse
face coordinates $t_-<t_+$. The identity
$U(x_{\rm fin})=a^2(t_+-t_-)/(\pi c)$ fixes the positive regular branch
used in Eq.~\eqref{main:endpoint-profile}.
The endpoint coefficient is also nonzero:
\begin{equation}
 \lim_{\lambda\uparrow q}
 \frac{2\pi U(\lambda)}{\sqrt{q^2-\lambda^2}}
 =\frac{2[b(b-1)(s-b)]^{1/4}}{H_{\rm g}(s)\sqrt{W_{\rm g}(s)}}>0.
 \label{eq:Sendpoint}
\end{equation}
With \(\delta=1/s\), \(m_{\rm e}=1-\delta\),
\(e=E(m_{\rm e})/K(m_{\rm e})\), the aspect
derivative is
\begin{equation}
 \frac{\dd(\pi q/c)}{\dd\delta}
 =-\frac{K(m_{\rm e})}{2\delta(1-\delta)}
                      \sqrt{e(e-\delta)(1-e)}<0.
\end{equation}
This fixes the real charge branch.

\begin{figure}[!ht]
 \centering
 \includegraphics[width=.63\textwidth]{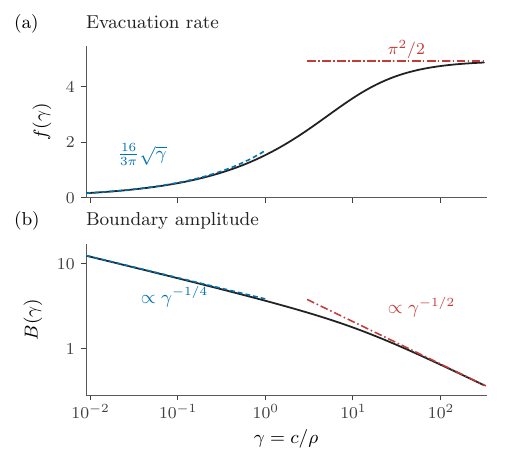}
 \caption{Analytic coefficient curves from the image solution:
\(f(\gamma)\) and \(B(\gamma)\), with their weak- and strong-coupling
limits. The curves are evaluated from the analytic formulas.
The five measurement couplings and their analytic
comparison values appear in Section~\ref{sec:numerics}.}
 \label{fig:Scoefficients}
\end{figure}

An evaluation starts with \(c/q\), solves the one-dimensional aspect
equation, and computes \(\Fc/q^2\). The Lieb equation~\eqref{main:lieb}
then supplies \(\rho/q\), hence \(\gamma\) and \(f\). The incomplete elliptic integrals use the parameter convention with
\(1-s<0\), rather than the modulus convention. At small \(c/q\)
the expression involves cancellations and benefits from increased
arithmetic precision. The accompanying data provide reference evaluations and the
192/384-node equilibrium-density comparison. At the six reference ratios
$c/q=0.2,0.5,1,2,4,10$, the relative density changes are below
$9\times10^{-14}$. The value of $B$ follows its relation to $f$.
These checks assess the numerical evaluation of the analytic coefficients.
The limiting forms \(\Fc/q^2\to1/8\) and
\(\Fc/q^2\sim[6\pi(c/q)]^{-1}\), together with
\(q\sim\pi\rho\) and \(q\sim2\sqrt{c\rho}\), respectively,
give the two physical limits stated in the Letter.

\FloatBarrier
\section{The collective return determinant}
\label{ret:section}

The completion in Sec.~\ref{main:charge} leaves a deterministic source
$\ii Lp_*$ and a Gaussian fluctuation $\eta$. This section evaluates
the determinant at that deterministic source. Section~\ref{gauss:section}
then controls the full average over $\eta$. Throughout, $c,q>0$ are fixed.
Let $U$ be the positive charge profile, $M_c=\int_{-q}^{q}U(\lambda)\,\dd\lambda=2F_c$,
and $\tau=LM_c/c$.  We prove
\begin{equation}
 -\log\det(I-\mathsf T_L)
 =\frac{2\zeta(3/2)}{\sqrt\pi}\sqrt\tau
  -\frac12\log\tau+O_{c,q}(1).
 \label{ret:result}
\end{equation}
The proof has two exact changes of representation, separated by two
controlled comparisons. The positive compression fixes the determinant
and its real logarithm; the shifted kernel exposes the two channels.
Removing its matrix dressing changes the logarithm by $O(\tau^{-1/2})$,
and replacing the distributed charge by an equal-mass point charge
changes it by $O(1)$. The resulting point-charge operator is exactly a
projected Bessel imaginary-time propagator. Its full spectrum determines
both growing terms in \eqref{ret:result}.

\subsection{The exact positive return operator}
Start on $\mathcal H_I=L^2([-q,q],\dd\lambda)$. Let
$\mathrm{Ext}:\mathcal H_I\to L^2(\mathbb R)$ extend a function by zero,
and let $P_\pm$ project onto the positive/negative Fourier subspaces
$\mathcal H_\pm$. With $D_\lambda=-\ii\partial_\lambda$, the multiplier
$e^{-c|D_\lambda|}$ has Fourier symbol $e^{-c|k|}$ and represents propagation
across the imaginary rapidity separation $c$. On the channel space
$\mathcal H_+\oplus\mathcal H_-$ define
\begin{equation}
 \begin{gathered}
 (\mathsf B_L f)=\bigl(P_+[e^{-\ii Lp_*/2}\mathrm{Ext}f],\,P_-[e^{\ii Lp_*/2}\mathrm{Ext}f]\bigr),
 \qquad \mathsf E_c=e^{-c|D_\lambda|}\oplus e^{-c|D_\lambda|},\\
 I+V_{\rm ref}[\ii Lp_*]=\mathsf B_L^\dagger\mathsf B_L=:\mathsf A_{\rm ref},
 \qquad I+V_c[\ii Lp_*]=\mathsf B_L^\dagger(I-\mathsf E_c)\mathsf B_L,\\
 \mathcal V=\mathsf B_L\mathsf A_{\rm ref}^{-1/2},\qquad
 \mathsf T_L=\mathcal V^\dagger\mathsf E_c\mathcal V.
 \end{gathered}
 \label{ret:compression}
\end{equation}
Both \(\mathsf B_L\) and \(\mathcal V\) map
\(\mathcal H_I\) into \(\mathcal H_+\oplus\mathcal H_-\), whereas
\(\mathsf A_{\rm ref}\) and \(\mathsf T_L\) act on \(\mathcal H_I\).
The symbol \(\mathcal V\) is the isometry used in End Matter C.
The two components keep the two scattering phases. Normalizing by
$\mathsf A_{\rm ref}^{-1/2}$ makes $\mathcal V$ an isometry,
and hence
\begin{equation}
 I+V_c[\ii Lp_*]
 =\mathsf A_{\rm ref}^{1/2}(I-\mathsf T_L)
                       \mathsf A_{\rm ref}^{1/2},
 \qquad
 \frac{\mathcal D_c[\ii Lp_*]}{\mathcal D_{\rm ref}[\ii Lp_*]}
 =\Det(I-\mathsf T_L).
 \label{derive:positive-quotient}
\end{equation}
The inverse square root is initially defined on its quadratic-form
domain. At fixed $L$, the finite interval and the propagation multiplier
make $\mathsf T_L$ trace class. Positivity of the forms in
\eqref{ret:compression} gives $0\le\mathsf T_L<I$. Thus the real
logarithm has the connected-cycle expansion
\begin{equation}
 -\log\Det(I-\mathsf T_L)
 =\sum_{m\ge1}\frac{\Tr\mathsf T_L^m}{m}.
 \label{derive:trace-cycles}
\end{equation}
One cycle consists of propagation followed by projection back into the
source-dependent channel subspace; $m$ counts repetitions of that cycle.
At $L=0$ the
phases disappear, $\mathsf A_{\rm ref}=I$ and $\mathsf T_0=P_c$, which
recovers the normalization $\mathcal D_c[0]$.

\subsection{The exact shifted-kernel representation}
The positive compression identifies the real logarithm and the repeated
returns. Its source-dependent subspace is inconvenient for large-$L$
estimates. A shifted Cauchy representation exposes
the finite-coupling propagation and the reference resolvent separately.
Determinant rearrangements of the form $\det(I-AB)=\det(I-BA)$ preserve
the determinant on their stated spaces; the subsequent removal of the
matrix dressing will instead require an asymptotic estimate.

Define the analytic charge potential and the normalized profile by
\begin{equation}
 h(z)=\ii\int_{-q}^{q}\frac{U(u)}{u-z}\,\dd u,
 \qquad
 \mu(\dd u)=\frac{c\,U(cu)}{M_c}\,\dd u.
 \label{ret:charge}
\end{equation}
Here \(h(z)\) is an analytic potential built from \(U\), distinct
from the response charge \(h_c=-U'\).
Thus $\mu$ is a probability measure supported in $[-q/c,q/c]$.
The exact shifted factorization \cite{suppItsKozlowski2013}, Propositions~2.1--2.3, gives
\begin{equation}
 \frac{\mathcal D_c[\ii Lp_*]}{\mathcal D_{\rm ref}[\ii Lp_*]}
 =\det(I-\mathsf T_L)=\det(I+\mathsf N_L).
 \label{ret:exact-factor}
\end{equation}
In the same line coordinates and channel ordering,
\(\mathsf N_L=\mathsf N_j|_{j=\ii Lp_*}\), where \(\mathsf N_j\)
is the general-source operator of Sec.~\ref{pos:source-proof}.
The gauge matrices obey \(\mathbf R_L=\mathbf R_{\ii Lp_*}\).
Generalized sine determinants also occur in dynamical correlators
\cite{suppGohmannKozlowskiMinin2026}. Here the source grows with $L$ and is
averaged over the Gaussian field, requiring the estimates developed below.

We use the Cauchy/Hardy realization on
$\mathcal H_-\oplus\mathcal H_+$, with the two components exchanged
relative to \eqref{ret:compression}. In line coordinates $x,y\in\mathbb R$,
set $\sigma_1=1$, $\sigma_2=-1$,
$z_i=x-\ii\sigma_i c/2$ and $w_j=y+\ii\sigma_jc/2$.
The exact kernel is
\begin{equation}
 (\mathsf N_L)_{ij}(x,y)=\frac{\sigma_i}{2\pi\ii}
 \frac{e^{-\sigma_iLh(z_i)+\sigma_jLh(w_j)}
 [\mathbf R_L(z_i)^{-1}\mathbf R_L(w_j)]_{ij}-\delta_{ij}}{z_i-w_j}.
 \label{ret:exact-kernel}
\end{equation}
For $i\ne j$ and $x=y$, take the removable divided difference before
separating the channels:
\begin{equation}
 (\mathsf N_L)_{ij}(x,x)=-\frac{\sigma_i}{2\pi\ii}
 e^{-2\sigma_iLh(z_i)}[\mathbf R_L(z_i)^{-1}\mathbf R_L'(z_i)]_{ij}.
 \label{ret:mixed-confluence}
\end{equation}
This generally nonzero block is retained in the relative comparison.
The auxiliary matrix $\mathbf R_L$ is the normalized unshifted matrix after extracting
the charge exponential.  The positive representation in
Eq.~\eqref{pos:kernel} gives the exterior bound \eqref{pos:exterior}:
$\mathbf R_L^{\pm1}$ are uniformly bounded away from the cut and equal
$I+O(z^{-1})$ at infinity, uniformly in $L$.

Section~\ref{pos:reference-proof} constructs the reference resolvent matrix
$\mathbf R_L$ and derives its exterior bounds. The subtraction
$-\delta_{ij}$ in \eqref{ret:exact-kernel} removes the source-independent
Cauchy background. The asymptotic comparison below includes this
subtraction and all mixed blocks.
\subsection{Comparison with two scalar channels}
\label{ret:reduction}
On the positive-frequency subspace of $L^2(\mathbb R)$, let
\begin{equation}
 \begin{gathered}
 D_x=-\ii\partial_x\ge0,\qquad \mathsf E=e^{-D_x},\qquad G_0=I-\mathsf E,\\
 w_{\mu,\tau}(z)=\exp\!\left[-\ii\tau\int\frac{\mu(\dd u)}{z-u}\right],
 \qquad d_{\mu,\tau}(x)=w_{\mu,\tau}(x+\ii/2),\\
 K_{\mu,\tau}=\mathsf E-M_d\mathsf EM_d^*,\qquad
 \mathsf A_{\mu,\tau}=I-K_{\mu,\tau}=G_0+M_d\mathsf EM_d^* .
 \end{gathered}
 \label{ret:model}
\end{equation}
Here $M_d$ is multiplication by $d$; stars denote Hilbert-space adjoints.
Positive frequencies extend analytically to the upper half-plane, and
$e^{-aD_x}f(x)=f(x+\ii a)$.  Consequently
$M_de^{-D_x/2}=e^{-D_x/2}M_w$ and
$K_{\mu,\tau}=e^{-D_x/2}(I-M_wM_w^*)e^{-D_x/2}\ge0$.
Its diagonal is integrable, so it is trace class.

The unitary rescaling \((\mathcal S_cf)(x)=\sqrt c\,f(cx)\)
gives \(\mathcal S_c(cD_\lambda)\mathcal S_c^{-1}=D_x\).
Thus \(\mathsf E\) is the positive-frequency restriction of the rescaled
single-channel multiplier; \(\mathsf E_c\) retains both channels.
The rescaling $\lambda=cx$ has put the propagation distance equal to one.
In these variables the source damping is
\begin{equation}
 |d_{\mu,\tau}(x)|^2
 =\exp\!\left[-\tau\int\frac{\mu(\dd u)}{(x-u)^2+1/4}\right].
 \label{derive:return-damping}
\end{equation}
For fixed charge support, its exponent is $-\tau/x^2+O(\tau/|x|^3)$
at large $|x|$. Damping is saturated in the inner region and crosses
over at $|x|\sim\sqrt\tau$. The corresponding rapidity scale is
$c\sqrt\tau=\sqrt{cLM_c}$. This explains the scale entering the
return calculation; the logarithm and the numerical coefficient still
require the spectral calculation below.

Replacing $\mathbf R_L$ by $I$ in \eqref{ret:exact-kernel}, followed by the unitary
rescaling $\lambda=cx$, gives two conjugate copies of $-K_{\mu,\tau}$.
Write $\mathsf N_L^0$ for that two-channel comparison kernel.
In the following relative comparison, both $\mathsf N_L$ and
$\mathsf N_L^0$ denote the operators after the same unitary rescaling
on both channels; we retain their symbols.
The superscript zero means \(\mathbf R_L\) has been replaced by \(I\);
it does not mean zero source. In particular, \(\mathsf N_L^0\) is
distinct from \(\mathsf N_0=\mathsf N_j|_{j=0}\), whose reference matrix
\(\mathbf R_0\) is not \(I\). We write
$\mathcal A=\overline{\mathsf A}_{\mu,\tau}\oplus
\mathsf A_{\mu,\tau}=I+\mathsf N_L^0$. Complex conjugation identifies the
negative-frequency copy with the positive-frequency one. We prove the relative
estimate
\begin{equation}
 E_L=\mathcal A^{-1/2}(\mathsf N_L-\mathsf N_L^0)\mathcal A^{-1/2},
 \qquad \|E_L\|_1\le C\tau^{-1/2},
 \qquad I+\mathsf N_L=\mathcal A^{1/2}(I+E_L)\mathcal A^{1/2}.
 \label{derive:relative-factorization}
\end{equation}
\paragraph{Damped Cauchy vectors.}
Write $\mathsf A=\mathsf A_{\mu,\tau}$, $\mathsf M=M_d$, and
$\mathsf Q=\mathsf A^{-1/2}\mathsf M$ locally, and set
$k_{u,n}(x)=(x-u+\ii/2)^{-n}$. Uniformly for probability measures
$\mu$ on the fixed support $[-q/c,q/c]$ and $u$ in that support, we prove
\begin{equation}
 \|\mathsf Qk_{u,1}\|_2^2\le C\tau^{-1/2},\qquad
 \|\mathsf Qk_{u,2}\|_2^2\le C\tau^{-3/2}.
 \label{ret:vectors}
\end{equation}
The geometric series for $(I-\mathsf E)^{-1}$ and analytic translation show
\begin{equation}
 \|G_0^{-1/2}\mathsf Mf\|_2^2
 =\sum_{m\ge0}\|\mathsf E^{m/2}\mathsf Mf\|_2^2
 \le\sum_{m\ge0}\|\mathsf E^{m/2}f\|_2^2.
 \label{ret:weighted-contraction}
\end{equation}
This is first a quadratic-form statement on the finite-norm domain.
It implies $\mathsf MG_0\mathsf M^*\le G_0$, hence
$\mathsf A\ge\mathsf M\mathsf M^*$ and $\|\mathsf Q\|\le1$.
At fixed $\tau$, $1/d$ is analytic and bounded by $e^{2\tau}$, so
$\mathsf A\ge e^{-4\tau}I$ and the inverse exists.
Also $\mathsf A\ge G_0$.  Comparing regularized inverses and then removing the
regularizer gives, whenever the right side is finite,
\begin{equation}
 \|\mathsf Qf\|_2^2\le\sum_{m\ge0}\int_{\mathbb R}
 |w(x+\ii y_m)|^2|f(x+\ii m/2)|^2\,\dd x,
 \qquad y_m=(m+1)/2.
 \label{ret:depth-bound}
\end{equation}
For all probability measures in the fixed support,
$|w(x+\ii y)|^2\le\exp[-\beta\tau y/(x^2+y^2)]$ for $y\ge1/2$.
With $f=k_{u,2}$, the substitution $x=y\tan\theta$ bounds the height integral
by $Cy^{-3}\min\{1,(y/\tau)^{3/2}\}$.  Summing the heights below and above
$\tau$ proves the second estimate in \eqref{ret:vectors}.
For $k_{u,1}$ the direct depth sum diverges.  Subtract
$g_b(x)=(x+\ii b)^{-1}$, $b\ge1$; at every depth,
\begin{equation}
 |(k_{u,1}-g_b)(x+\ii m/2)|
 \le\frac{Cb}{x^2+y_m^2}.
 \label{ret:pole-subtraction}
\end{equation}
Apply \eqref{ret:depth-bound} to the difference and $\|\mathsf Q\|\le1$ to $g_b$.
Since $\|g_b\|_2^2=\pi/b$,
$\|\mathsf Qk_{u,1}\|_2^2\le C(b^2\tau^{-3/2}+b^{-1})$.
Taking $b=\sqrt\tau$ proves the first estimate.
The same subtraction works for analytic vector functions with a uniformly
bounded $1/z$ coefficient and $O(z^{-2})$ remainder at every outward depth.

\paragraph{Cauchy contour factorization.}
Because $h(\bar z)=-\overline{h(z)}$, the row and column exponential
multipliers are conjugates.  For a fixed counterclockwise contour $\Gamma$
around the cut, contained in $|\Im z|<c/4$,
\begin{equation}
 \frac{\mathbf R_L(w)-\mathbf R_L(z)}{z-w}
 =\frac1{2\pi\ii}\oint_\Gamma
 \frac{\mathbf R_L(\zeta)-I}{(z-\zeta)(w-\zeta)}\,\dd\zeta.
 \label{ret:contour}
\end{equation}
Thus $\mathsf N_L-\mathsf N_L^0$ is an integral of separated row and column vectors.
The left vector contains $\mathbf R_L(z)^{-1}/(z-\zeta)$. The exterior
bound \eqref{pos:exterior} controls this full factor and supplies the
uniform large-$z$ expansion at every translated depth needed above.
After the unitary rapidity rescaling, both channel vectors obey the first
estimate of \eqref{ret:vectors}.  The trace norm of a rank-one operator is
the product of the two vector norms. Multiplying these bounds and
integrating over the fixed contour proves the $\|E_L\|_1\le C\tau^{-1/2}$
estimate in Eq.~\eqref{derive:relative-factorization}, including all mixed blocks.
The divided difference in \eqref{ret:contour} also retains the finite
mixed-channel value when $z=w$.
At fixed $\tau$, $\mathcal A^{1/2}-I$ is trace class. Taking the
determinant of Eq.~\eqref{derive:relative-factorization} and using
$|\log\det(I+E_L)|\le\|E_L\|_1/(1-\|E_L\|)$ for large $\tau$ gives

\begin{equation}
 \log\Det(I-\mathsf T_L)=2\log\Det \mathsf A_{\mu,\tau}+O(\tau^{-1/2}),
 \label{ret:relative-conclusion}
\end{equation}
Equation~\eqref{ret:exact-factor} fixes the real logarithm.
For a return eigenvalue $t$ close to one, the logarithm responds to a
perturbation with weight $(1-t)^{-1}$. The relative factors in
Eq.~\eqref{derive:relative-factorization} control this amplification,
which an absolute kernel bound alone would miss.
\subsection{Equal-mass point-charge comparison}
The residual-field estimate in Sec.~\ref{gauss:section} uses the original
regular profile $U$. In the return determinant, replacing the distributed
charge by a point charge of equal mass changes the logarithm by a bounded
amount:
\begin{equation}
 \log\Det \mathsf A_{\mu,\tau}=\log\Det \mathsf A_{\delta_0,\tau}+O(1).
 \label{ret:point-reduction}
\end{equation}
To control the determinant beyond the far-field expansion, use the interpolation
$\mu_s=(1-s)\delta_0+s\mu$. Set
$\mathsf A_s=\mathsf A_{\mu_s,\tau}$,
$\mathsf Q_s=\mathsf A_s^{-1/2}M_{d_s}$,
$k_u(x)=(x-u+\ii/2)^{-1}$ and
$\vartheta_s(u)=\|\mathsf Q_sk_u\|_2^2$.
Differentiating the exact Cauchy kernel in \eqref{ret:model} yields
\begin{equation}
 \partial_s \mathsf A_s=-\frac{\tau}{2\pi}M_{d_s}
 \left[\int |k_{u,1}\rangle\langle k_{u,1}|\,
                  (\mu-\delta_0)(\dd u)\right]M_{d_s}^*.
 \label{ret:charge-derivative}
\end{equation}
Taking the logarithmic trace and using equal mass gives
\begin{equation}
 \partial_s[-\log\det\mathsf A_s]
 =\frac{\tau}{2\pi}\int[\vartheta_s(u)-\vartheta_s(0)]\,\mu(\dd u),
 \qquad |\vartheta_s'(u)|\le C\tau^{-1}.
 \label{derive:point-comparison-mechanism}
\end{equation}
For the derivative bound, hold $\mathsf Q_s$ fixed when differentiating
in $u$. Cauchy--Schwarz and Eq.~\eqref{ret:vectors}, applied to $\mu_s$,
give $|\vartheta_s'(u)|\le2\|\mathsf Q_sk_{u,1}\|_2
\|\mathsf Q_sk_{u,2}\|_2\le C\tau^{-1}$, uniformly in $s\in[0,1]$.
Equal mass produces the difference inside the integral. Its absolute
value is at most $C\tau^{-1}|u|$, uniformly along the interpolation.
The prefactor $\tau$ cancels this bound, and integrating over
$0\le s\le1$ proves \eqref{ret:point-reduction} because the support is
fixed. No centering assumption is needed for this bounded comparison.
The detailed shape of $U$ can change the constant term; it cannot change
the growing terms at this accuracy.

\subsection{Point charge, Bessel propagation, and all return cycles}

\label{ret:bessel}

For $\mu=\delta_0$, the bounded analytic function
$w(z)=e^{-\ii\tau/z}$ has unit modulus on the real boundary almost
everywhere. Multiplication by $w$ is therefore an isometry on
$\mathcal H_+$. Its range complement has orthogonal projection
$\Pi_w=I-M_wM_w^*$, and
$K_{\delta_0,\tau}=e^{-D_x/2}\Pi_we^{-D_x/2}$.
The identity $\det(I-AB)=\det(I-BA)$ now gives an exact restriction:
\begin{equation}
 \det\mathsf A_{\delta_0,\tau}
 =\det{}_{\operatorname{Ran}\Pi_w}
       (I-\Pi_we^{-D_x}\Pi_w).
 \label{derive:model-space}
\end{equation}

The unitary inversion $\mathcal I f(z)=z^{-1}f(-1/z)$ changes the
multiplier to $e^{\ii\tau z}$. Under the positive-frequency Fourier
transform this multiplier shifts the frequency support from $(0,\infty)$
to $(\tau,\infty)$. Its range complement is therefore $L^2(0,\tau)$.
On the real boundary $z=x$, the same transformations act on the generator as
\begin{equation}
 D_x\ \xrightarrow{\ \mathcal I\ }\ x^2D_x-\ii x
 \ \xrightarrow{\ \text{Fourier}\ }\ -\partial_s(s\partial_s).
 \label{derive:Bessel-generator}
\end{equation}
For the last change of variables let $x=2\sqrt s$ and
$({\cal U}f)(x)=\sqrt{x/2}\,f(x^2/4)$, which is unitary from
$L^2(\dd s)$ to $L^2(\dd x)$. Direct differentiation yields
$\mathcal U[-\partial_s(s\partial_s)]\mathcal U^{-1}
=-\partial_x^2-1/(4x^2)$, and hence
\begin{equation}
 \begin{gathered}
 \det \mathsf A_{\delta_0,\tau}=\det(I-T_{\XB}),\qquad \XB=2\sqrt\tau,\\
 T_{\XB}=P_{\XB}e^{-\mathsf{H}_{\mathrm{B}}}P_{\XB},\qquad
 \mathsf{H}_{\mathrm{B}}=-\partial_x^2-\frac{1}{4x^2},\\
 p_t(x,y):=e^{-t\mathsf{H}_{\mathrm{B}}}(x,y)
 =\frac{\sqrt{xy}}{2t}e^{-(x^2+y^2)/(4t)}I_0\!\left(\frac{xy}{2t}\right).
 \end{gathered}
 \label{ret:heat-model}
\end{equation}
\(\mathsf{H}_{\mathrm{B}}\) acts on \(L^2(0,\infty;\dd x)\), and
\(T_{\XB}\) is its projected propagator on \(L^2(0,\XB;\dd x)\).
The length \(\XB=2N_*\) belongs to the auxiliary Bessel problem,
distinct from the physical coordinate \(X\) in Sec.~\ref{phys:response}.
$P_{\XB}$ restricts to $(0,\XB)$. The two local zero-energy branches at the
origin are $\sqrt x$ and $\sqrt x\log x$. The transformation from the
regular radial generator selects the first: $f(x)/\sqrt x$ has no
logarithmic branch. This regularity condition fixes the Bessel boundary problem at the origin.

In the Hankel-transform representation, $T_{\XB}$ is the truncated
Bessel operator of order zero with symbol $\chi(k)=e^{-k^2}$.
Basor and Ehrhardt established determinant asymptotics for regular symbols
\cite{suppBasorEhrhardt2003}; their theorem does not directly cover the
quadratic zero of $1-\chi(k)$ at $k=0$.
Belitsky and Korchemsky proposed a Fisher--Hartwig extension
\cite{suppBelitskyKorchemsky2021}, Eq.~(5.11).
With Bessel order zero, singularity exponent $\beta=1$, and $2g=\XB$,
that conjecture predicts the linear-in-$\XB$ and logarithmic terms
in Eq.~\eqref{ret:scalar-asymptotic} below.
Here the heat comparison supplies the required $O(1)$ remainder
independently. Its connection to the Lieb--Liniger EFP also requires
the preceding charge and relative-operator comparisons, the original
rapidity endpoints, and the residual-field estimate.

The operator $T_{\XB}$ propagates on the half-line and projects after
one unit of imaginary time, allowing excursions beyond $\XB$ between
projections. A permanent absorbing wall removes those paths.

Let $\mathsf{H}_{\mathrm{B},\XB}^{\mathrm{D}}$ also impose absorption at $\XB$.
Projection at integer times removes fewer paths than continuous absorption.
We prove the uniform estimate
\begin{equation}
 0\le\Delta_{\XB}(m):=\operatorname{Tr}T_{\XB}^m-
       \operatorname{Tr}e^{-m\mathsf{H}_{\mathrm{B},\XB}^{\mathrm{D}}}
 \le C\frac{\sqrt{\log(2+m)}}{\sqrt m},
 \qquad \XB\ge1,\quad m\ge1.
 \label{ret:cycle-bound}
\end{equation}
The difference counts paths that leave $(0,\XB)$ and return between two
integer observation times.

\paragraph{One-step bridge loss and all-cycle control.}
OpenAI's GPT-6 Astra suggested the deep-pole subtraction in
Eq.~\eqref{ret:pole-subtraction} and the one-step planar-bridge localization
used below.
We prove \eqref{ret:cycle-bound} directly by localizing missed crossings
near the absorbing wall. Write $X=\XB$ and
$D_Y(m)=\Tr e^{-m\mathsf H_{\mathrm B,Y}^{\mathrm D}}$.
The regular radial realization at the origin is unchanged.
The angular integral for $I_0$ gives
$I_0(z)\leq Ce^z/\sqrt{1+z}$ for $z\geq0$, so \eqref{ret:heat-model} implies
\[
 p_t(x,y)\leq Ct^{-1/2}e^{-(x-y)^2/(4t)}.
\]
Move the absorbing wall to $X+a$, $a\geq1$. A planar unit-time bridge from
$xe_1$ to $ye^{\ii\theta}$ has its chord inside radius $\max(x,y)$.
To reach the wall its centered fluctuation must exceed
$a+X-\max(x,y)$. The Gaussian bridge tail, integrated over the angle,
therefore bounds the one-step loss. With $w_X(x)=e^{-\kappa(X-x)^2}$,
\begin{equation}
 \begin{split}
 E_{X,a}(x,y)&:=p_1(x,y)-p_1^{\mathrm D,X+a}(x,y)\geq0,\\
 E_{X,a}(x,y)&\leq Ce^{-\kappa[a+X-\max(x,y)]^2}p_1(x,y)\\
 &\leq Ce^{-\kappa a^2}[w_X(x)+w_X(y)]p_1(x,y),\qquad 0<x,y<X.
 \end{split}
 \label{ret:localized-loss}
\end{equation}
All constants are independent of $X,a,m$. The bridge representation
follows from the planar heat kernel by angular integration and the
positive radial weight in \eqref{ret:heat-model}.

Loops remaining below $X+a$ contribute at most $D_{X+a}(m)$.
Every other loop crosses that wall in at least one of its $m$ unit steps.
A union bound over those steps, with the other projections removed,
gives, for $m\geq2$,
\begin{equation}
 \begin{split}
 \Tr T_X^m-D_{X+a}(m)
 &\leq m\int_0^X\!\!\int_0^X
 E_{X,a}(x,y)p_{m-1}(y,x)\,\dd x\,\dd y\\
 &\leq Cm e^{-\kappa a^2}\int_0^Xw_X(x)p_m(x,x)\,\dd x
 \leq C\sqrt m\,e^{-\kappa a^2}.
 \end{split}
 \label{ret:missed-cycles}
\end{equation}
Nonnegativity permits the kernel comparisons. The second line uses
symmetry and the semigroup identity
$\int_0^\infty p_1(x,y)p_{m-1}(y,x)\dd y=p_m(x,x)$;
then $p_m(x,x)\leq C/\sqrt m$ and $\int_0^Xw_X\dd x\leq C$.
For $m=1$, the diagonal integral of \eqref{ret:localized-loss} gives
\eqref{ret:missed-cycles} directly.

The regular eigenfunctions at a wall $Y$ are proportional to
$\sqrt x\,J_0(kx)$; absorption at $Y$ gives $k=j_{0,n}/Y$,
where $j_{0,n}$ are the positive zeros of $J_0$. Hence
\[
 D_Y(m)=\sum_{n\geq1}e^{-m j_{0,n}^2/Y^2},\qquad
 \partial_YD_Y(m)=\frac{2m}{Y^3}\sum_{n\geq1}
 j_{0,n}^2e^{-m j_{0,n}^2/Y^2}\leq\frac C{\sqrt m}.
\]
For the last bound use $j_{0,n}\geq c_*n$,
$u e^{-u}\leq Ce^{-u/2}$, and the Gaussian sum bounded by its integral.
Continuous absorption at $X$ removes more paths than integer-time testing,
so $D_X(m)\leq\Tr T_X^m$. Integrating the derivative bound and using
\eqref{ret:missed-cycles} yields
\begin{equation}
 0\leq\Delta_X(m)
 \leq C\left[\frac a{\sqrt m}+\sqrt m\,e^{-\kappa a^2}\right].
 \label{ret:buffer-bound}
\end{equation}
Taking $a=A\sqrt{\log(2+m)}$ with fixed $A$ sufficiently large proves
Eq.~\eqref{ret:cycle-bound} for all $X\geq1$ and integer $m\geq1$.

The bound is uniform in both the interval
size and the cycle length, and
$\sum_{m\ge1}\Delta_{\XB}(m)/m\le
C\sum_{m\ge1}\sqrt{\log(2+m)}\,m^{-3/2}<\infty$.
This uniform summability permits expansion of the connected logarithm
and summation over the continuously absorbed modes, giving
\begin{equation}
 -\log\det(I-T_{\XB})
 =\sum_{n\ge1}-\log\!\left(1-e^{-j_{0,n}^2/\XB^2}\right)+O(1),
 \label{ret:dirichlet-comparison}
\end{equation}
Thus the eigenvalues in \eqref{ret:dirichlet-comparison} include the
boundary condition inherited from the exact return operator.

\paragraph{The bulk spacing and the endpoint shift.}

Only the leading spacing and endpoint shift of these zeros are needed.
The large-argument expansion of $J_0$ gives
$j_{0,n}=\pi(n-\tfrac14)+O(n^{-1})$.
Replacing $j_{0,n}^2$ by $\pi^2(n-\tfrac14)^2$ in
\eqref{ret:dirichlet-comparison} costs $O(1)$: their difference is bounded,
and the derivative of $-\log(1-e^{-v/\XB^2})$ has magnitude at most $1/v$,
so the errors are bounded by $C\sum_n n^{-2}$.
Set $a=\pi/\XB$ and $\mathfrak b(y)=-\log(1-e^{-y^2})$.
Near zero, $\mathfrak b(y)=-2\log y+O(y^2)$. After removing this
singularity, $\mathfrak b(y)+2\log y\,\mathbf1_{y\le1}$ is integrable
with finite total variation, so its Riemann-sum error is bounded.
For the removed part, as $a\downarrow0$, let
$N=\lfloor a^{-1}+1/4\rfloor$. The endpoint
shift appears explicitly in the Gamma product:
\begin{equation}
 \begin{split}
 -2\sum_{n=1}^{N}\log[a(n-\tfrac14)]
 &=-2N\log a-2\log\Gamma(N+\tfrac34)
                      +2\log\Gamma(\tfrac34)\\
 &=\frac{2}{a}+\frac12\log a+O(1).
 \end{split}
 \label{derive:boundary-Gamma-product}
\end{equation}
The last line is Stirling's formula with $N=a^{-1}+O(1)$.
The coefficient of $\log a$ comes from the $3/4$ offset, whereas the
integral of $\mathfrak b$ gives the bulk term. Restoring the regular
part of the sum yields
\begin{equation}
 \begin{split}
 \sum_{n\ge1}\mathfrak b\!\left(a(n-\tfrac14)\right)
 &=\frac1a\int_0^\infty \mathfrak b(y)\,\dd y+\frac12\log a+O(1),\\
 \int_0^\infty \mathfrak b(y)\,\dd y
 &=\sum_{m\ge1}\frac1m\int_0^\infty e^{-my^2}\,\dd y
 =\frac{\sqrt\pi}{2}\zeta(3/2).
 \end{split}
 \label{ret:riemann-sum}
\end{equation}
Equations~\eqref{ret:heat-model}--\eqref{ret:riemann-sum} yield
\begin{equation}
 \log\det \mathsf A_{\delta_0,\tau}
 =-\frac{\zeta(3/2)}{\sqrt\pi}\sqrt\tau
   +\frac14\log\tau+O(1).
 \label{ret:scalar-asymptotic}
\end{equation}
For one channel, $\XB=2\sqrt\tau$ and
$\log a=\log\pi-\log2-\frac12\log\tau$. These conversions give the
$\sqrt\tau$ and $\frac14\log\tau$ terms in
\eqref{ret:scalar-asymptotic}; the conjugate channel doubles both.
Summing all cycles in \eqref{derive:trace-cycles} produces $\zeta(3/2)$.
Related polylogarithms occur in exclusion-process current statistics
\cite{suppDerridaGerschenfeld2009}, for a different observable; here the
cycle sum determines the void boundary cost.
Combining this with \eqref{ret:relative-conclusion} and
\eqref{ret:point-reduction} proves \eqref{ret:result}. In this calculation
the return range grows as $\sqrt L$; after the quadratic charge response
has been extracted there is no return term proportional to $L$.
The original endpoint and Gaussian factors must still be checked before
concluding $A=0$ for the full probability.

\subsection{The original rapidity endpoint contribution}
\label{ret:endpoint}

The charge solution in Eq.~\eqref{main:endpoint-profile} has
\begin{equation}
 v(x):=g_{p_*}(x)=2\pi U(x)
 =\sqrt{q^2-x^2}\,u(x),\qquad 0<u_{\min}\leq u(x)\leq u_{\max}<\infty.
 \label{pos:profile}
\end{equation}
For $j=\ii Lp_*$ the symmetrized operator in \eqref{pos:kernel} is
$I-\mathsf K_*^{1/2}M_{1-e^{-Lv}}\mathsf K_*^{1/2}\geq I/2$.
Its log determinant decreases as $v$ increases: the trace derivative
along a positive interpolation has a fixed sign.  Hence
$\mathcal W_{\rm ref}[\ii Lu_{\max}x]
\leq\mathcal W_{\rm ref}[\ii Lp_*]
\leq\mathcal W_{\rm ref}[\ii Lu_{\min}x]$.
The two bounds are ordinary sine-kernel gap determinants with their
quadratic exponent removed, at $s=qLu_{\min}/2$ and $s=qLu_{\max}/2$.  Their known asymptotic
\cite{suppDeiftItsKrasovskyZhou2006} gives
\begin{equation}
 r_{\rm end}(L):=\log\mathcal W_{\rm ref}[\ii Lp_*]
 =-\frac14\log(qL)+O_{c,q}(1).
 \label{pos:endpoint}
\end{equation}
The constants $u_{\min},u_{\max}$ only change the bounded term. This logarithm belongs
to the original occupied-rapidity endpoints. It is separate from the
Bessel endpoint shift generated by the return comparison. Their
contributions to $S=-\log P_0$ are respectively $+\frac14\log L$ and
$-\frac12\log L$; Sec.~\ref{sec:assembly} combines them after controlling
the remaining Gaussian average.

\section{The remaining Gaussian average}
\label{gauss:section}
The deterministic calculation in Sec.~\ref{ret:section} evaluated
$\mathcal W_c[\ii Lp_*]$. The completed probability contains
$\E_\Sigma\mathcal W_c[\ii Lp_*+\eta]$ instead. We compare these quantities by integrating the full Gaussian field at
each finite auxiliary-source window. Positivity of the auxiliary source
measure reduces the comparison to a bound on a nonnegative scalar random
variable.
\subsection{Finite-rank source shifts and the positive source functional}
\label{pos:source-proof}

Keeping both shifted scattering channels yields the normalized auxiliary generating functional
\begin{equation}
 \displaystyle
 \Phi_c[j]=\frac{\mathcal W_c[j]}{\mathcal D_c[0]}
 =\mathcal W_{\rm ref}[j]
 \frac{\Det(I+\mathsf N_j)}{\Det(I+\mathsf N_0)}.
 \label{pos:characteristic}
\end{equation}
Here $\mathsf N_j$ is the complete shifted kernel and
$\Det(I+\mathsf N_0)=\mathcal D_c[0]$. We construct this identity below
as a limit of positive finite-window ensembles, retaining mixed-channel
minors and the source-independent Cauchy subtraction.

\paragraph{Shifted factorization and finite-rank source shifts.}
OpenAI's GPT-6 Astra suggested the finite-rank source-shift construction
below, which expresses the paired principal terms, including the reference
factor, as Fourier transforms of finite positive measures.
For $k=1,2$, put $\sigma_k=3-2k$,
$z_k(x)=x-\ii\sigma_kc/2$ and $w_k(x)=\overline{z_k(x)}$.
The exact shifted factorization \cite{suppItsKozlowski2013} uses
$\mathsf N_j=\mathsf M_j-\Cshift$, where
\begin{equation}
 \Cshift_{kl}(x,y)=
 \frac{\sigma_k\delta_{kl}}{2\pi\ii[z_k(x)-w_l(y)]},\qquad
 \mathsf M_{j,kl}(x,y)=
 \frac{\sigma_k[\chi_j(z_k(x))^{-1}\chi_j(w_l(y))]_{kl}}
 {2\pi\ii[z_k(x)-w_l(y)]}.
 \label{pos:shift-kernel}
\end{equation}
Here $0\leq\Cshift\leq I$; coincident mixed-channel quotients have their
removable values.  The subtraction $\Cshift$ removes the source-independent Cauchy background.

For finitely many paired points with labels $k_i$, define the matrices
\begin{gather}
 (\mathbf A_j)_{ij}=
 \frac{[\chi_j(z_i)^{-1}\chi_j(w_j)]_{k_i k_j}}{z_i-w_j},\qquad
 (\mathbf A_{\mathrm C})_{ij}=\frac{\delta_{k_i k_j}}{z_i-w_j},\quad
 \notag\\
 \ell_S(x)=\sum_i\sigma_{k_i}\log\frac{x-w_i}{x-z_i}.
 \label{pos:finite-matrices}
\end{gather}
The shifted determinant factorization is supplied by
Ref.~\cite{suppItsKozlowski2013}, Sec.~2. We now derive the finite-rank
source identity needed for the positive measure:
\begin{equation}
 \mathcal D_{\rm ref}[j]\det\mathbf A_j
 =\mathcal D_{\rm ref}[j+\ell_S]\det\mathbf A_{\mathrm C}.
 \label{pos:source-shift}
\end{equation}
To derive this identity from the integrable resolvent, proceed as follows.
Group $r_k(x)=\prod_{i:k_i=k}(x-w_i)/(x-z_i)$.
Rational interpolation gives
\begin{equation}
 \frac{r_k(y)/r_k(x)-1}{x-y}
 =\sum_{a,b:k_a=k_b=k}
 \frac{[(\mathbf A_{\mathrm C,k})^{-1}]_{ab}}{(x-w_a)(y-z_b)}.
 \label{pos:interpolation}
\end{equation}
This is a finite-rank update of the unshifted kernel.  With
$f_i(x)=e^{-\sigma_{k_i}j(x)/2}/(x-w_i)$,
$h_i(x)=e^{\sigma_{k_i}j(x)/2}/(x-z_i)$ and
$d_i=\sigma_{k_i}/(2\pi\ii)$, its Cauchy resolvent reads
$(\mathbf A_j)_{ij}=(\mathbf A_{\mathrm C})_{ij}
+d_jh_i(I+V_{\rm ref}[j])^{-1}f_j$.
Order the paired points by channel. Then $\mathbf A_{\mathrm C}$
is block diagonal, but the resolvent correction has entries for every
pair $i,j$, including $k_i\ne k_j$. Applying the finite-rank determinant
lemma to the complete update gives
\[
 \frac{\mathcal D_{\rm ref}[j+\ell_S]}{\mathcal D_{\rm ref}[j]}
 =\det\!\left[I+\mathbf A_{\mathrm C}^{-1}
                      (\mathbf A_j-\mathbf A_{\mathrm C})\right]
 =\frac{\det\mathbf A_j}{\det\mathbf A_{\mathrm C}},
\]
which is \eqref{pos:source-shift}. The block-diagonal background does
not split this determinant into independent channel factors: its
mixed resolvent entries remain inside the same determinant. Pairings
are bilinear; a Hermitian adjoint would be incorrect for complex $j$.
The calculation is first made for distinct paired points and invertible
reference determinant; confluent cases follow by continuity and the
undivided identity by analytic continuation.

For conjugate pairs in the appropriate half-planes, $\ell_S=\ii p_S$
with $p_S$ real.  Combining \eqref{pos:source-shift} and
\eqref{pos:unshifted-law} yields
\begin{equation}
 \mathcal W_{\rm ref}[j]\det\mathsf M_j(S,S)
 =\det\Cshift(S,S)e^{-Q[p_S]}
 \E_*\!\left[e^{-J_*(p_S)}
 e^{\ii\{J_*(j)+\vartheta_S(j)\}}\right].
 \label{pos:paired}
\end{equation}
The matrix $\Cshift(S,S)$ is a positive Gram matrix.  The real linear
functional $\vartheta_S$ is the cross term
$Q[j+\ii p_S]-Q[j]-Q[\ii p_S]=\ii\vartheta_S(j)$.
Each pair at depth $a/2$ contributes the positive Poisson average
\begin{equation}
 \vartheta_{x,a}(j)
 =\frac1\pi\int_{-q}^q
 \frac{(a/2)g_j(t)}{(t-x)^2+(a/2)^2}\,\dd t.
 \label{pos:poisson}
\end{equation}
Thus every paired principal term in \eqref{pos:paired} is the Fourier
transform of a finite positive measure, also for unequal depths and mixed
channels.  This property concerns the complete minors, not separate channel
determinants.

\paragraph{Removing the background by a relative limit.}
At zero source, substitution of \eqref{pos:Fzero} gives
\begin{equation}
 \mathsf M_0=\Cshift-\mathsf B_{\rm C}\mathsf B_{\rm C}^*\geq0,
 \qquad
 (\mathsf B_{\rm C})_k(x,t)=\frac{\sigma_k}{2\pi[z_k(x)-t]},
 \quad -q\leq t\leq q.
 \label{pos:zero-source}
\end{equation}
The Cauchy map \(\mathsf B_{\rm C}\) joins the interval to the
shifted line channels. It is distinct from \(\mathsf B_L\) in
Eq.~\eqref{ret:compression}, including its value at \(L=0\).
The channel decomposition gives $\mathsf B_{\rm C}\mathsf B_{\rm C}^*\leq\Cshift$.
Furthermore $\mathsf B_{\rm C}^*\mathsf B_{\rm C}=P_c$, the interval restriction of
$e^{-c|D_\lambda|}$, is a compact strict contraction.  Hence
$H_0:=I+\mathsf N_0=I-\mathsf B_{\rm C}\mathsf B_{\rm C}^*\geq\epsilon_0 I$
for some fixed $\epsilon_0>0$.

Retain finitely many enlarged channels:
\[
 \mathsf L_J(f_0,\ldots,f_J)=\sum_{n=0}^{J}\Cshift^{n/2}f_n,
 \qquad G_J:=\mathsf L_J\mathsf L_J^*=\sum_{n=0}^{J}\Cshift^n.
\]
The Cauchy kernels translate the rows and columns of
$\widetilde{\mathsf M}_j=\mathsf L_J^*\mathsf M_j\mathsf L_J$
to depths $(n+1)c/2$. Let $P_\Lambda$ restrict each enlarged channel to
$|x|\leq\Lambda$. Equation~\eqref{pos:paired} makes
\begin{equation}
 \Phi_{\srcwin}[j]=\mathcal W_{\rm ref}[j]
 \frac{\Det(I+P_\Lambda\widetilde{\mathsf M}_jP_\Lambda)}
      {\Det(I+P_\Lambda\widetilde{\mathsf M}_0P_\Lambda)},
 \qquad \srcwin=(J,\Lambda),
 \label{pos:finite-law}
\end{equation}
a normalized characteristic functional: its numerator is a sum of positive
source measures, including mixed channels, and its denominator is their
total mass. The functional $J_{\srcwin}(j)$ is a sum of evaluations
$g_j(t)$ and positive averages \eqref{pos:poisson}.

Although $G_J$ need not converge, its inverse does. For $0\leq t\leq1$,
$[\sum_{n=0}^{J}t^n]^{-1}-(1-t)
=t^{J+1}/\sum_{n=0}^{J}t^n\leq1/(J+1)$. Functional calculus gives
\begin{equation}
 G_J^{-1}=I-\Cshift+R_J,
 \qquad 0\leq R_J\leq\frac{I}{J+1}.
 \label{pos:inverse-limit}
\end{equation}

The infinite background is not trace class, but
$\Delta_j:=\mathsf M_j-\mathsf M_0=\mathsf N_j-\mathsf N_0$ is:
its Cauchy resolvent factors through two square-integrable kernels joining
the interval to the shifted lines. Form the relative determinant in
\eqref{pos:finite-law} before removing the window. At fixed $J$,
$\mathsf L_J^*\Delta_j\mathsf L_J$ is trace class, and
$(I+P_\Lambda\widetilde{\mathsf M}_0P_\Lambda)^{-1}$ converges strongly
with norm at most one. The relative perturbation therefore converges in
trace norm. Using
\[
 \mathsf L_J(I+\mathsf L_J^*\mathsf M_0\mathsf L_J)^{-1}\mathsf L_J^*
 =(G_J^{-1}+\mathsf M_0)^{-1}=(H_0+R_J)^{-1}
\]
and cyclicity of the relative determinant gives
\begin{equation}
 \lim_{\Lambda\to\infty}\Phi_{J,\Lambda}[j]
 =\mathcal W_{\rm ref}[j]\Det[I+(H_0+R_J)^{-1}\Delta_j].
 \label{pos:relative-limit}
\end{equation}
Since $H_0+R_J\geq\epsilon_0I$, the resolvent identity gives
\[
 \|[(H_0+R_J)^{-1}-H_0^{-1}]\Delta_j\|_1
 \leq\frac{\|\Delta_j\|_1}{\epsilon_0^2(J+1)}.
\]
Thus $J\to\infty$ gives \eqref{pos:characteristic}, with
$\Det(I+\mathsf N_0)=\Det(I-P_c)=\mathcal D_c[0]>0$.
The order is $\Lambda\to\infty$ at fixed $J$, then $J\to\infty$;
only normalized relative determinants are used for the infinite background.

At each window $\srcwin$, the construction gives
$\Phi_{\srcwin}[j]=\E_{\srcwin}e^{\ii J_{\srcwin}(j)}$, with
$J_{\srcwin}$ real and linear for real sources. The window regulates
this auxiliary positive representation, not the number of Gaussian field modes.

The positivity required for the Gaussian calculation is explicit already
at finite $\srcwin$.  With $v=g_{p_*}>0$ from \eqref{pos:profile}, its point
charges and Poisson averages define a finite positive weighted measure
$\mu_{\srcwin}$ such that
\begin{equation}
 J_{\srcwin}(j)=\int_{-q}^q\frac{g_j(t)}{v(t)}\,\dd\mu_{\srcwin}(t),
 \qquad Y_{\srcwin}:=J_{\srcwin}(p_*)=\mu_{\srcwin}([-q,q])\geq0.
 \label{pos:weighted-measure}
\end{equation}
For analytic $j$ the ratio extends to the endpoints.

\subsection{Completion and the tilted source measure}
For a fixed auxiliary-source configuration, averaging over $\eta$ is
an ordinary Gaussian characteristic integral. Its linear phase has zero
mean and produces a nonnegative quadratic cost. Denote the source
amplitude and that cost by
\begin{equation}
 Y_{\srcwin}=J_{\srcwin}(p_*)\ge0,
 \qquad H_{\srcwin}=\frac12\Var_\Sigma J_{\srcwin}(\eta),
 \label{gauss:variables}
\end{equation}
where $\eta$ is the full real Gaussian field of covariance $\Sigma$.
Use the positive measure already defined in
\eqref{pos:weighted-measure}. The inequality
$|J_{\srcwin}(\eta)|\le Y_{\srcwin}\|g_\eta/v\|_\infty$ gives
\begin{equation}
 Y_{\srcwin}=\int\dd\mu_{\srcwin},
 \qquad J_{\srcwin}(\eta)=\int\frac{g_\eta(\lambda)}{v(\lambda)}\,\dd\mu_{\srcwin}(\lambda),
 \qquad H_{\srcwin}\le K Y_{\srcwin}^2,
 \quad K=\frac12\E_\Sigma\|g_\eta/v\|_\infty^2<\infty.
 \label{gauss:quadratic-bound}
\end{equation}
The ratio extends to the endpoints: if $\eta(q\cos\theta)=\sum a_n\cos(n\theta)$, then
$g_\eta=\sum a_n\sin(n\theta)$ and $|\sin(n\theta)|\le n|\sin\theta|$.
Analytic coefficient decay, the strict field moment margin, and positive $u$ in \eqref{main:endpoint-profile} make $K$ finite and independent of the window and $L$. Explicitly,
$\|g_\eta/v\|_\infty\le(q u_{\min})^{-1}\sum n|a_n|$.
Under $\dd P_\Sigma=e^{Q}\dd P_C/Z_C$, H\"older's inequality with the strict
moment margin and analytic Gaussian moments bounds its second moment.
The bound uses the actual regular profile $U$ and its endpoint behavior.
Constants may depend on fixed $c,q$, the analytic strip and $u_{\min}$;
uniformity as $\gamma\to0$ is not asserted.

At finite $\srcwin$, insert the characteristic functional into the field
integral and apply \eqref{derive:correlated-characteristic} configuration
by configuration. Integrating that identity over the positive source law
gives
\begin{equation}
 \E_C\!\left[e^{Q[\phi+\ii L\lambda]}\Phi_{\srcwin}[\phi+\ii L\lambda]\right]
 = Z_Ce^{-F_cL^2}\E_{\srcwin} e^{-LY_{\srcwin}-H_{\srcwin}}.
 \label{gauss:complete}
\end{equation}
The exchange is dominated by $\E_Ce^{Q[\phi]}$: the elementary source
$u_0(\lambda)=\lambda$ has $g_{u_0}=\sqrt{q^2-\lambda^2}\ge0$,
so $J_{\srcwin}(u_0)\ge0$ and the extra Laplace factor has modulus at most
one. The cross covariance produces $Y_{\srcwin}=J_{\srcwin}(p_*)$ at the
dressed source, while the remaining characteristic integral produces
$H_{\srcwin}$, giving the completion used in Sec.~\ref{main:charge}.

The finite auxiliary-source construction makes the completed integral
\eqref{round1:full-completion} equivalent to the following real-source identity:
\begin{equation}
 P_0(R)=Z_Ce^{-F_cL^2}
 \lim_{\srcwin}\E_{\srcwin} e^{-LY_{\srcwin}-H_{\srcwin}},
 \qquad Y_{\srcwin}\ge0,\quad 0\le H_{\srcwin}\le K Y_{\srcwin}^2.
 \label{main:master}
\end{equation}

Separate the deterministic Laplace weight from the residual field cost:
\begin{equation}
 \begin{aligned}
 Z_{\srcwin}(L)&=\E_{\srcwin} e^{-LY_{\srcwin}}, &
 I_{\srcwin}(L)&=-\log Z_{\srcwin}(L),\\
 G_{\srcwin}(L)&=\E_{\srcwin,L}e^{-H_{\srcwin}}, &
 \dd\nu_{\srcwin,L}&=Z_{\srcwin}(L)^{-1}e^{-LY_{\srcwin}}\dd\nu_{\srcwin}.
 \end{aligned}
 \label{gauss:tilt}
\end{equation}
The law $\nu_{\srcwin,L}$ is tilted toward configurations with smaller
$Y_{\srcwin}$ by the positive factor $e^{-LY_{\srcwin}}$. At each fixed $L$,
the source-window limits give $Z_{\srcwin}\to Z$ and $G_{\srcwin}\to Z_{\rm G}$.
To justify them, use the joint real variables
$(J_{\srcwin}(\eta),J_{\srcwin}(p))$ with either $p=u_0$ or $p=p_*$.
Since $g_{u_0}\ge0$ and $g_{p_*}>0$, the weak limit retains a nonnegative
second coordinate; its bounded Fourier--Laplace transform agrees with
the entire Fredholm source function by the real-source identity.
The choice $p=u_0$ identifies the original numerator in
\eqref{main:gaussian-input}; $p=p_*$ treats the completed average.
Moreover $|\Phi_{\srcwin}[\eta+\ii Lp_*]/Z_{\srcwin}(L)|\le1$.
Dominated convergence therefore passes the full Gaussian integral at
fixed $L$, without assuming a limiting random functional in an
infinite-dimensional dual space. The large-$L$ estimate is taken only
after this limit.

\subsection{Tail and moment bounds}
Under the tilted measure in \eqref{gauss:tilt},
\begin{equation}
 \Pr_{\srcwin,L}(Y_{\srcwin}>y)
 \le\min\{1,e^{I_{\srcwin}(L)-Ly}\},
 \qquad
 \E_{\srcwin,L}Y_{\srcwin}^2
 \le\frac{I_{\srcwin}(L)^2+2I_{\srcwin}(L)+2}{L^2}.
 \label{gauss:tail}
\end{equation}
Indeed, the tilted numerator of this tail is at most $e^{-Ly}$ and its
normalization is $Z_{\srcwin}=e^{-I_{\srcwin}}$. Integrating $2y$ times the
resulting bound and splitting at $y=I_{\srcwin}(L)/L$ gives the second
inequality. Jensen's inequality and \eqref{gauss:quadratic-bound} then give
\begin{equation}
 0\le-\log G_{\srcwin}(L)\le\E_{\srcwin,L}H_{\srcwin}
 \le K\frac{I_{\srcwin}(L)^2+2I_{\srcwin}(L)+2}{L^2}.
 \label{gauss:jensen}
\end{equation}
Take the window limits at fixed $L$. The deterministic source determinant gives
\begin{equation}
 Z(L)=\frac{\mathcal W_c[\ii Lp_*]}{\mathcal D_c[0]}
 =\frac{e^{r_{\rm end}(L)}\Det(I-\mathsf T_L)}{\mathcal D_c[0]},
 \qquad I(L)=-\log Z(L)=O(\sqrt L).
 \label{gauss:deterministic}
\end{equation}
Only after these limits do we increase $L$. Equations~\eqref{gauss:jensen} and \eqref{gauss:deterministic} prove
\begin{equation}
 \displaystyle
 0\le-\log Z_{\rm G}(L)
 \le K\frac{I(L)^2+2I(L)+2}{L^2}=O(L^{-1}).
 \label{gauss:bound}
\end{equation}
The deterministic cost $I(L)=O(\sqrt L)$ and the tilt $e^{-LY}$ imply a second-moment bound
of order $L^{-1}$ for $Y$. The remaining Gaussian cost is at most
$KY^2$, so its average changes $-\log P_0$ by $O(L^{-1})$.
The covariance has already entered the leading action through $F_c$;
\eqref{gauss:bound} concerns only the factor left after that exact
completion. It contributes neither a growing term nor a logarithm.

\section{Assembly, physical response, and limits}
\label{sec:assembly}
\subsection{Assembly and normalization}

Combining the endpoint and return estimates with \eqref{main:factorization} gives
\begin{equation}
 S(\ell;\gamma)=F_cL^2+
 \frac{2\zeta(3/2)}{\sqrt\pi}\sqrt\tau
 -\frac12\log\tau+\frac14\log(qL)+O_{c,q}(1).
 \label{gauss:assembly}
\end{equation}
Equation~\eqref{main:scales} yields \eqref{main:result}. The two return channels supply $-\tfrac12\log\ell$, while the original endpoints supply $+\tfrac14\log\ell$. There is no term proportional to $L$.

The absence of a linear term is a conclusion about all factors together:
the quadratic charge action has been extracted exactly, the saturated
return region has width $\sqrt L$, the reference endpoints contribute
only a logarithm, and the remaining field factor is bounded as above.
Together, these estimates exclude a linear term. The square-root
coefficient is fixed by the polarization moment already determined by
the image solution, without an independent boundary parameter.

At zero source, $\Det(I-\mathsf T_0)=\mathcal D_c[0]$, $r_{\rm end}(0)=0$, and the exact field identity gives $Z_CZ_{\rm G}(0)=1$. Hence $P_0(0)=1$ with the same normalization used at large $L$. The stated asymptotics take the thermodynamic limit first and then $L\to\infty$ at fixed $c,\rho>0$.
\subsection{Physical response and measurement conditions}
\label{phys:response}

\paragraph{Moving the empty boundary.}
The void probability also determines the cost of extending an existing empty
interval.  Define the response scales
\begin{equation}
 n_*=f\rho^2R,\qquad d_*=(cn_*)^{-1/2},\qquad N_*=n_*d_*.
 \label{phys:scales}
\end{equation}
These definitions use the scalar rate; no local density profile is assumed.
Using $M_c=2F_c$ and $L=2R$ gives
$n_*=LM_c$, $d_*^{-1}=\sqrt{cLM_c}$ and $N_*^2=\tau$.
The inverse response length $d_*^{-1}$ therefore equals the growing rapidity scale
found in \eqref{derive:return-damping}. This equality of scales does not
identify the auxiliary charge profile $U(\lambda)$ with a physical
density profile in position space.
Let $G_\pm$ be the distances from the two empty endpoints to the nearest
particles.  Translation invariance gives, exactly,
\begin{equation}
 \Pr\!\left(G_-\ge d_*u_-,\ G_+\ge d_*u_+
       \mid [-R,R]\text{ empty}\right)
 =\frac{P_0\!\left(R+\tfrac12d_*(u_-+u_+)\right)}{P_0(R)}.
 \label{phys:gap-ratio}
\end{equation}
Only $S(\ell;\gamma)=f\ell^2+B\sqrt\ell+o(\sqrt\ell)$ is needed below.
For bounded $u_\pm\ge0$, its quadratic term changes by
$N_*(u_-+u_+)+O(d_*^2)$; the remaining difference is $o(N_*)$.
Thus the negative logarithm of \eqref{phys:gap-ratio}, divided by $N_*$,
converges locally uniformly to $u_-+u_+$.  This is a large-deviation statement
for additional gaps of order $d_*$, rather than a distributional limit on
the smaller scale $n_*^{-1}=d_*/N_*$.

Averaging the boundary density over the conditioning radius gives a
further consequence.  Let $n_r(x)$ denote the density
conditioned on $[-r,r]$ being empty.  Moving the two boundaries in the
configuration integral gives $\rho\,\partial_\ell S(\rho r;\gamma)=2n_r(r^+)$ almost everywhere.
This is the one-dimensional void/conditional-intensity identity
\cite{suppTorquatoEtAl2008}: in the notation of that reference,
$E_V(r)=P_0(r)$, $s_1(r)=2$, and $\rho G_V(r)=n_r(r^+)$.
Indeed, stationarity implies
$0\le P_0(r)-P_0(r+\epsilon)\le2\rho\epsilon$, so $S$ is locally absolutely
continuous wherever $P_0>0$.  With $d_*$ fixed at the base radius $R$, put
$R_u=R+d_*u/2$ and $H_R(u)=[S(\rho R_u;\gamma)-S(\ell;\gamma)]/N_*$.
The same finite-difference argument gives $H_R\to u$ locally uniformly.
Consequently, for every continuous compactly supported test function $\chi$,
\begin{equation}
 \int \chi(u)\frac{n_{R_u}(R_u^+)}{n_*}\,\dd u
 \longrightarrow \int\chi(u)\,\dd u.
 \label{phys:contact-average}
\end{equation}
For smooth $\chi$, this follows by integration by parts from
$H_R'=n_{R_u}(R_u^+)/n_*$; positivity and locally bounded mass extend the result
to continuous tests.

\paragraph{Convexity of the emptiness action.}
OpenAI's GPT-6 Astra suggested the Robin heat-flow argument below for
establishing log-concavity of the ground-state wavefunction on the ordered
alcove.
We give the finite-volume argument before taking the thermodynamic limit.
Let $M\ge2$ bosons occupy a ring of circumference $\mathcal L$, and
normalize the positive periodic ground wave $\Psi_0$ on $[0,\mathcal L)^M$.
After removing the center of mass, the ordered relative alcove is
\[
 \mathcal A=\{x_1<\cdots<x_M<x_1+\mathcal L,\ \textstyle\sum_i x_i=0\}.
\]
The contact cusp gives $\partial_n\Psi_0=-\alpha\Psi_0$ on every face,
with outward unit normal $n$ and $\alpha=c/\sqrt2$.
Its incenter $C$ has equally spaced coordinates and inradius
$r=\mathcal L/(M\sqrt2)$. Thus $g=\exp[-a|x-C|^2]$,
$a=\alpha/(2r)$, satisfies the same Robin conditions.

\emph{Robin heat evolution.}
Set $w=e^{t\Delta_{\rm Robin}/2}g>0$ and $z=w/g$.
Then $\partial_n z=0$, $z(0)=1$, and
\[
 z_t=\tfrac12\Delta z-2a(x-C)\cdot\nabla z+
       [2a^2|x-C|^2-a(M-1)]z.
\]
The Neumann heat semigroup $T_N(t)$ on this affine-Weyl alcove is
Gaussian convolution of the reflected extension. The folding map is
$1$-Lipschitz, so, for Lipschitz $h$,
\[
 \|\nabla T_N(t)h\|_\infty\le\operatorname{Lip}(h),\qquad
 \|D^2T_N(t)h\|_\infty\le C_Mt^{-1/2}\operatorname{Lip}(h).
\]
The second bound follows by subtracting $h(x)$ in the differentiated
Gaussian integral. The source in the Duhamel equation for $z$ has
Lipschitz norm bounded by a constant times $\|z\|_{C^2}$.
Short-time iteration therefore gives $\nabla z=O(t)$,
$D^2z=O(\sqrt t)$, and $D^2\log w=-2aI+O(\sqrt t)<0$.

For $t>0$ the heat solution is smooth up to the alcove vertices.
To justify this point, use the complete type-A Bethe eigenbasis
\cite{suppEmsiz2010} and the type-A Gaudin--Korepin norm formula
\cite{suppKorepin1982,suppKBI1993}. These are separate inputs: completeness
supplies the spectral expansion, while the norm formula controls its
normalized coefficients. For real Bethe momenta $\boldsymbol k$, set
$\kappa_c(p)=2c/(c^2+p^2)$. The Gaudin matrix is
\[
 G_{ij}=\delta_{ij}\!\left[\mathcal L+
        \sum_{a=1}^M\kappa_c(k_i-k_a)\right]-\kappa_c(k_i-k_j).
\]
For any real vector $\boldsymbol v$,
\[
 \boldsymbol v^{\mathsf T}G\boldsymbol v
 =\mathcal L\sum_i v_i^2+
   \sum_{i<j}\kappa_c(k_i-k_j)(v_i-v_j)^2
 \ge\mathcal L\|\boldsymbol v\|^2,
 \qquad \det G\ge\mathcal L^M.
\]
The center-of-mass eigenvalue is $\mathcal L$; the restriction to the
relative subspace has $M-1$ eigenvalues bounded below by $\mathcal L$. The plane-wave amplitudes have equal modulus;
their common modulus cancels against the corresponding factor in the
norm formula. The remaining normalization contains $\det G$, up to
fixed volume and permutation factors. Thus each normalized eigenfunction
has derivatives of order $j$ bounded by
$C_{M,\mathcal L,c,j}(1+|\boldsymbol k|)^j$.
The bounded scattering phases in the Bethe equations give polynomial
spectral counting. The differentiated heat series consequently converges
uniformly on the closed alcove for times bounded away from zero.

\emph{Hessian maximum principle.}
Put $F=\log w$ and $H=D^2F$. In the interior,
\[
 H_t=\tfrac12\Delta H+\nabla F\cdot\nabla H+H^2.
\]
On a flat face, $\partial_nF=-\alpha$ gives $H_{nv}=0$ and
$\partial_nH_{vv'}=0$ for tangent vectors $v,v'$.
Differentiating $F_t=(\Delta F+|\nabla F|^2)/2$ normally also gives
$\partial_nH_{nn}=2\alpha H_{nn}$.
For the positive spectral part $H_+$, let $Z=\Tr(H_+^2)$.
The convex spectral chain rule, understood by smooth approximation,
yields
\[
 Z_t\le\tfrac12\Delta Z+\nabla F\cdot\nabla Z+2\|H\|Z,
 \qquad \partial_nZ\le4\alpha Z.
\]
Multiply by $Z$ and integrate over $\mathcal A$.
The trace inequality
$\|Z\|_{L^2(\partial\mathcal A)}^2\le
\epsilon\|\nabla Z\|_{L^2(\mathcal A)}^2+
C_\epsilon\|Z\|_{L^2(\mathcal A)}^2$
absorbs the face terms; smoothness on the closed alcove excludes a
vertex flux. On any compact positive-time interval, Gronwall's
inequality propagates $Z=0$ from the short-time result.
Ground-state projection as $t\to\infty$ therefore proves
$D^2\log\Psi_0\le0$ on $\mathcal A$.
Translation invariance restores the center-of-mass direction with zero
curvature. This is a special-alcove argument, not a claim for arbitrary
Robin domains.

\emph{Marginalization and the thermodynamic limit.}
For a void of full length $0<L<\mathcal L$, write
\[
 P_M(L)=M!\!\int_{0<x_1<\cdots<x_M<\mathcal L-L}
                  \Psi_0(x)^2\,\dd^Mx,\qquad S_M(L)=-\log P_M(L).
\]
The integration domain is jointly convex in $(x,L)$ and lies in the
ordered alcove with its center-of-mass direction restored.
The integrand, extended by zero outside this domain, is jointly
log-concave. Pr\'ekopa's marginal theorem \cite{suppPrekopa1973} makes
$P_M(L)$ log-concave and hence $S_M(L)$ convex.
Pointwise thermodynamic convergence at fixed $L,c,\rho=M/\mathcal L$
preserves convexity. Since $L=2\ell/\rho$, this proves convexity of
the action $S(\ell;\gamma)$ used in the Letter.

The endpoint response also has an ordinary derivative limit.
At finite volume, differentiating the two occupied endpoints separately
gives $n_M=S_M'$ and $S_M''=n_M^2-C_{{\rm opp},M}$, where
$C_{{\rm opp},M}\ge0$ is the conditional factorial pair density at
opposite endpoints. Thus $0\le S_M''\le n_M^2$.
Convex secants and convergence of $S_M$ bound $n_M$ on compact
subintervals; this differential inequality makes the sequence locally
equicontinuous. It follows that $n_M$ converges locally uniformly to
the derivative of the limiting action.

To identify this derivative with the ordinary spatial boundary trace,
keep $L=2R$ fixed and let $\Omega$ be the occupied arc on the ring.
For $x\in\Omega$, define the conditional density and derivative norm
\[
 \nu_M(x)=\frac{M}{P_M(L)}\int_{\Omega^{M-1}}
             |\Psi_0(x,X)|^2\,\dd X,\qquad
 k_M(x)=\frac{M}{P_M(L)}\int_{\Omega^{M-1}}
             |\partial_x\Psi_0(x,X)|^2\,\dd X.
\]
The spectator domain is independent of $x$. Enlarging it to the full
ring and using translation invariance of the unconditioned state gives
\[
 \nu_M(x)\le\frac{\rho_M}{P_M(L)},\qquad
 k_M(x)\le\frac{2T_M}{\mathcal L P_M(L)},\qquad
 |\partial_x\nu_M(x)|\le2\sqrt{\nu_M(x)k_M(x)},
\]
where $\rho_M=M/\mathcal L$ and $T_M$ is the unconditioned total
kinetic energy. The derivative inequality holds almost everywhere by
Cauchy--Schwarz and gives a continuous one-sided representative.
Positivity of the interaction and the normalized constant trial wave
give
\[
 \frac{T_M}{\mathcal L}\le\frac{E_M}{\mathcal L}
 \le\frac{cM(M-1)}{2\mathcal L^2},\qquad
 |\partial_x\nu_M(x)|\le
       \frac{2\sqrt{c\rho_M^3}}{P_M(L)}.
\]
Here no derivative acts on the empty-interval projector: the bounds
refer to the occupied side. At each fixed $R$, the thermodynamic limit
$P_M(2R)\to P_0(R)>0$ makes this a volume-independent Lipschitz bound.
With the occupied coordinate increasing from $R$, it implies
$|\nu_M(R+s)-n_M(2R)|\le C_Rs$ for sufficiently small $s\ge0$.
Consequently any thermodynamic density limit has the same continuous
boundary trace as the limit of $n_M$; a shrinking spatial boundary
layer cannot change that trace. The constant $C_R$ need not be uniform
as $R\to\infty$, since the thermodynamic limit is taken first.
Thus, in the Letter's variables,
$n_R(R^+)=(\rho/2)\partial_\ell S$.

\paragraph{Pointwise contact from convexity.}
Weyl-alcove log-concavity and Pr\'ekopa's theorem imply convexity of the
thermodynamic emptiness action, as proved above. For $0<\varepsilon<1$,
\begin{equation}
 \frac{S(\ell)-S((1-\varepsilon)\ell)}{\varepsilon\ell}
 \le S'_-(\ell)\le S'_+(\ell)
 \le\frac{S((1+\varepsilon)\ell)-S(\ell)}{\varepsilon\ell}.
 \label{phys:secants}
\end{equation}
Divide by $2f\ell$, take $\ell\to\infty$, and then
$\varepsilon\downarrow0$. Both one-sided derivatives converge to
$2f\ell$ in relative units. The boundary-response identity therefore yields
\begin{equation}
 \frac{n_R(R^+)}{f\rho^2R}\longrightarrow1.
 \label{phys:pointwise-contact}
\end{equation}
This leading pointwise limit uses only the quadratic rate and convexity.
A termwise contact expansion through order $\ell^{-1}$ additionally
requires a differentiable asymptotic remainder $r_\gamma$ with
$r_\gamma'(\ell)=o(\ell^{-1})$. With that regularity, the expansion is
\begin{equation}
 n_R(R^+)=f\rho^2R+
 \frac{\zeta(3/2)}{2\sqrt\pi}\sqrt{\frac f\gamma}
 \frac{\rho}{\sqrt\ell}-\frac{\rho}{8\ell}+o(\ell^{-1}).
 \label{phys:contact-smooth-remainder}
\end{equation}
Here $r_\gamma$ denotes the bounded remainder in the full action
expansion; Eq.~\eqref{phys:pointwise-contact} is independent of this
additional derivative condition.

\paragraph{A weak-coupling scale for the physical field.}
Set $\lambda_*=d_*^{-1}$ and
$\psi_*(X)=\psi(R+d_*X)/\sqrt{n_*}$.  The definitions and the canonical
Bose commutator give
\begin{equation}
 \begin{gathered}
 N_*=\lambda_*/c,\qquad
 \gamma_*:=c/n_*=\frac{\gamma}{f\ell}=N_*^{-2},\\
 [\psi_*(X),\psi_*^\dagger(Y)]=N_*^{-1}\delta(X-Y).
 \end{gathered}
 \label{phys:weak-coupling}
\end{equation}
The small commutator is $N_*^{-1}=\sqrt{\gamma_*}$.
These scales motivate a classical-field description of the boundary region.
Determining a spatial profile and identifying it with the Lieb--Liniger
mean requires additional control of the exterior sources and the local
field limit; the scalar asymptotics alone do not supply that identification.

\paragraph{Detection, trapping, and transverse confinement.}
Independent detection with efficiency $0<\eta_{\rm det}<1$ measures
$P_{\eta_{\rm det}}(0)=\langle(1-\eta_{\rm det})^{\widehat N_{[-R,R]}}\rangle$.
Since $\langle\widehat N_{[-R,R]}\rangle=2\ell$, Jensen's inequality gives
\begin{equation}
 -\log P_{\eta_{\rm det}}(0)\le -2\ell\log(1-\eta_{\rm det}).
 \label{phys:efficiency}
\end{equation}
Fixed nonunit efficiency therefore loses the quadratic hard-gap rate as
$\ell\to\infty$.  A necessary condition for retaining that rate is
$-\log(1-\eta_{\rm det})\gtrsim f\ell/2$; sufficiency also requires bounds on
nonzero-count probabilities.  The thinned sine-process crossover
\cite{suppBothnerDeiftItsKrasovsky2018} concerns a distinct counting observable.

Near a symmetric, slowly varying trap center, the scale estimate
$\delta S=O\!\left(\ell^2(R/R_{\rm trap})^2\right)$ requires
$\rho R_{\rm trap}\gg\ell^{7/4}$ to resolve the $\sqrt\ell$ correction.
The homogeneous result does not fix the curvature coefficient.
In an atomic waveguide \cite{suppOlshanii1998}, the interaction energy at the
response scale must remain below transverse excitation.  Restoring units,
the applicable scale conditions are
\begin{equation}
 \max(c,q)\ll\lambda_*\ll\sqrt{m\omega_\perp/\hbar},
 \qquad
 \ell\ll\frac{m\omega_\perp}{\hbar\rho^2\gamma f(\gamma)}.
 \label{phys:waveguide}
\end{equation}
These conditions delimit access to the one-dimensional asymptotic regime;
they are separate from quadrature and auxiliary Gaussian sampling errors.

\section{Finite-window Gaussian determinant measurements}
\label{sec:numerics}

We measure finite-interval probabilities by sampling the QISM
ordinary-Gaussian representation. The analytic coefficient evaluations
are described separately in Section~\ref{sec:charge}. All measurements
use \(\rho=1\), hence \(c=\gamma\), \(R=\ell\), and \(L=2\ell\).
The calculation solves the Lieb equation at the specified coupling
and discretizes the thermodynamic rapidity integral on \([-q,q]\).
Its node count \(n\) is a quadrature size, not a particle number.
The adopted identity, its Gaussian realization and order of limits are specified in
Sec.~\ref{main:framework}, including the convergence proof in
Sec.~\ref{env:section}.

\subsection{Estimator, sampling, and numerical controls}

Let \(\lambda_j,w_j\) be Lebesgue Gauss--Legendre nodes and weights.
The denominator is evaluated separately:
\begin{equation}
 \mathcal N_c=\det\!\left[
 \delta_{jk}-\frac{\sqrt{w_jw_k}\,c}
 {\pi[c^2+(\lambda_j-\lambda_k)^2]}\right].
 \label{eq:SnumGaudin}
\end{equation}
Its continuum limit is \(\det(I-P_c)\).
For \(m=1,\ldots,M_{\rm feat}\), define the two real features
\begin{equation}
 b_{2m-1}(\lambda)=\sqrt{2/m}\,\Re w_c(\lambda)^m,
 \qquad
 b_{2m}(\lambda)=\sqrt{2/m}\,\Im w_c(\lambda)^m,
 \quad w_c(\lambda)=\frac{\lambda}{c-\ii\lambda}.
 \label{eq:Snumfeatures}
\end{equation}
The \(M_{\rm feat}\) feature pairs provide \(2M_{\rm feat}\) real Gaussian
coordinates and realize the truncated anchored covariance in
\eqref{main:covariance}. Field values and analytic derivatives use
the same coordinates. With a fixed real contour guide \(\mathbf{a}\)
and a real sample \(\mathbf{z}\), write
\(r_j=\sum_\nu b_\nu(\lambda_j)z_\nu\) and
\(p_j=\sum_\nu b_\nu(\lambda_j)a_\nu\), and similarly \(r'_j,p'_j\).
For \(j\ne k\), put \(d=\lambda_j-\lambda_k\),
\(u=(r_j-r_k)/2\), \(\theta=(Ld+p_j-p_k)/2\). The full matrix is
\begin{align}
 \mathsf M_{jk}&=-\frac{\sqrt{w_jw_k}\,c}{\pi d(d^2+c^2)}
 \left\{\cosh u[d\cos\theta+c\sin\theta]\right.\nonumber\\
 &\hspace{33mm}\left.+\ii\sinh u[d\sin\theta-c\cos\theta]\right\},
 \nonumber\\
 \mathsf M_{jj}&=1-\frac{w_j}{2\pi}(L+p'_j+2/c)
                       +\ii\frac{w_j}{2\pi}r'_j.
 \label{eq:SnumMatrix}
\end{align}
The two exponential scattering channels are both present: the braces
come from the sum of \(e^{u+\ii\theta}/(d+\ii c)\) and
\(e^{-u-\ii\theta}/(d-\ii c)\). The diagonal is their analytic
confluent limit, including the derivative and \(2/c\) term.
The matrix is complex symmetric under transpose, not Hermitian.
We use packed symmetric elimination to store and evaluate this matrix,
with matched-field comparisons against general LU recorded separately.

If the proposal precision is \(\mathsf H=\mathsf T\mathsf T^{\mathsf T}\)
with lower triangular \(\mathsf T\), draw a standard normal
\(\mathbf{y}\) and solve \(\mathbf{z}=\mathsf T^{-\mathsf T}\mathbf{y}\).
Untransformed additional coordinates have identity precision. The weight is
\begin{align}
 W(\mathbf{y})&=\frac{\det\mathsf M(\mathbf{z},\mathbf{a})}{\mathcal N_c}
    e^{\|\mathbf{a}\|^2/2-\ii\mathbf{a}\cdot\mathbf{z}}
    J(\mathbf{y}),\nonumber\\
 J(\mathbf{y})&=\frac1{\det\mathsf T}
       \exp\!\left[-\frac{\|\mathbf{z}\|^2-\|\mathbf{y}\|^2}{2}\right].
 \label{eq:SnumWeight}
\end{align}
The probability estimator is the sample mean of \(\Re W\), including
the full complex determinant, phase and density correction. The weight
is used directly, without taking its modulus or self-normalizing the
importance ratio. We monitor the imaginary mean as a diagnostic.
Exact compensation leaves the expectation independent of the guide,
which affects the variance. The sampled observable uses no asymptotic
coefficient as input.
The numerical evaluation uses logarithmic scaling to handle exponentially
small probabilities. The accompanying package documents the stored-weight
scaling and reproduces the fits, tables and Fig.~2 from the measured costs
and empirical covariance matrices.

\paragraph{Measurement set and covariance.}

We evaluate the probability at $\gamma=0.25,0.5,1,2,10$ and
$\ell=64,80,96,112,128,160,192,224,256$. Table~\ref{tab:numsettings} summarizes the observation counts and numerical
parameters. The accompanying data provide per-window observation counts,
measured costs, empirical covariance matrices, and the numerical settings
used at each coupling. The free and conditional fits use this common
measurement set.

\begin{table}[htbp]
\centering\small
\caption{Measurement settings across the nine windows. Observation counts are per window; shared fields contribute to cross-window covariance. Per-window counts and the settings used at each coupling are supplied with the data. Matched-field refinement tests use separate settings.}
\label{tab:numsettings}
\begin{tabular}{rrrrr}
\toprule
$\gamma$ & Observations/window & Nodes $n$ & Pairs $M_{\rm feat}$ & Precision (bits)\\
\midrule
0.25 & 5--52 & 1152--4608 & 768, 1024 & 1024--4096\\
0.5 & 9--228 & 768--3072 & 192 & 1024--2560\\
1 & 10--104 & 512--2048 & 96 & 768--2304\\
2 & 22--96 & 384--1536 & 96 & 512--2048\\
10 & 32--48 & 320--1280 & 48 & 640--2176\\
\bottomrule
\end{tabular}
\end{table}

Field identity is determined by the Gaussian stream and its position.
Repeated evaluations of the same field for numerical controls do not add
independent observations. At $\gamma=0.25,2,10$, different windows use
independent field pools. At $\gamma=0.5,1$, shared fields require the full
cross-window covariance. The number of observations summed over windows
therefore need not equal the number of distinct Gaussian fields.

\begin{samepage}
For an availability block $h$ containing $n_h\ge2$ common fields, let
$V_{hi}$ denote the rescaled real weight at window $i$, $N_i$ its total
sample count, and $\bar V_i$ its mean. The delta-method contribution is
\begin{equation}
 (\mathsf V_S)_{ij}^{(h)}
 =n_h\,\widehat{\operatorname{Cov}}_h
 \left(\frac{V_{hi}}{N_i\bar V_i},
       \frac{V_{hj}}{N_j\bar V_j}\right).
 \label{eq:SnumCovariance}
\end{equation}
\end{samepage}
The block sum is positive semidefinite. For independent per-window pools
we use the pooled IID variance. For a linear contrast
$\mathbf{w}^{\mathsf T}\mathbf{S}$, the propagated variance is
$\mathbf{w}^{\mathsf T}\mathsf V_S\mathbf{w}$.

\paragraph{Numerical controls and error scope.}

Matched-field comparisons vary the rapidity quadrature, arithmetic
precision, Gaussian feature count and normalization quadrature while
holding the Gaussian input fixed. They compare the complete compensated
complex weights in Eq.~\eqref{eq:SnumWeight}. For example, at
$\gamma=0.25$, $\ell=256$, a joint refinement from $(n,M_{\rm feat},b)
=(4608,1024,4096)$ to $(5120,1536,4608)$, where $b$ is the arithmetic
precision in bits, also increases the equation and normalization
quadratures from 96 to 128 nodes. On two matched fields the maximum
relative complex-weight difference is below $9.4\times10^{-20}$.
At $\gamma=10$, separate node and precision comparisons use two matched
fields at each primary window, with maximum relative differences below
$9\times10^{-34}$ and $3\times10^{-42}$, respectively.

For the Gaussian field alone, put $r=q/\sqrt{c^2+q^2}<1$. The omitted
feature variance satisfies
\begin{equation}
 \sup_\lambda\operatorname{Var}\delta\phi_{M_{\rm feat}}(\lambda)
 \le\frac{2r^{2M_{\rm feat}+2}}{(M_{\rm feat}+1)(1-r^2)}.
 \label{eq:SnumFieldTail}
\end{equation}
This bound controls the omitted Gaussian variance.
At fixed finite $n,M_{\rm feat},L$, a sufficient condition for the
$p$th absolute importance moment is $pI-(p-1)\mathsf H>0$: the
determinant is an exponential polynomial with derivative factors, and
this positive quadratic form dominates its linear exponential growth.
Gaussian-input precision and determinant arithmetic precision are
separate numerical parameters.

The quoted statistical errors propagate the empirical MC covariance.
Matched-field refinements characterize quadrature and precision sensitivity,
and the window comparisons below characterize sensitivity to the finite-window
fitting form.

\subsection{Free coefficient fits}

To compare the leading coefficients without fixing them to theory,
we fit all nine windows to
\begin{equation}
 S_{\rm fit}(\ell)=f\ell^2+A\ell+B\sqrt\ell+
 C_{\log}\log\ell+C_0+C_1/\sqrt\ell.
 \label{eq:SnumPrimary}
\end{equation}
All six coefficients are free. We apply Cholesky whitening with the complete
empirical covariance, scale the design columns, and solve the resulting
least-squares problem by 80-digit QR. No covariance floor or
residual-based rescaling of the statistical errors is applied.
Table~\ref{tab:freefits} gives the results. Strong coefficient
correlations and three residual degrees of freedom limit their use as
asymptotic estimates.

\begin{table}[htbp]
\centering\small
\caption{Analytic coefficients and nine-window fits with all six coefficients free. The analytic linear coefficient is zero. Parentheses give empirical MC standard errors conditional on the fit form and covariance. The full fit, including the other three coefficients, is supplied with the data.}
\label{tab:freefits}
\begin{tabular}{rrrrrrr}
\toprule
$\gamma$ & $f_{\rm th}$ & $\widehat f$ & $\widehat A$ & $B_{\rm th}$ & $\widehat B$ & $\chi^2/\mathrm{dof}$\\
\midrule
0.25 & 0.808708919 & $0.80870957(99)$ & $-0.0017(28)$ & 5.301719520 & $5.38(13)$ & $4.11/3$\\
0.5 & 1.119352695 & $1.11935290(24)$ & $-0.00057(71)$ & 4.410513022 & $4.437(34)$ & $2.45/3$\\
1 & 1.533113484 & $1.533113567(61)$ & $-0.00024(17)$ & 3.649873293 & $3.6606(80)$ & $7.6/3$\\
2 & 2.064860990 & $2.064860832(42)$ & $0.00047(12)$ & 2.995167042 & $2.9729(57)$ & $1.59/3$\\
10 & 3.588505462 & $3.5885054537(25)$ & $0.0000324(73)$ & 1.765824109 & $1.76390(35)$ & $6.9/3$\\
\bottomrule
\end{tabular}
\end{table}

The analytic comparison values in Table~\ref{tab:freefits} are
computed from the image solution and homogeneous Lieb normalization,
without fitting the measured probabilities. The Lieb quadrature was
checked at 64, 128, 192 and 384 nodes, and the elliptic evaluation at
80 and 120 decimal digits, providing convergence checks of both inputs.

\subsection{Logarithmic extraction and window dependence}

We hold the analytic $f$, $A=0$ and $B$ fixed and fit
\begin{equation}
 S(\ell;\gamma)-f\ell^2-B\sqrt\ell
 =C_{\log}\log\ell+C_0+C_1/\sqrt\ell
 \label{eq:logwindow160}
\end{equation}
at $\ell=160,192,224,256$. The three coefficients on the right are
free, leaving one residual degree of freedom. The fit uses the same
measurement set and covariance as the nine-window analysis, restricted
to these four windows. This conditional fit isolates the logarithmic
coefficient after subtracting the analytic bulk and square-root terms.

\begin{table}[htbp]
\centering\small
\caption{Conditional logarithmic estimates with analytic $f,A=0,B$ fixed. The primary fit uses $\ell=160,192,224,256$; its MC standard error and residual statistic are shown. The final column spans the three alternative window selections specified in the text and is a sensitivity diagnostic, not an uncertainty interval.}
\label{tab:logprimary}
\begin{tabular}{rrrrr}
\toprule
$\gamma$ & $\widehat C_{\log}$ & MC standard error & $\chi^2/\mathrm{dof}$ & Alternative-window range\\
\midrule
0.25 & $-0.2445783$ & $0.00307$ & $0.007447/1$ & $[-0.2437012,-0.2412334]$\\
0.5 & $-0.2450564$ & $0.00112$ & $0.208/1$ & $[-0.2459388,-0.2458638]$\\
1 & $-0.2483998$ & $1.7\times10^{-5}$ & $0.817/1$ & $[-0.2484914,-0.2484179]$\\
2 & $-0.2493038$ & $9.82\times10^{-5}$ & $2.934/1$ & $[-0.2490374,-0.2488180]$\\
10 & $-0.2491556$ & $5.37\times10^{-6}$ & $16.35/1$ & $[-0.2490631,-0.2489346]$\\
\bottomrule
\end{tabular}
\end{table}

The errors in Table~\ref{tab:logprimary} are conditional on the fixed
leading coefficients and fitting form. At $\gamma=10$, the residual is
large relative to its MC error despite the fitted center being near $-1/4$.

The final column of Table~\ref{tab:logprimary} summarizes three alternative
window selections, $[128,256]$, $[112,256]$ and $[112,224]$, retaining
all measured windows between the endpoints. The first two test the influence
of shorter intervals; the third also removes the largest window.
The fit form, analytic leading coefficients and covariance prescription
remain fixed. The shared observations make these comparisons correlated.
The individual estimates, MC errors and residual statistics are supplied
with the data.

At $\gamma=10$, the four-window estimate of $C_{\log}$ has propagated
MC standard error $5.37\times10^{-6}$. This statistical precision makes
finite-window corrections visible: extending the range to $[112,256]$
gives $\chi^2/\mathrm{dof}\simeq1006/3$ for the three-parameter residual
form in Eq.~\eqref{eq:logwindow160}. The window dependence therefore
resolves structure beyond that truncation. The reported logarithmic
comparison uses $\ell=160,192,224,256$; the alternative windows quantify
its sensitivity to the retained asymptotic terms.

\FloatBarrier
\renewcommand{\refname}{References}